\documentclass[11pt,reqno]{article}
\usepackage{amsmath}
\usepackage{amsthm}
\usepackage{amssymb}
\usepackage{amsfonts}
\usepackage{cases}
\usepackage{graphicx}
\usepackage{xcolor}
\usepackage[margin=1in]{geometry}
\usepackage[utf8]{inputenc}
\usepackage[margin=1in]{geometry}
\usepackage{verbatim}
\usepackage{caption,subcaption}
\usepackage{algorithm}
\usepackage{algpseudocode}
\usepackage[normalem]{ulem}
\makeatletter
\renewcommand{\fnum@algorithm}{\fname@algorithm}
\makeatother
\usepackage[colorlinks,citecolor=blue,urlcolor=blue]{hyperref}
\usepackage{bm}

\numberwithin{equation}{section}
\newtheorem{definition}{Definition}[section]
\newtheorem{remark}{Remark}[section]
\newtheorem{theorem}{Theorem}[section]
\newtheorem{lemma}{Lemma}[section]
\newtheorem{proposition}{Proposition}[section]
\newtheorem{corollary}{Corollary}[section]

\newcommand{\be}{\begin{equation}}
\newcommand{\ee}{\end{equation}}
\newcommand{\bee}{\begin{equation*}}
\newcommand{\eee}{\end{equation*}}
\newcommand{\bi}{\begin{itemize}}
\newcommand{\ei}{\end{itemize}}

\usepackage{algorithm}
\usepackage{algpseudocode}
\usepackage{soul}%delete line

\def \E{\mathbb{E}}
\def \F{\mathbb{F}}
\def \N{\mathbb{N}}
\def \P{\mathbb{P}}
\def \Q{\mathbb{Q}}
\def \R{\mathbb{R}}

\def \Fc{{\mathcal F}}

\newcommand{\EE}{\mathbb{E} }

\newcommand{\PP}{\mathbb{P} }

\newcommand{\cF}{\mathcal{F} }

\newcommand{\eal}{\end{align}}
\newcommand{\bal}{\begin{align}}
\newcommand{\ea}{\end{eqnarray}}
\newcommand{\ba}{\begin{eqnarray}}
\newcommand{\ean}{\end{eqnarray*}}
\newcommand{\ban}{\begin{eqnarray*}}

\newcommand{\en}{\end{equation}}

\usepackage{graphicx}
\graphicspath{{./fig/}}

\usepackage{bm}

\title{Partial Information in a Mean-Variance Portfolio Selection Game\thanks{We would like to thank the Associate Editor and two anonymous reviewers, whose comments helped improve the quality of this paper significantly.}}%{Partial Information Breeds Systemic Risk}
\author{Yu-Jui Huang\thanks{
Department of Applied Mathematics, University of Colorado, Boulder, CO 80309-0526, USA, email: \texttt{yujui.huang@colorado.edu}. Partially supported by National Science Foundation (DMS-2109002).
}
\and 
Li-Hsien Sun\thanks{
           Graduate Institute of Statistics, National Central University, Taiwan, email: \texttt{lihsiensun@ncu.edu.tw}. Partially supported by National Science and Technology Council (NSTC-112-2628-M-008-002-MY3).
}
}
\begin{document}

\maketitle
\begin{abstract}
This paper considers finitely many investors who perform mean-variance portfolio selection under relative performance criteria. That is, each investor is concerned about not only her terminal wealth, but how it compares to the average terminal wealth of all investors. % (i.e., the mean field). 
At the {\it inter-personal} level, each investor selects a trading strategy in response to others' strategies. % which affect the mean field. 
This selected strategy additionally needs to yield an equilibrium {\it intra-personally}, so as to resolve time inconsistency among the investor's current and future selves (triggered by the mean-variance objective). A Nash equilibrium we look for is thus a tuple of trading strategies under which every investor achieves her intra-personal equilibrium simultaneously. We derive such a Nash equilibrium explicitly in the idealized case of full information (i.e., the dynamics of the underlying stock is perfectly known) and semi-explicitly in the realistic case of partial information (i.e., the stock evolution is observed, but the expected return of the stock is not precisely known). The formula under partial information consists of the myopic trading and intertemporal hedging terms, both of which depend on an additional state process that serves to filter the true expected return and whose influence on trading is captured by a degenerate Cauchy problem. %, one of which depends on the other. %, whose solutions are constructed by elliptic regularization and a stability analysis of the state process. 
Our results identify that relative performance criteria can induce {\it downward self-reinforcement} of investors' wealth---if every investor suffers a wealth decline simultaneously, then everyone's wealth tends to decline further. This phenomenon, as numerical examples show, is negligible under full information % (whence unobserved in the literature), it is 
but pronounced under partial information. %This suggests that fund managers' consideration of relative performance may exacerbate fire-sale spillovers.  
%indicate that partial information {\it alone} can reduce investors' wealth significantly, thereby causing or aggravating systemic risk. Intriguingly, in two different scenarios of the expected return (i.e., it is constant or alternating between two values), our Nash equilibrium formula spells out two distinct manners systemic risk materializes.  
\end{abstract}

\textbf{MSC (2020):} 
91G45, %Financial networks (including contagion, systemic risk, regulation)
91A80, %Applications of game theory
93E11 %Filtering in stochastic control theory [See also 60G35]
%91G80.
%60G40,  	%Stopping times; optimal stopping problems; gambling theory
%60J45,  	%Probabilistic potential theory
%93E20,  % Optimal stochastic control
%60H10, %Stochastic ordinary differential equations
%94A17 %Measures of information, entropy
%91A13, % Games with infinitely many players
%49K21. % Optimality conditions:	Problems involving relations other than differential equations
%60J05,  %	Discrete-time Markov processes on general state spaces
\smallskip


\section{Introduction}
Investors in practice make trading decisions under partial information: they observe the evolution of a stock, but do not know the precise stock dynamics. When evaluating their trading decisions, investors commonly compare their investment performance relatively to others'. This paper strives to elucidate how the blending of these two, partial information and relative evaluation, changes each investor's trading decision and affects the overall wealth accumulation of all investors.

We consider $N\in\N$ investors (e.g., fund managers) who trade a stock $S$ on a finite time horizon $T>0$, subject to a {\it relative performance criterion} in the spirit of Espinosa and Touzi \cite{Touzi2015} and Lacker and Zariphopoulou \cite{Lacker-Zari2017}. That is, when choosing her trading strategy $\pi_i$, investor $i$ ($i=1,...,N$) is concerned about not only her terminal wealth $X_i(T)$, but also how it compares to the average wealth of all investors $\overline X(T):=\frac{1}{N} \sum_{i=1}^N X_i(T)$ (i.e., the relative performance), thereby considering the mixed performance $\mathcal P_i(T) := (1-\lambda_i)X_i(T)+\lambda_i(X_i(T)-\overline X(T))= X_i(T) -\lambda_i\overline X(T)$ for some $\lambda_i\in[0,1)$. In \cite{Touzi2015} and \cite{Lacker-Zari2017}, each investor chooses $\pi_i$ by maximizing expected utility of $\mathcal P_i(T)$ given others' strategies $\{\pi_j\}_{j\neq i}$, and a resulting Nash equilibrium $(\pi^*_1,\pi^*_2,...,\pi^*_N)$ for this $N$-player game is derived. By contrast, we assume that each investor selects $\pi_i$ under a mean-variance objective associated with $\mathcal P_i(T)$; see \eqref{MV_model} below. 

%Demarzo et al. \cite{DKK2008} 
%Markowitz \cite{Mark1952} where the relative concerns parameters can be varied in the mean and variance respectively. 

As a mean-variance objective induces time inconsistency, what constitutes a Nash equilibrium is nontrivial. As explained in Huang and Zhou \cite{HZ22}, in a game where players have time-inconsistent preferences, two levels of game-theoretic reasoning are intertwined. At the {\it inter-personal} level, each player selects an action in response to other players' chosen strategies. The selected action, importantly, has to be an equilibrium at the {\it intra-personal} level, so as to resolve time inconsistency among this player's current and future selves. Hence, we say $(\pi_1^*,\pi^*_2,...,\pi^*_N)$ is a Nash equilibrium, if for every $i=1,...,N$, $\pi^*_i$ is investor $i$'s intra-personal equilibrium given others' strategies $\{\pi^*_j\}_{j\neq i}$---namely, every player achieves her  intra-personal equilibrium simultaneously (Definition~\ref{def:E}). 
%This amounts to a ``{\it soft} inter-personal equilibrium'' in \cite{HZ22}, while the stronger notion ``{\it sharp} inter-personal equilibrium'' therein cannot be easily defined for a mean-variance objective; see the discussion below Definition~\ref{def:E}. 

We focus on investigating how a Nash equilibrium $(\pi^*_1,\pi^*_2,...,\pi^*_N)$ changes from the  idealized case of full information (where investors know the precise dynamics of $S$) to the realistic one of partial information (where investors only observe the evolution of $S$). %Note that it is the latter case that reflects the reality: investors in practice do not know the precise dynamics of $S$, but attempt to infer that from the observed evolution of $S$. As we will see, this change from an idealized case to a realistic one can drastically enlarge systemic risk.  
For concreteness, we assume that under partial information, investors do not know precisely the expected return $\mu$ of $S$, except that it has two possible values $\mu_1,\mu_2\in\R$ (with $\mu_1>\mu_2$). Our analysis breaks into two distinct scenarios: (i) $\mu\in\R$ is a fixed constant, and investors attempt to infer the true value of $\mu$ between $\mu_1$ and $\mu_2$; %(Section~\ref{sec:constant mu}); 
(ii) $\mu$ is a continuous-time Markov chain, and investors attempt to infer the recurring changes of $\mu(t)$ between $\mu_1$ and $\mu_2$. % (Section~\ref{sec:alternating mu}).
The first scenario applies to the stock of a company with unreported innovation (Koh and Reeb \cite{KR15}),  for which investors need to infer if an innovation has raised the expected  return from $\mu_2$ to $\mu_1$. The second scenario, on the other hand, models the repeated shifts between a bull and a bear market.

As baselines of our analysis, Theorem~\ref{thm:E mu known} (resp.\ Theorem~\ref{thm:E M known}) presents an explicit Nash equilibrium $(\pi^*_1,\pi^*_2,...,\pi^*_N)$ under full information (i.e., when $\mu$ is perfectly known). These results are already new to the literature. The former generalizes the extended Hamilton-Jacobi-Bellman (HJB) equation in Bj\"{o}rk et al.\ \cite[Section 10]{BKM2016}, derived for the classical mean-variance problem of one single agent, to the current multi-player setting. The latter further integrates the regime-switching framework of \cite{Zhou2003} into our multi-player extended HJB equation. 

When $\mu\in\R$ is an unknown constant, investors compute the probability of $\mu=\mu_1$ conditioned on the observed evolution of $S$ by the current time $u\in [0,T]$; see \eqref{p_j} below. By the theory of nonlinear filtering, this posterior probability, denoted by $P(u)$, satisfies a stochastic differential equation (SDE), i.e., \eqref{P} below, that involves only known model parameters (such as $\mu_1,\mu_2\in\R$); see Lemma~\ref{lem:hW}. This allows the original dynamics of $S$, which involves the unknown $\mu$, to be expressed equivalently in terms of the observable process $P$, so that portfolio selection can be made based on the joint observation $(s,p)=(S(u), P(u))$. The additional state process $P$ complicates the search for a Nash equilibrium. To fully capture the dependence on $p=P(u)$, we need the solutions to two Cauchy problems, one of which depends on the derivative of the other; see \eqref{Cauchy} and \eqref{Cauchy'} below (with $\eta\equiv 0$). As the first Cauchy problem is degenerate, we approximate it by uniformly elliptic equations, following Heath and Schweizer \cite{HS00}, to show that it has a unique classical solution with a concrete stochastic representation. By analyzing the $L^2$-derivative of the process $P$ with respect to (w.r.t.) its initial value $P(u)=p$ (introduced in Friedman \cite{Friedman-book-75}), we further find that the partial derivative of the unique classical solution w.r.t.\ $p$ is bounded and also admits a concrete stochastic representation; see Lemma~\ref{lem:Cauchy} and its proof in Section \ref{subsec:proof of lem:Cauchy}. For the second Cauchy problem, also degenerate, a unique solution and its stochastic representation are constructed by similar arguments (Corollary~\ref{lem:Cauchy'}). Ultimately, by solving the extended HJB equation for the $N$ investors, which is a coupled system of $2N$ partial differential equations (PDEs), we obtain a Nash equilibrium $(\pi^*_1,\pi^*_2,...,\pi^*_N)$ and the corresponding value functions in semi-explicit forms, in terms of the solutions to the two Cauchy problems; see Theorem~\ref{thm:E mu unknown} for details. 

In the second scenario where $\mu$ alternates between $\mu_1$ and $\mu_2$ behind the scenes, investors compute the probability of $\mu(u)=\mu_1$ conditioned on the observed evolution of $S$ by the current time $u\in [0,T]$; see \eqref{p_j'} below. Arguments for the previous scenario can then be similarly carried out. Specifically, the above posterior probability, denoted again by $P(u)$, also satisfies an SDE---in fact, it is the one in the previous scenario plus a drift term (i.e., \eqref{P'} below); see Lemma~\ref{lem:tW}. This drift term shows up in the aforementioned Cauchy problems, as an additional coefficient of the first derivative w.r.t.\ $p$. %nonzero $\eta$ function (given by \eqref{eta} below) in the aforementioned Cauchy problems \eqref{Cauchy} and \eqref{Cauchy'}. %, as an additional nonzero $\eta$ function given by \eqref{eta} below. 
Solving the $N$-player extended HJB equation %for the present case $\mu=\mu(M(t))$ 
then yields the same formulas as before (for a Nash equilibrium and the corresponding value functions), %as in the previous scenario, %as in Theorem~\ref{thm:E mu unknown} (for the case of a constant $\mu$), 
except that the involved Cauchy problems contain a new term induced by the added drift; %nonzero $\eta$ function; 
see Theorem~\ref{THM_1} for details. 

Our study admits several interesting financial implications. First, our results extend the interpretation of {\it myopic trading} and {\it intertemporal hedging} in Basak and Chabakauri \cite{BC10}, where the mean-variance problem is studied for a single investor with full information, to the case of multiple investors (connected via relative performance criteria) with partial information. At any time $u\in [0,T]$, the current value $p=P(u)$ of the posterior probability process gives an estimate of the unknown $\mu$, i.e., $\theta(p):=p \mu_1+(1-p)\mu_2$. Myopic trading is an investment decision that (blindly) treats the estimate $\theta(p)$ as the true value of $\mu$, while intertemporal hedging serves to make corrections by hedging against the fluctuations in the estimate (or, the fluctuations in $P(\cdot)$). These two kinds of trading correspond precisely to the two terms of $\pi^*_i$ in \eqref{NE-1} below, the equilibrium trading strategy we derive. Relative performance criteria further modify the size of myopic trading and intertemporal hedging by reshaping investors' risk appetite: each investor now considers not only her own risk aversion coefficient but all other investors'. See the discussions below Remark~\ref{rem:NE-1'} as well as Remark~\ref{re_1} for details.   

Second, the above interpretation suggests that intertemporal hedging is dominant over myopic trading (or vice versa) if the posterior probability process $P(\cdot)$ is volatile (or stable). The two scenarios of $\mu$ offer perfect illustrations. In the case of a constant $\mu$, as $P(\cdot)$ satisfies a driftless SDE (i.e., \eqref{P} below), which may oscillate wildly between 0 and 1, a large intertemporal hedging term is expected to hedge against the strong fluctuations in $P(\cdot)$. In the case of an alternating $\mu$, $P(\cdot)$ instead satisfies a mean-reverting SDE (i.e., \eqref{P'} below). A small intertemporal hedging term is then expected, as the oscillation of $P(\cdot)$ is dampened by its tendency to return to a mean level between 0 and 1. Numerical examples in Section~\ref{subsec:discuss} confirm our intuition: when $\mu$ is constant (resp.\ alternating), intertemporal hedging (resp.\ myopic trading) dominantly dictates the wealth process $X_i$. 
Besides numerical illustrations, we also prove that the intertemporal hedging term contains the aforementioned $L^2$-derivative of $P(\cdot)$ w.r.t.\ the initial value $p=P(t)$, which measures how intensely $P$ will oscillate on $[t,T]$. This allows us to quantitatively estimate intertemporal hedging from a given dynamics of $P(\cdot)$; see the discussion below Remark~\ref{re_2}. 

Third, considering relative performance encourages ``{downward self-reinforcement}'' of investors' wealth, and it is amplified under partial information. As discussed in Section~\ref{subsec:DSR}, the attempt to reduce the variance of $X_i(T)-\lambda_i \overline X(T)$, with $\lambda_i>0$, encourages comovement of $X_i$ and $\overline X$, which in turn suggests {\it self-reinforcement} of investors' wealth: if all wealth processes $\{X_i\}_{i=1}^N$ simultaneously fall (resp.\ rise), such that so does $\overline X$, all of $\{X_i\}_{i=1}^N$ will tend to fall (resp.\ rise) further. Since a sharp simultaneous drop of all investors' wealth is more likely when there is a misbelief of the $\mu$ value among investors, partial information is expected to make {\it downward} self-reinforcement more frequent and intense. Indeed,  Figure~\ref{fig:mu=mu_1_relative} shows that under partial information, investors' wealth can fall sharply simultaneously and the consideration of relative performance further reduces all investors’ wealth significantly; under full information, such sharp simultaneous declines of wealth do not exist and considering relative performance has a negligible effect. To the best of our knowledge, the way partial information and relative performance criteria jointly impact investors' wealth has not been identified before. % spelled out and demonstrated. 

Finally, there are implications for systemic risk worth further investigation. As explained in Shleifer and Vishny \cite{SV11} and Duarte and Eisenbach \cite{DE21}, systemic risk commonly unfolds as fire-sale spillovers. In particular, a mutual fund or a hedge fund resorts to fire sales (i.e., quickly selling securities held in its portfolio) when it faces significant capital withdrawals by its shareholders; see Coval and Stafford \cite{CS07}. 
%The discussion just made on ``downward self-reinforcement'' then suggests that fund managers' consideration of relative performance may exacerbate fire-sale spillovers. Indeed, 
As trading decisions are practically made under partial information, fund managers' consideration of relative performance may markedly reduce fund values due to ``downward self-reinforcement'' just discussed. %, as explained and numerically illustrated in Section~\ref{subsec:DSR}. 
With lower fund values, an initial fire sale can occur more easily, when shareholders of a fund rush to redeem their investment once they see the fund value fall below a certain threshold; the ensuing spillovers are likely more devastating, as other funds are more vulnerable now due to smaller balances. This seemingly suggests that %leads to a somewhat unexpected conclusion: 
fund managers' consideration of relative performance may exacerbate fire-sale spillovers. Certainly, a thorough examination is needed and we point out several future directions at the end of Section~\ref{subsec:systemic risk}. 

The rest of the paper is organized as follows. Section~\ref{sec:setup} introduces the model setup.   Section~\ref{sec:constant mu} focuses on constant investment opportunities (i.e., $\mu$ is constant). For $N$ investors who consider mean-variance objectives under relative performance criteria, we derive a Nash equilibrium under full information (Section~\ref{subsec:full constant mu}) and under partial information (Section~\ref{subsec:partial constant mu}). Section~\ref{sec:alternating mu} focuses on alternating investment opportunities  (i.e., $\mu$ is alternating) and a Nash equilibrium for the $N$ investors is again derived under full information (Section~\ref{subsec:full alternating mu}) and under partial information (Section~\ref{subsec:partial alternating mu}). Section~\ref{sec:discuss} discusses the  financial implications, including myopic trading and intertemporal hedging (Section~\ref{subsec:discuss}), downward self-reinforcement (Section~\ref{subsec:DSR}), and systemic risk (Section~\ref{subsec:systemic risk}). The appendices collect proofs.

%%%%%%%%%%%%%%%%%%%%%%%%%%%%%%%
%%%%%%%%%%%%%%%%%%%%%%%%%%%%%%%

\section{The Setup}\label{sec:setup}
Let $(\Omega,{\cal{F}}, \mathbb F= \{\mathcal F_t\}_{t\ge 0}, {\P})$ be a filtered probability space that supports a standard Brownian motion $W$ and a process $\mu:\Omega\times [0,\infty)\to B$ (which are in general $\F$-adapted), where $B\subseteq\R$ is a Borel set. 
%an $\Fc_0$-measurable $\mu:\Omega\to\R$. 
%is equipped with a filtration $\{\mathcal F_t\}_{t\ge 0}$ satisfying the usual conditions and supports a standard Brownian motion $W$. 
Consider a financial market with a riskfree rate $r\ge0$ and a stock price process $S$ given by  
\ba 
\label{risky_0} dS(u)=\mu(u) S(u)du+\sigma S(u) dW(u),\ \ u\ge 0,\quad S(0)=s>0,%\\
%\label{riskless_1}dS_0(u)&=&rS_0(u)du,\; S_0(t)=s_0,
\ea
where $\sigma, s>0$ are given constants. Let %$\mathbb F^\mu := \{\Fc^\mu_t\}_{t\ge 0}$, 
$\mathbb F^S := \{\Fc^S_t\}_{t\ge 0}$ (resp.\ $\mathbb F^{\mu,S} := \{\Fc^{\mu,S}_t\}_{t\ge 0}$) be the natural filtration generated by $S$ (resp.\ by both $\mu$ and $S$). 

Given a fixed time horizon $T>0$, suppose that there are $N\in\N$ investors (e.g., fund managers) trading the stock $S$. For each $i=1,...,N$, investor $i$ decides the amount of wealth $\pi_i(u)$ to invest in $S$ at every time $u\in[0,T]$, based on her current information (either $\Fc^{\mu,S}_u$ or $\Fc^S_u$). The resulting wealth process at any time $t\in [0,T)$ is then
\begin{align}\label{wealth full}
 dX_i(u) &=  r \big(X_i(u)-\pi_i(u)\big)du + \pi_i(u) \frac{dS(u)}{S(u)}\notag\\
 &=\big(rX_i(u)+\pi_i(u)(\mu(u)-r)\big)du+\pi_i(u)\sigma d W(u),\ \ u\in[t,T],\quad X_i(t)=x_i\in\R.%\\
%\nonumber dp(u)&=&\eta(p(u))du+\beta(p(u))d\widehat W(u),\quad p(t)=p,
\end{align}
For convenience, we will often write 
\[
\bm x=(x_1,...,x_N)\in\R^N,\quad \bm\pi = (\pi_1,...,\pi_N),\quad \hbox{and}\quad \bm X = (X_1,...,X_N).
\]
By taking $\overline \pi(u) := \frac{1}{N}\sum_{j=1}^N\pi_j(u)$ and $\overline x :=  \frac{1}{N}\sum_{j=1}^N x_j$, the average wealth $\overline X := \frac{1}{N}\sum_{j=1}^NX_j$ satisfies
\be\label{bc_ave}
d\overline X(u) =\big(r\overline X(u)+\overline\pi(u)(\mu(u)-r)\big) du+\overline \pi(u)\sigma  dW(u),\ \ u\in[t,T],\quad \overline X(t)=\overline x.
\en

When investors have access to $\F^{\mu,S}$, the dynamics of $S$ in \eqref{risky_0} %(including $\mu(u)$ and $dW(u)$) 
is fully known. %; indeed, in view of \eqref{risky_0}, $W$ is $\F^{\mu,S}$-adapted. In this case, we define 
An admissible trading strategy is defined as follows.

\begin{definition}\label{def:admissibility}
$\pi:\Omega\times[0,T]\to \R$ is an {\it admissible} trading strategy under full information, if it is $\mathbb F^{\mu,S}$-progressively measurable and satisfies %$\EE^{t,\bm x}\bigg[\int_t^T\left|\pi_i(u)\right|^2du\bigg]<\infty$ for all $t\in[0,T]$ and $x\in\R^N$, where the superscript ``$t,\bm x$'' denotes conditioning on $\bm X(t) = x$. 
\begin{equation}\label{integrability}
\EE^{t,\bm x,b}\bigg[\int_t^T\left|\pi(u)\right|^2du\bigg]<\infty,\quad \forall (t,\bm x,b)\in[0,T]\times\R^N\times B,
\end{equation}
where the superscript ``$t,\bm x,b$'' denotes conditioning on $\bm X(t) = \bm x$ and $\mu(t) = b$. 
%for all $t\in[0,T]$. Similarly, $\pi:\Omega\times[0,T]\to \R$ is an {\it admissible} trading strategy under partial information, if it is $\mathbb F^{S}$-progressively measurable and satisfies $\EE\big[\int_t^T\left|\pi_i(u)\right|^2du \mid \Fc_t^S\big]<\infty$ for all $t\in[0,T]$. 
We denote by $\mathcal A_{\mu,S}$ %(resp.\ $\mathcal A_S$) 
the set of all admissible trading strategies under full information. 
\end{definition}

On the other hand, when investors have access to only $\F^S$, $\mu(u)$ in \eqref{risky_0} remains unknown and can only be estimated by observations of $S$.\footnote{It is not unrealistic to assume that $\mu$ is uncertain but $\sigma$ is known. As explained in \cite[footnote 2]{KR12}, high-frequency data readily give good estimators for $\sigma$, while estimating $\mu$ is much more challenging statistically.}  As a concrete illustration, we assume in this case that $\mu(u)$ takes two possible values $\mu_1$ and $\mu_2$ (with $\mu_1>\mu_2$), i.e., $B=\{\mu_1,\mu_2\}$. Given a (subjective) prior probability $p_0\in(0,1)$ of the event $\{\mu(0) = \mu_1\}$, we consider the posterior probability process
\be\label{p_j_0}
{\mathfrak{p}}_j(u) :=\PP(\mu(u)=\mu_j\mid \mathcal F^S_u),\quad j=1,2,
%\quad \hbox{with}\ \ \mathcal{S}(u) :=\sigma\{S(v):t\leq v\leq u\},
\quad \forall u> 0.\footnote{At time 0, investors have no information (i.e., $\Fc_0^S =\emptyset$) and can only subjectively assign a probability $p_0\in (0,1)$ to the event $\{\mu(0) = \mu_1\}\in \Fc^\mu_0$. The process ${\mathfrak{p}}_1(u)$ then continuously updates this initial belief $p_0$ based on observations of $S$ over time. That is, ${\mathfrak{p}}_1(u)$ depends on the initial subjective probability $p_0$, although such dependence will not be explicitly expressed in most parts of this paper for notational convenience.}
\ee

\begin{definition}\label{def:admissibility'}
$\pi:\Omega\times[0,T]\to \R$ is an {\it admissible} trading strategy under partial information, if it is $\mathbb F^{S}$-progressively measurable and satisfies %$\EE^{t,\bm x}\bigg[\int_t^T\left|\pi_i(u)\right|^2du\bigg]<\infty$ for all $t\in[0,T]$ and $x\in\R^N$, where the superscript ``$t,\bm x$'' denotes conditioning on $\bm X(t) = x$. 
\begin{equation}\label{integrability'}
\EE^{t,\bm x, p}\bigg[\int_t^T\left|\pi(u)\right|^2du\bigg]<\infty,\quad \forall (t,\bm x,p)\in[0,T]\times\R^N\times [0,1],
\end{equation}
where the superscript ``$t,\bm x, p$'' denotes conditioning on $\bm X(t) = \bm x$ and ${\mathfrak{p}}_1(t)=p$. 
%for all $t\in[0,T]$. Similarly, $\pi:\Omega\times[0,T]\to \R$ is an {\it admissible} trading strategy under partial information, if it is $\mathbb F^{S}$-progressively measurable and satisfies $\EE\big[\int_t^T\left|\pi_i(u)\right|^2du \mid \Fc_t^S\big]<\infty$ for all $t\in[0,T]$. 
We denote by $\mathcal A_{S}$ %(resp.\ $\mathcal A_S$) 
the set of all admissible trading strategies under partial information. 
\end{definition}

This paper will mainly focus on admissible strategies that are {\it Markov}, as defined below.

\begin{definition}\label{def:Markov}
$\bm \pi=(\pi_1,...,\pi_N)\in\mathcal A_{\mu,S}^N$ is Markov %w.r.t.\ an $\F^{\mu, S}$-adapted (or $\F^S$-adapted) $Y:[0,T]\times\Omega\to\R^\ell$, with $\ell\in\N$, 
%for the $N$-player game \eqref{MV_model}, 
if for any $i=1,...,N$, there is a Borel measurable $\xi_i:[0,T]\times\R^N\times\R\to \R$ such that $\pi_i(t) =\xi_i (t,\bm X(t),\mu(t))$ for a.e.\ $t\in[0,T]$ a.s.
%$\xi_i:[0,T]\times\R^d\to \R$ such that $\pi_i(t) =\xi_i (t,\bm X(t))$ for a.e.\ $t\in[0,T]$ a.s. 
Similarly, $\bm \pi=(\pi_1,...,\pi_N)\in\mathcal A_{S}^N$ is Markov %w.r.t.\ an $\F^{\mu, S}$-adapted (or $\F^S$-adapted) $Y:[0,T]\times\Omega\to\R^\ell$, with $\ell\in\N$, 
%for the $N$-player game \eqref{MV_model}, 
if for any $i=1,...,N$, there is a Borel measurable $\xi_i:[0,T]\times\R^N\times[0,1]\to \R$ such that $\pi_i(t) =\xi_i (t,\bm X(t),{\mathfrak{p}}_1(t))$ for a.e.\ $t\in[0,T]$ a.s.
We will write $\pi_i$ and $\xi_i$ interchangeably. 
\end{definition}

\begin{remark}
Markov strategies should be understood through SDE systems. For instance, if 
$\bm \pi=(\pi_1,...,\pi_N)\in\mathcal A_{\mu, S}^N$ is Markov, %w.r.t.\ $\bm X=(X_1,...,X_N)$, 
it inherently means that for any $(t,\bm x,b)\in[0,T)\times\R^N\times B$, the coupled system
\begin{align}\label{system}
dX_i(u) & =\Big(rX_i(u)+\xi_i\big(u,X_1(u),...,X_N(u),\mu(u)\big)(\mu(u)-r)\Big)du\notag \\
&\hspace{1.5in}+\xi_i\big(u,X_1(u),...,X_N(u),\mu(u)\big)\sigma d W(u),\qquad i=1,...,N, 
\end{align}
has a solution on $[t,T]$ given $\bm X(t)=\bm x$ and $\mu(t)=b$. 
\end{remark}

Given two Markov $\bm\pi, \bm\theta\in\mathcal A_{\mu,S}^N$ %(or $\mathcal A_S^N$) w.r.t.\ an $\R^\ell$-valued process $Y$ %for \eqref{MV_model} 
and $s\in (0,T)$, we can define the concatenation of $\bm \theta$ and $\bm \pi$ at time $s$, denoted by $\bm\theta\otimes_s\bm\pi = (\theta_1\otimes_s\pi_1,...,\theta_N\otimes_s\pi_N)$, as
\be\label{concatenate}
(\theta_i\otimes_s\pi_i)(t,\bm x,b) := 
\begin{cases}
\theta_i(t,\bm x,b),\quad &\hbox{for}\ 0\le t<s,\ (\bm x,b)\in\R^N\times B,\\
\pi_i(t,\bm x,b),\quad &\hbox{for}\ s\leq t\le T,\ (\bm x,b)\in\R^N\times B, 
\end{cases} 
\quad \forall i=1,...,N, 
\ee
The concatenation of $\bm \theta, \bm \pi\in\mathcal A_S^N$ at time $s$ can be defined similarly, with ``$b\in B$''  in \eqref{concatenate} replaced by ``$p\in[0,1]$.'' 
Note that $\bm\theta\otimes_s\bm\pi\in\mathcal A_{\mu,S}^N$  (or $\mathcal A_S^N$) is again Markov by construction.

%%%%%%%%%%%%%%%%%%%%%%%%%%%%%%%
%%%%%%%%%%%%%%%%%%%%%%%%%%%%%%%

\section{The Case of Constant Investment Opportunities}\label{sec:constant mu}
In this section, we take $\mu:\Omega\times [0,\infty)\to\R$ to be a constant process, i.e., $B=\{b\}$ for some $b\in\R$. 
%$\mu(t) \equiv\mu(0)$ for all $t\ge 0$. 
For simplicity, we will write $\mu\in\R$ for the constant process, so that \eqref{risky_0} takes the form 
%As $\mu(0)$ is $\Fc^{\mu}_0$-measurable, investors under full information (i.e., with access to $\F^{S,\mu}$) know the realization of $\mu(0)$ and thus the whole process $\mu$. We then simply write $\mu\in\R$ for the constant process $\mu$,
\ba 
\label{risky_1} dS(u)=\mu S(u)du+\sigma S(u) dW(u),\ \ u\ge 0,\quad S(0)=s>0,%\\
%\label{riskless_1}dS_0(u)&=&rS_0(u)du,\; S_0(t)=s_0,
\ea
%Definition~\ref{def:admissibility} stipulates that investors only has access to $\F^S$, but not $\F$. As a result, they can observe the evolution of $S$ in \eqref{risky_1}, but do not know the realization of $\mu$ (which is $\mathcal F_0$-measurable). 

%%%%%%%%%%%%%%%%%%%%

\subsection{Analysis under Full Information $\F^{\mu,S}$}\label{subsec:full constant mu}
Assume that investors have access to $\F^{\mu,S}$. The dynamics of $S$ in \eqref{risky_1}, including $\mu\in\R$ and $dW(u)$, is then fully known; note that $W$ is $\F^{\mu,S}$-adapted in view of \eqref{risky_1}. 
%When the $\Fc_0$-measurable $\mu:\Omega\to\R$ in \eqref{risky_1} is a constant function, we will simply write $\mu\in\R$ and it is trivial that $\mu$ is $\mathcal F^S_t$-measurable for all $t\ge 0$. As $\mu\in\R$ is now fully known to the investors, \eqref{risky_1} indicates that $\F^S=\F^W$, i.e., $W$ is $\F^S$-adapted. 
%This corresponds to the case of full information, where investors know the dynamics of $S$ in \eqref{risky_1} precisely (including $\mu\in\R$, $\sigma>0$, and $W$). 

\begin{remark}\label{rem:Markov mu constant known}
With the knowledge of $\mu\in\R$, the condition \eqref{integrability} for an admissible trading strategy $\pi\in\mathcal A_{\mu,S}$ reduces to 
\begin{equation}\label{integrability mu constant known}
\EE^{t,\bm x}\bigg[\int_t^T\left|\pi(u)\right|^2du\bigg]<\infty,\quad \forall (t,\bm x)\in[0,T]\times\R^N. 
\end{equation}
Also, in the concatenation \eqref{concatenate} of two Markov $\bm\mu,\bm\theta\in\mathcal A_{\mu,S}^N$, the variable $b\in B$ can be dropped, as it is now a fixed known constant (i.e., $b=\mu\in\R$). 
\end{remark}
%Hence, for any $i=1,...,N$ and $\pi_i\in\mathcal A_{\mu,S}$, 

Suppose that each investor considers mean-variance portfolio selection under a relative performance criterion. Specifically, in line with \cite{Touzi2015,Lacker-Zari2017}, investor $i$, for all $i=1,...,N$, is concerned about not only the terminal wealth $X_i(T)$ but also how it compares relatively to the average wealth of all investors $\overline X(T)$, thereby considering a mixed performance criterion
\begin{equation}\label{mixed}
(1-\lambda_i)X_i(T)+\lambda_i(X_i(T)-\overline X(T))= X_i(T)-\lambda_i \overline X(T), %=\left(1-\frac{\lambda_i}{N}\right)X_i(T)-{\lambda_i}\overline X_{(-i)}(T),
\end{equation}
where $\lambda_i\in [0,1)$ is the weight for the relative component $X_i(T)-\overline X(T)$ assigned by investor $i$. 

%\begin{remark}\label{rem:rewrite}
%For each $i=1,2,...,N$, define $\overline x_{(-i)}:=\frac{1}{N}\sum_{j=1,...,N, j\neq i} x_j$,
%\[
%{\overline X}_{(-i)}(u) := \frac{1}{N}\sum_{j=1,...,N,j\neq i} X_j(u)\quad\hbox{and}\quad {\overline \pi}_{(-i)}(u) := \frac{1}{N}\sum_{j=1,...,N,j\neq i} \pi_j(u),\quad \forall u\in [0,T].
%\]  
%Then, the mixed performance criterion \eqref{mixed} can be rewritten as
%\be\label{mixed'}
%\left(1-\frac{\lambda_i}{N}\right)X_i(T)-{\lambda_i}\overline X_{(-i)}(T)
%\ee
%and the dynamics of ${\overline X}_{(-i)}$, similarly to \eqref{bc_ave}, is given by
%\be\label{bc_ave_-i}
%d\overline X_{(-i)}(u) =\big(r\overline X_{(-i)}(u)+\overline\pi_{(-i)}(u)(\mu-r) \big)du+\overline \pi_{(-i)}(u)\sigma  dW(u),\ \ u\in[t,T],\quad \overline X(t)=\overline x_{(-i)}.
%\en
%The formulations \eqref{mixed'}-\eqref{bc_ave_-i} will be useful for proving some of our main results; see Appendix~\ref{sec:proof of main}. 
%\end{remark}

Now, for each $i=1,...,N$, given the current time $t\in[0,T]$ and wealth levels $\bm x=(x_1,\cdots,x_N)\in\R^N$, as well as the trading strategies of the other $N-1$ investors (i.e., $\pi_j\in\mathcal A$ for all $j\neq i$), investor $i$ looks for a trading strategy $\pi_i\in\mathcal A$ that maximizes the mean-variance objective % $U_i(t)=U_i(t,\bm X(t) ,p(t))$ is 
\ba
J_i\big(t, \bm x, \{\pi_j\}_{j\neq i}, \pi_i\big)&:=&
\EE^{t,\bm x}\left[X_i(T)-\lambda^M_i\overline X(T)\right]-\frac{\gamma_i}{2}\mathrm{Var}^{t,\bm x}\left[X_i(T)-\lambda^V_i\overline X(T)\right],
%\EE^{t,\bm x}\left[\left(1-\frac{\lambda^M_i}{N}\right)X_i(T)-\lambda^M_i\overline X_{(-i)}(T)\right]\\ %\ \bigg |\ \bm X(t)=\bm x\right]\\
%&&-\frac{\gamma_i}{2}\mathrm{Var}^{t,\bm x}\left[\left(1-\frac{\lambda^V_i}{N}\right)X_i(T)-\lambda^V_i\overline X_{(-i)}(T)\right], %\ \bigg |\ \bm X(t)= \bm x\right],
\label{MV_model}
\ea
where the superscript ``${t,\bm x}$'' denotes conditioning on $\bm X(t)=\bm x$ and $\gamma_i>0$ is the risk aversion parameter for the $i$-th investor. We allow the possibility of two distinct weights $\lambda^M_i,\lambda^V_i\in[0,1)$ for the relative component---one for the mean part and the other for the variance part. 

Our goal is to find a Nash equilibrium $\bm \pi=(\pi_1,...,\pi_N)\in\mathcal A^N$ for this $N$-player game. Because a mean-variance objective is known to induce time inconsistency, how a Nash equilibrium should be defined requires a deeper thought. As elaborated in \cite{HZ22}, in a dynamic game where players have time-inconsistent preferences, there are two intertwined levels of game-theoretic reasoning. At the {\it inter-personal} level, each player selects an action in response to other players' chosen strategies. The selected action, importantly, has to be an equilibrium at the {\it intra-personal} level (i.e., among the player's current and future selves), so as to resolve time inconsistency psychologically within the player. %between her current and future selves. 
%With this in mind, let us first introduce Markov trading strategies.
By recalling Markov strategies (Definition~\ref{def:Markov}) and Remark~\ref{rem:Markov mu constant known}, %the concatenation of Markov strategies in \eqref{concatenate}, 
we define a Nash equilibrium as follows.

\begin{definition}\label{def:E}
We say $\bm \pi^*=(\pi^*_1,...,\pi^*_N)\in\mathcal A^N_{\mu,S}$ is a Nash equilibrium for the $N$-player game \eqref{MV_model}, subject to the dynamics $\bm X=(X_1,...,X_N)$ in \eqref{wealth full} with $\mu(u)\equiv\mu\in\R$, if $\bm \pi^*$ is Markov %w.r.t.\ $\bm X$ % for \eqref{MV_model} 
and for any $(t,\bm x)\in[0,T)\times\R^N$ and Markov $\bm\pi=(\pi_1,...,\pi_N)\in\mathcal A^N_{\mu,S}$ %w.r.t.\ $\bm X$, %for \eqref{MV_model}, 
\be\label{intra E}
\liminf_{h\downarrow 0}\frac{J_i\big(t,\bm{x},\{\pi^*_j\}_{j\neq i},\pi_i^*\big)-J_i\big(t,\bm{x},\{\pi^*_j\}_{j\neq i},{\pi_i\otimes_{t+h}\pi^*_i}\big)}{h}\geq 0,\quad \forall i=1,...,N.
\en
\end{definition}

%\begin{remark}
Condition \eqref{intra E} extends the standard definition of an intra-personal equilibrium for one single agent (see e.g., Bj\"{o}rk et al. \cite{BKM2016, BKM2017}) to a setup with multiple agents. When the other $N-1$ players' strategies $(\pi^*_1,...,\pi^*_{i-1}, \pi^*_{i+1}, \pi^*_N)$ are fixed, \eqref{intra E}  states that at any time and state $(t,\bm x)$, as long as the $i$-th investor's future selves will employ the strategy $\pi^*_i$, her current self cannot be better off by using any other strategy $\pi_i$. That is, $\pi^*_i$ is an equilibrium {\it intra-personally} for the $i$-th investor.  
%\end{remark}
A Nash equilibrium under Definition~\ref{def:E}, as a result, stipulates that every investor achieves her own intra-personal equilibrium {\it simultaneously}, given the other players' strategies. This corresponds precisely to the notion ``{\it soft} inter-personal equilibrium'' in \cite[Definition 2.3]{HZ22}.  

There is a stronger notion of a Nash equilibrium, i.e., ``{\it sharp} inter-personal equilibrium'' in \cite[Definition 2.6]{HZ22}, for a game where players have time-inconsistent preferences. It requires every player to attain her {\it optimal} intra-personal equilibrium (instead of an arbitrary one, as is required by a soft inter-personal equilibrium) simultaneously. In \cite{HZ22}, a sharp inter-personal equilibrium is shown to exist in a Dynkin game, relying on the definition and characterization of an optimal intra-personal equilibrium under optimal stopping in \cite{HZ19, HZ20, HW21}. For mean-variance portfolio selection, there is no consensus on how to compare different intra-personal equilibria and define an optimal one accordingly. We thus stay with the notion ``{\it soft} inter-personal equilibrium'' in Definition~\ref{def:E}. 

To precisely characterize a Nash equilibrium as in Definition~\ref{def:E}, we consider the constants
\be\label{kappa}
\kappa_i := \frac{1}{\gamma_i}\left(1-\frac{\lambda^V_i}{N}\right)^{-1} \left(1-\frac{\lambda^M_i}{N}\right)>0\quad \hbox{$i=1,\cdots,N$},
\ee
\be\label{kappa bar}
\overline \kappa :=\frac{1}{N}\sum_{j=1,...,N} \kappa_j\quad\hbox{and}\quad \overline\lambda^V:=\frac{1}{N}\sum_{j=1,...,N}\lambda^V_j. 
\ee

\begin{theorem}\label{thm:E mu known}
A Nash equilibrium $\bm \pi^*=(\pi^*_1,...,\pi^*_N)\in\mathcal A^N_{\mu,S}$ for the $N$-player game \eqref{MV_model}, subject to the wealth dynamics \eqref{wealth full} with $\mu(u)\equiv\mu\in\R$, is given by
\be\label{NE_constant}
\pi_i^*(t)=e^{-r(T-t)}\frac{\mu-r}{\sigma^2}\left(\kappa_i+\frac{\lambda^V_i}{1-\overline\lambda^V}\overline  \kappa\right),\quad \forall i=1,...,N. 
\en
Moreover, the value function under the Nash equilibrium $\bm \pi^*$ is
\be\label{value_constant}
V_i(t,\bm x) = e^{r(T-t)}\left(x_i- {\lambda_i^M}\overline x\right)+(T-t) N_i,\quad \forall i=1,...,N. 
\en
where %$N_i$ is a constant defined by
\be\label{N_i const}
N_i:=\left(\frac{\mu-r}{\sigma}\right)^2 \bigg\{\left(\kappa_i+\frac{\lambda^V_i-\lambda_i^M}{1-\overline\lambda^V}\overline  \kappa\right)-\frac{\gamma_i\kappa_i^2}{2} \bigg\}.
\en
\end{theorem}
The proof of Theorem~\ref{thm:E mu known} is relegated to Section~\ref{subsec:proof of thm:E mu known}. 

\begin{remark}
As $\pi^*_i$ in \eqref{NE_constant} does not depend on $\bm x=(x_1,...,x_N)\in\R^N$, the system \eqref{system} becomes decoupled and each $X_i$ can be solved independently. This simple structure is motivated by the intra-personal equilibrium strategy in Basak and Chabakauri \cite{BC10} for the classical mean-variance problem of a single agent, which does not depend on the agent's wealth.  
\end{remark}

\begin{remark}\label{rem:1/gamma to kappa}
If $\lambda^M_i=\lambda^V_i=0$ (i.e., relative performance is disregarded), $\frac{\mu-r}{\sigma^2}\big(\kappa_i+\frac{\lambda^V_i}{1-\overline\lambda^V}\overline \kappa\big)$ in \eqref{NE_constant} reduces to %the Merton ratio 
$\frac{\mu-r}{\sigma^2\gamma_i}$, i.e., $\pi^*_i(t)$ becomes the intra-personal equilibrium strategy in \cite{BC10} for the classical mean-variance problem of a single investor. 
In other words, the constant $1/\gamma_i$ in the single-agent case is replaced by a linear combination of $\{1/\gamma_j\}_{j=1}^N$ in our multi-agent case, where the involved constant coefficients depend on $\{\lambda_j^M\}_{j=1}^N$ and $\{\lambda_j^V\}_{j=1}^N$.  
%%Theorem~\ref{thm:E mu known} is in line with the results in \cite{GH2022,ZCZ2019}
%%In addition, there are several studies related to investment and reinsurance problem using the mean-variance criterion with relative performance. See \cite{GH2022,ZCZ2019} for instance.  
\end{remark}

\begin{remark}
When $\lambda^M_i=\lambda^V_i=\lambda$ for all $i=1,...,N$ and $r=0$, $\pi^*_i(t)$ in \eqref{NE_constant} simplifies to 
\be\label{NE_constant_r=0}
\pi_i^* %= \frac{\mu-r}{\sigma^2\gamma_i} + \frac{\lambda }{1-\lambda}\frac{\mu-r}{\sigma^2} \overline\gamma^{(-1)}_N 
= \frac{1}{1-\lambda}\frac{\mu-r}{\sigma^2}\left((1-\lambda)\gamma_i^{-1}+{\lambda}\overline\gamma^{(-1)}_N\right),\quad \forall i=1,...,N, 
\en
with $\overline\gamma^{(-1)}_N:=\frac{1}{N}\sum_{j=1}^N\gamma_j^{-1}$. This is comparable to \cite[Proposition 3.6]{Touzi2015}, where
\be\label{NE_constant_r=0'}
\hat \pi_i=\frac{1}{1-\lambda}\frac{\mu-r}{\sigma^2}\bigg(\bigg(1-\frac{\lambda N}{N+\lambda-1}\bigg)\gamma_i^{-1}+\frac{\lambda N}{N+\lambda-1} \overline\gamma^{(-1)}_N\bigg),\quad \forall i=1,...,N, 
\en
is an equilibrium trading strategy of investor $i$, who aims at maximizing her expected utility under an exponential utility function, i.e., $\E[-\exp\{-\gamma_i (X_i(T) - \lambda \overline{X}(T))\}]$. While $\pi^*_i$ and $\hat\pi_i$ both take the form of $\frac{1}{1-\lambda}\frac{\mu-r}{\sigma^2}$ multiplied by a convex combination of $\gamma_i^{-1}$ and $\overline\gamma^{(-1)}_N$, the coefficients of the convex combination in \eqref{NE_constant_r=0'} depend on $N$, the number of investors, and those in \eqref{NE_constant_r=0} do not.\footnote{Fund managers usually compare their performance with peers within the same fund category and have access to the average category performance computed readily by the financial industry. For instance, Morningstar, Inc.\ places U.S.\ stock funds in nine categories based on the size (large, medium, or small capitalization) and style (value, blend, or growth) of the stocks typically owned by a fund and computes the average performance for each category. There are two possible consequences: (i) as each category tends to represent specific risk tolerance (e.g., ``Large Value'' attracts most risk averse shareholders, while ``Small Growth'' does the opposite), the risk aversion coefficients $\{\gamma_i\}_{i=1}^N$ of fund managers within one category should not vary too much, if managers truthfully represent shareholders' preferences; (ii) fund managers in practice evaluate and improve relative performance based on Morningstar's category average and don't actually think of (or bother with) checking the number of funds $N$ in a category (which in fact changes constantly over time). This suggests that $\pi^*_i$ can have practical advantages over $\hat\pi_i$---as the former has less dependence on $N$ (only through $\overline\gamma^{(-1)}_N=\frac{1}{N}\sum_{j=1}^N\gamma_j^{-1}$, which should be stable w.r.t.\ $N$ by (i)), it is easier to implement and closer to fund managers' actual handling of relative performance. 
} 
%Hence, $\pi^*_i$ and $\hat\pi_i$ are different for any fixed $N>1$, but share the same limit as $N\to\infty$ (because $\frac{\lambda N}{N+\lambda-1}\to\lambda$ as $N\to\infty$).
\end{remark}

%\begin{remark}
%Theorem~\ref{thm:E mu known} is in line with the results in \cite{GH2022,ZCZ2019}
%%In addition, there are several studies related to investment and reinsurance problem using the mean-variance criterion with relative performance. See \cite{GH2022,ZCZ2019} for instance.  
%\end{remark}

%%%%%%%%%%%%%%%%%%%

\subsection{Analysis under Partial Information $\F^S$}\label{subsec:partial constant mu}
Assume that investors have access to only $\F^{S}$, i.e., they observe the evolution of $S$ in \eqref{risky_1} but do not know $\mu\in\R$ and $dW(u)$. As specified above \eqref{p_j_0}, we assume that there are two possible values $\mu_1$ and $\mu_2$ (with $\mu_1>\mu_2$) for $\mu\in\R$ in \eqref{risky_1} and investors do not know which one is the true value. %(i.e., the $\Fc_0$-measurable $\mu$ takes the form $\mu:\Omega\to\{\mu_1,\mu_2\}$), while the volatility coefficient $\sigma>0$ in \eqref{risky_1} is known to all investors.
Our setup is suitable for modeling companies with unreported innovation: even when it is known that an innovation can improve a company's expected return from $\mu_2$ to $\mu_1$, investors usually have no idea the progress of the innovation or whether the company has taken the initiative at all. Indeed, more than half of the NYSE-listed firms do not report research and development \cite{KR15} and European firms file patents for less than 36\% (resp.\ 25\%) of product innovations (resp.\ process innovations) \cite{AK98}. 

As the true value of $\mu$ is unknown, the Nash equilibrium formula in Theorem~\ref{thm:E mu known} is no longer of use. To overcome this, we will rely on the posterior probability \eqref{p_j_0}, which now takes the form: given a (subjective) prior probability $p_0\in(0,1)$ of the event $\{\mu = \mu_1\}$, 
\be\label{p_j}
{\mathfrak{p}}_j(u) =\PP(\mu=\mu_j\mid \mathcal F^S_u),\quad j=1,2,\quad \forall u> 0.
\ee

\begin{lemma}\label{lem:p to 1}
If $\mu=\mu_1$ (resp.\ $\mu=\mu_2$), then ${\mathfrak  p}_1(u)$ in \eqref{p_j} converges to 1 (resp.\ 0) as $u\to\infty$. %If $\mu=\mu_2$, then ${\mathfrak  p}_1(u)\to 0$ as $u\to\infty$. 
\end{lemma}

By Lemma~\ref{lem:p to 1} (whose proof is relegated to Section~\ref{subsec:proof of lem:p to 1}), the true value of $\mu$ can be learned from the long-time limit of ${\mathfrak  p}_1(\cdot)$. When one is restricted to a finite time horizon $T>0$, the long-time limit is not necessarily available. It is then important to study the actual evolution of ${\mathfrak{p}}_1(\cdot)$  over time. In fact, ${\mathfrak p}_1(\cdot)$ can be characterized as the unique strong solution to an SDE, as shown in the next result (whose proof is relegated to Section~\ref{subsec:proof of lem:hW}). 

\begin{lemma}\label{lem:hW}
Fix any $t\ge 0$. 
\begin{itemize}
\item [(i)] Let $B$ be any standard Brownian motion. For any $p \in (0,1)$, the SDE 
\be\label{P eta}
dP(u)= \frac{\mu_1-\mu_2}{\sigma}P(u)(1-P(u))  dB(u),\ \ u\ge t, \quad P(t) =p,
\ee
has a unique strong solution, which satisfies $P(u)\in(0,1)$ for all $u\ge t$ a.s.
\item [(ii)] Given $S$ in \eqref{risky_1} and $\mathfrak{p}_1$ in \eqref{p_j}, the process 
\be\label{Inno_W}
\widehat W(u) := \frac1\sigma \bigg[\log\bigg(\frac{S(u)}{S(t)}\bigg)-(\mu_1-\mu_2)\int_t^u {\mathfrak p}_1(s) ds -\left(\mu_2-\frac{\sigma^2}{2}\right)(u-t)\bigg],\quad u\ge t,
\en
is a standard Brownian motion adapted to $\{\mathcal F^S_u\}_{u\ge t}$. Moreover, $\{{\mathfrak p}_1(u)\}_{u\ge t}$ %in \eqref{p_j} 
is the unique strong solution to 
\be\label{P}
dP(u)=\frac{\mu_1-\mu_2}{\sigma}P(u)(1-P(u))  d\widehat W(u),\ \ u\ge t,\quad P(t) = {\mathfrak p}_1(t) \in (0,1), 
\ee
with ${\mathfrak p}_1(u)\in(0,1)$ for all $u\ge t$ a.s. 
Hence, $S$ in \eqref{risky_1} can be expressed equivalently as
\ba 
\label{risky_1'} dS(u)= \big((\mu_1-\mu_2)P(u)+\mu_2\big) S(u)du+\sigma S(u) d\widehat W(u),\ \ u\ge t,\quad S(t)=s>0,%\\
%\label{riskless_1}dS_0(u)&=&rS_0(u)du,\; S_0(t)=s_0,
\ea
where $P$ is the unique strong solution to \eqref{P}. 
\end{itemize}
\end{lemma}
%The proof of Lemma~\ref{lem:hW} is relegated to Section~\ref{subsec:proof of lem:hW}.  

The expression \eqref{risky_1'} is important: $S$ in \eqref{risky_1}, which involves the unknown $\mu$, is now expressed alternatively in terms of the known constants $\mu_1$, $\mu_2$, $\sigma$ and the observable process $P(\cdot) = {\mathfrak p}_1(\cdot)$.
%$\widehat W(u)$  is the innovation process which is a standard Brownian motion discussed in \cite{LS2013,Oksendal2003} 
When investors view the stock $S$ as \eqref{risky_1'}, their wealth processes  can also be expressed in terms of $P$ in \eqref{P} and $\widehat W$ in \eqref{Inno_W}, such that the dynamics is fully observable. Specifically, for any $i=1,...,N$ and $\pi_i\in\mathcal A_S$, the wealth process \eqref{wealth full} of investor $i$ can be equivalently expressed as
\be\label{wealth full'}
 dX_i(u) = \left[rX_i(u)+\pi_i(u)\big((\mu_1-\mu_2)P(u)+\mu_2-r\big)\right]du+\pi_i(u)\sigma d\widehat W(u),\ u\in[t,T],\ X_i(t)=x_i,
\ee
where $P$ is the unique strong solution to \eqref{P}. The average wealth $\overline X$ in \eqref{bc_ave} thus takes the form
\be\label{bc_ave P}
d\overline X(u) =\left[r\overline X(u)+\overline\pi(u)\big((\mu_1-\mu_2)P(u)+\mu_2-r\big)\right]du+\overline \pi(u)\sigma  d\widehat W(u),\quad \overline X(t)=\overline x.
\en
%Similarly, $\overline X_{(-i)}$ in \eqref{bc_ave_-i} now fulfills
%\be\label{bc_ave_-i P}
%d\overline X_{(-i)}(t) =r\overline X_{(-i)}(t)+\overline\pi_{(-i)}(t)\Big((\mu_1-\mu_2)P(t)+\mu_2-r\Big) dt+\overline \pi_{(-i)}(t)\sigma  d\widehat W(t),\quad \overline X(0)=\overline x_{(-i)}.
%\en
In line with \eqref{MV_model}, for each $i=1,2,...,N$, given the current time $t\in[0,T]$, wealth levels $\bm x=(x_1,\cdots,x_N)\in\R^N$, and posterior probability $p\in(0,1)$ of the event $\{\mu=\mu_1\}$, as well as the trading strategies of the other $N-1$ investors (i.e., $\pi_j\in\mathcal A_S$ for all $j\neq i$), investor $i$ looks for a trading strategy $\pi_i\in\mathcal A_S$ that maximizes the mean-variance objective % $U_i(t)=U_i(t,\bm X(t) ,p(t))$ is 
%the $i$-th investor's mean-variance objective \eqref{MV_model} then becomes
\begin{align}
J_i\big(t, \bm x, p, \{\pi_j\}_{j\neq i}, \pi_i\big)&:= \EE^{t,\bm x, p}\left[X_i(T)-\lambda^M_i\overline X(T)\right] -\frac{\gamma_i}{2}\mathrm{Var}^{t,\bm x, p}\left[X_i(T)-\lambda^V_i\overline X(T)\right]
%\EE^{t,\bm x, p}\left[\left(1-\frac{\lambda^M_i}{N}\right)X_i(T)-\lambda^M_i\overline X_{(-i)}(T)\right] \\%\ \bigg |\ \bm X(t)=\bm x,\ P(t)=p\right]\\
%&\hspace{0.2in}-\frac{\gamma_i}{2}\mathrm{Var}^{t,\bm x, p}\left[\left(1-\frac{\lambda^V_i}{N}\right)X_i(T)-\lambda^V_i\overline X_{(-i)}(T)\right], %\ \bigg |\ \bm X(t)= \bm x,\ P(t)=p\right].
\label{MV_model P}
\end{align}
where the superscript ``${t,\bm x,p}$'' denotes conditioning on $\bm X(t)=\bm x$ and $P(t)=p$. 

%Due to the additional state process $P$, Markov trading strategies (Definition \ref{def:Markov}) need to be modified accordingly.

%\begin{definition}\label{def:Markov'}
%We say $\bm \pi=(\pi_1,...,\pi_N)\in\mathcal A^N$ is Markov for the $N$-player game \eqref{MV_model P}, if for any $i=1,...,N$, there exists a Borel measurable $\xi_i:[0,T]\times\R^d\times (0,1)\to \R$ such that $\pi_i(t) = \xi_i(t,\bm X(t), P(t))$ for a.e.\ $t\in[0,T]$ a.s. We will write $\pi_i$ and $\xi_i$ interchangeably.
%\end{definition}

%Similarly to \eqref{concatenate}, given two Markov $\bm\pi, \bm\theta\in\mathcal A^N$ for \eqref{MV_model P} and $s\in (0,T)$, we can define the concatenation of $\bm \theta$ and $\bm \pi$ at time $s$, denoted by $\bm\theta\otimes_s\bm\pi = (\theta_1\otimes_s\pi_1,...,\theta_N\otimes_s\pi_N)$, as
%\[
%(\theta_i\otimes_s\pi_i)(t,\bm x,p) := 
%\begin{cases}
%\theta_i(t,\bm x,p),\quad &\hbox{for}\ 0\le t<s,\ \bm x\in\R^N,\ p\in(0,1),\\
%\pi_i(t,\bm x,p),\quad &\hbox{for}\ s\leq t\le T,\ \bm x\in\R^N,\ p\in (0,1), 
%\end{cases} 
%\quad \forall i=1,...,N. 
%\]
%Note that $\bm\theta\otimes_s\bm\pi\in\mathcal A^N$ is Markov for \eqref{MV_model P} by construction. We now define a Nash equilibrium for \eqref{MV_model P} in line with Definition~\ref{def:E}. 

\begin{definition}\label{def:E'}
We say $\bm \pi^*=(\pi^*_1,...,\pi^*_N)\in\mathcal A^N_S$ is a Nash equilibrium for the $N$-player game \eqref{MV_model P}, subject to the dynamics $\bm X=(X_1,...,X_N)$ in \eqref{wealth full'} and $P$ in \eqref{P}, if $\bm\pi^*$ is Markov %w.r.t.\ $(\bm X, P)$ % for \eqref{MV_model P} 
and for any $(t,\bm x,p)\in[0,T)\times\R^N\times (0,1)$ and Markov $\bm\pi=(\pi_1,...,\pi_N)\in\mathcal A^N_S$ %w.r.t.\ $(\bm X,P)$, %for \eqref{MV_model P}, 
\be\label{intra E'}
\liminf_{h\downarrow 0}\frac{J_i\big(t,\bm{x},p,\{\pi^*_j\}_{j\neq i},\pi_i^*\big)-J_i\big(t,\bm{x},p,\{\pi^*_j\}_{j\neq i},{\pi_i\otimes_{t+h}\pi^*_i}\big)}{h}\geq 0,\quad \forall i=1,...,N.
\en
\end{definition}
To find such a Nash equilibrium, we will generalize the extended HJB equation in Bj\"ork et al.\ \cite{BKM2016, BKM2017}, derived for one single agent with a time-inconsistent preference, to our present multi-agent case. 
%for the problem \eqref{MV_model P}, which is a coupled system of $2N$ partial differential equations (PDEs)---two PDEs for each investor. 
%To precisely state a Nash equilibrium in Theorem~\ref{thm:E mu unknown} below, we need to introduce a set of two Cauchy problems, one depending on the other. 
First, for notational convenience, define $\theta, \beta:[0,1]\to \R$ by
\be\label{theta}
\theta(p):=(\mu_1-\mu_2)p+\mu_2\quad \hbox{and}\quad %\eta(p) := -(q_1+q_2)p+q_2,\quad 
\beta(p):=\frac{\mu_1-\mu_2}{\sigma}p(1-p). 
\ee
For each $i=1,...,N$, %by recalling the formulations in Remark~\ref{rem:rewrite}, 
we can rewrite \eqref{MV_model P} as
\begin{equation}\label{J_i'}
J_i\big(t, \bm x,p, \{\pi_j\}_{j\neq i}, \pi_i\big) =\EE^{t,\bm x,p}\left[F_i(\bm X(T))\right]+\frac{\gamma_i}{2}\E^{t,\bm x,p}[H_i(\bm X(T))]^2,
\end{equation}
where 
\begin{align*}
H_i(\bm x) := \left(1-\frac{\lambda^V_i}{N}\right)x_i - \lambda^V_i\overline x_{(-i)},\quad F_i(\bm x)&:=\left(1-\frac{\lambda^M_i}{N}\right)x_i-\lambda^M_i\overline x_{(-i)}-\frac{\gamma_i}{2} H_i(\bm x)^2,  
\end{align*}
with $\overline x_{(-i)}:=\frac{1}{N}\sum_{j=1,...,N, j\neq i} x_j$. 
% $g_i(t,\bm x)=\EE\left[X(T)\left(1-\frac{\lambda^V_i}{N}\right)X_i(T)-\lambda^V_i\overline X_{(-i)}(T)\bigg |{\mathcal{F}}(t)\right]$. 
Now, to derive the extended HJB equation for a Nash equilibrium $\bm \pi^*=(\pi^*_1,...,\pi^*_N)\in\mathcal A^N_S$ and the corresponding value functions $\{V_i(t,\bm x,p):=  J_i(t, \bm x,p, \{\pi^*_j\}_{j\neq i}, \pi^*_i) \}_{i=1}^N$, we simply apply the one-agent analysis in Bj\"ork et al.\ \cite{BKM2017} for every investor. Assuming that the other investors' strategies $\{\pi^*_j\}_{j\neq i}$ are all given, investor $i$ strives to find a strategy $\pi^*_i$ that fulfills \eqref{intra E'}. This is essentially a (one-agent) time-inconsistent control problem studied in Bj\"ork et al.\ \cite{BKM2017} with a more general state process $(X_1,...,X_N,P)$ in \eqref{wealth full'} and \eqref{P}, where the dynamics of $P$ and $X_j$, with $j\neq i$, are fixed (as $\pi_j =\pi^*_j$ is given) and investor $i$ can only influence $X_i$ through the choice of $\pi_i$. The same derivation in \cite{BKM2017}, under the dynamics of $(X_1,...,X_N,P)$, then yields
%Following the derivation in Bj\"ork et al.\ \cite[Section 10]{BKM2016}, %and \cite[]{BKM2017}, 
%the extended HJB equation for a Nash equilibrium $\bm \pi^*=(\pi^*_1,...,\pi^*_N)\in\mathcal A^N$ and the corresponding equilibrium value functions $\{V_i(t,\bm x,p)\}_{i=1}^N$ takes the form: for any $i=1,...,N$, 
\begin{align}
\nonumber \partial_tV_i+\sup_{\pi_i}\bigg\{&\sum_{j\neq i}\left(rx_j+(\theta(p)-r)\pi^*_j\right)\partial_{x_j}V_i+\left(rx_i+(\theta(p)-r)\pi_i\right)\partial_{x_i}V_i\\
\nonumber&+\frac{\sigma^2}{2}\sum_{j\neq i}\sum_{k\neq i}\pi_j^*\pi^*_k\partial_{x_jx_k}V_i+\frac{\sigma^2}{2}\pi_i^2\partial_{x_ix_i}V_i+{\sigma^2}\pi_i\sum_{j\neq i}\pi_j^*\partial_{x_ix_j}V_i\\
\nonumber&+\eta(p)\partial_pV_i+\frac{\beta(p)^2}{2}\partial_{pp}V_i+ \sigma\beta(p)\sum_{j\neq i}\pi^*_j\partial_{x_jp}V_i+\sigma\pi_i\beta(p)\partial_{x_ip}V_i\\
\nonumber &-\frac{\gamma_i\sigma^2}{2}\sum_{j\neq i}\sum_{k\neq i}{\pi_j^*\pi_k^*}\partial_{x_j}g_i\partial_{x_k}g_i\\
\nonumber &-\frac{\gamma_i\sigma^2}{2}\pi_i^2 (\partial_{x_i}g_i)^2-\gamma_i\sigma^2\pi_i\sum_{j\neq i}\pi_j^*\partial_{x_i}g_i\partial_{x_j}g_i\\
&-\frac{\gamma_i\beta(p)^2}{2} (\partial_{p}g_i)^2-\gamma_i\sigma\pi_i\beta(p)\partial_{x_i}g_i\partial_{p}g_i-\gamma_i\sigma\beta(p)\sum_{j\neq i}\pi_j^*\partial_{p}g_i\partial_{x_j}g_i\bigg\}=0
\label{V_i}
\end{align}
with the terminal condition $V_i(T,x,p)=(1-\frac{\lambda^M_i}{N})x_i-\lambda^M_i\overline x_{(-i)}$, where the function $g_i(t,\bm x,p):= \E^{t,\bm x,p}[H_i(\bm X(T))]$ (first appeared on the right-hand side of \eqref{J_i'}) is characterized by
\ba
\nonumber \partial_t g_i&+&\sum_{j=1,...,N}\left(rx_j+(\theta(p)-r)\pi^*_j\right)\partial_{x_j}g_i+\frac{\sigma^2}{2}\sum_{j=1,...,N}\sum_{k=1,...,N}\pi_j^*\pi^*_k\partial_{x_jx_k}g_i\\
&+&\eta(p)\partial_pg_i+\frac{\beta(p)^2}{2}\partial_{pp}g_i+ \sigma\beta(p)\sum_{j= 1,...,N}\pi^*_j\partial_{x_jp}g_i=0\label{g_i}
\ea
with the terminal condition $g_i(T,\bm x,p)=(1-\frac{\lambda^V_i}{N})x_i-\lambda^V_i\overline x_{(-i)}$.\footnote{When $J_i$ in \eqref{J_i'} is placed in the context of \cite[Definition 4.1]{BKM2017}, the extended HJB equation (i.e., \cite[(4.1)-(4.3)]{BKM2017}) takes a specific form: (i) the two terms containing $f$ in \cite[(4.1)]{BKM2017} cancel out because $\EE^{t,\bm x,p}\left[F_i(\bm X(T))\right]$ in \eqref{J_i'} has no additional dependence on $(\bm x,p)$ other than $(\bm X(t), P(t))=(\bm x,p)$; (ii) as \cite[(4.1)]{BKM2017}) no longer contains $f$, \cite[(4.2)]{BKM2017}) is not needed; (iii) $G$ and $g$ therein are $G(y)=\frac{\gamma_i}{2}y^2$ and $g(t,\bm x, p) =\E^{t,\bm x,p}[H_i(\bm X(T))]$; (iv) thanks to \cite[Section 7.4]{BKM2016}, the terminal condition in \cite[(4.3)]{BKM2017} should be modified to $g(T,\bm x, p) = H_i(\bm x)$. Considering all this yields \eqref{V_i}-\eqref{g_i}. } Note that $\eta:[0,1]\to\R$ in \eqref{V_i}-\eqref{g_i}  is a generic Lipschitz function that allows for suitable flexibility of our model.\footnote{For our current analysis in Section~\ref{subsec:partial constant mu}, this $\eta$ function is not required and we can simply take $\eta\equiv 0$. In Section~\ref{subsec:partial alternating mu}, $\eta$ will need to take a specific non-zero form; see \eqref{eta}.} That is, in our multi-agent setting, the extended HJB equation is \eqref{V_i}-\eqref{g_i} for $i=1,...,N$, which is a coupled system of $2N$ PDEs---two PDEs for each investor. For each $i=1,...,N$, we would like $\pi_i^*$ to attain the supremum in \eqref{V_i}, which yet involves a function $g$ that depends on $\pi^*_i$ through \eqref{g_i}. It is this ``fixed point'' property that makes solving the PDE system challenging.  
%While deriving this system is not hard from the analysis in \cite{BKM2016}, the real challenge is how the $2N$ coupled equations can be solved jointly. 
To this end, 
%To solve the extended HJB equation \eqref{V_i}-\eqref{g_i} for $i=1,...,N$, 
we take up the ansatz 
\be\label{sol. form}
\begin{split}
V_i(t,\bm x,p)&= A_i(t)x_i+B_i(t)\overline x_{(-i)}+C_i(t,p),\\
g_i(t,\bm x,p)&= a_i(t)x_i+b_i(t)\overline x_{(-i)}+c_i(t,p),
\end{split}
\ee
for some functions $A_i$, $B_i$, $C_i$, $a_i$, $b_i$, and $c_i$ to be determined. As detailed in Section~\ref{subsec:extended HJB}, by plugging this ansatz into \eqref{g_i} and matching the coefficients on both sides, one can solve for $a_i$ and $b_i$ explicitly and the remaining terms in the equation form the Cauchy problem
%Given $i=1,...,N$, recall $\kappa_i>0$ defined in \eqref{kappa}. Now, by setting $Q:= [0,T)\times (0,1)$, we consider, for any $\eta:[0,1]\to\R$, the first Cauchy problem 
\begin{equation}\label{Cauchy}
\begin{cases}
\partial_t c+\left(\eta(p)- \beta(p)\frac{\theta(p)-r}{\sigma}\right)\partial_pc+\frac{\beta(p)^2}{2}\partial_{pp}c+ \kappa_i\left(\frac{\theta(p)-r}{\sigma}\right)^2=0&\hbox{for}\ (t,p)\in Q,\\
c(T,p)=0,&\hbox{for}\ p\in (0,1),%\label{terminal} 
\end{cases}
\end{equation}
where $\kappa_i>0$ is defined in \eqref{kappa} and we set $Q:= [0,T)\times (0,1)$. 
As $\beta(p)^2$ is {\it not} bounded away from 0 on $(0,1)$, \eqref{Cauchy} is degenerate, i.e., the {\it uniform ellipticity} condition fails. Hence, standard results of parabolic equations (e.g., \cite[Chapter 1]{Friedman-book-64} and \cite[Chapter 6]{Friedman-book-75}) cannot be readily applied and one needs the approximation result in \cite{HS00}. %Our strategy is to approximate \eqref{Cauchy} by a sequence of uniformly elliptic equations, each of which admits a classical solution by the standard results, and show that the limiting solution $c$ fulfills \eqref{Cauchy} by an analytic argument. 
By probabilistic arguments that involve taking derivatives of stochastic processes (see Remark~\ref{rem:P'}), we further obtain concrete stochastic representations of $c$ and $\partial_p c$. All this yields the next result, whose proof is relegated to Section~\ref{subsec:proof of lem:Cauchy}. 

\begin{lemma}\label{lem:Cauchy}
%Set $Q:= [0,T)\times (0,1)$ and 
Suppose that $\eta:[0,1]\to\R$ is Lipschitz and for any $t\ge 0$ and $p\in (0,1)$,
\be\label{P'''}
dP(u)=\eta(P(u))du+\beta(P(u))d W(u),\ \ u\ge t, \quad P(t)=p,
\ee
has a unique strong solution with $P(u)\in (0,1)$ for all $u\ge t$ a.s. Define a probability measure on $(\Omega,\mathcal F_T)$ by
\begin{equation}\label{Q}
\Q(A) := \E[1_A  Z(T)]\quad \forall A\in\mathcal F_T,
\end{equation}
with 
\begin{equation}\label{Z}
Z(u):= \exp\bigg(-\frac12\int_t^u\left(\frac{\theta(P(s))-r}{\sigma}\right)^2 ds -\int_t^u \frac{\theta(P(s))-r}{\sigma} dW(s)\bigg)\quad t\le u\le T, 
\end{equation}
as well as the process
\begin{equation}\label{W_Q}
W_\Q(u) := W(u) + \int_t^u \frac{\theta(P(s))-r}{\sigma}ds,\quad t\le u\le T. 
\end{equation}
Then, for any $i=1,...,N$, 
\begin{itemize}
\item [(i)] the Cauchy problem \eqref{Cauchy} has a unique solution $c\in C^{1,2}([0,T)\times (0,1))$ that is continuous up to the boundary $\{T\}\times (0,1)$. The solution $c$ is bounded on $[0,T]\times (0,1)$ and admits the stochastic representation 
\be\label{c=E}
c(t,p) = \kappa_i \E_\Q^{t,p}\bigg[\int_t^{T}\left(\frac{\theta(P(u))-r}{\sigma}\right)^2 du\bigg],\quad \forall (t,p)\in[0,T]\times (0,1), 
\ee
where $\E_\Q$ denotes taking expectation under $\Q$. Specifically, $P$ in \eqref{P'''}, when viewed under $\Q$, becomes the unique strong solution to 
\be\label{P''}
dP(u)=\left(\eta(P(u))-\beta(P(u)) \frac{\theta(P(u))-r}{\sigma}\right)du+\beta(P(u))d W_\Q(u),\ \ u\in[t, T],\ \ P(t)=p.
\ee
%where $B$ is an (arbitrary) standard Brownian motion. 
\item [(ii)] $\partial_p c$ is bounded on $[0,T]\times (0,1)$ and admits the stochastic representation 
\begin{align}\label{c_p=E}
\partial_p c(t,p) &=  \frac{2\kappa_i}{\sigma^2} (\mu_1-\mu_2) \E_\Q^{t,p}\bigg[\int_t^T \zeta(u) \Big( \theta\left(P(u)\right) -r\Big) du \bigg],
\end{align}
where %$P$ is the unique strong solution to \eqref{P''} and 
$\zeta$ is the unique strong solution to the SDE with random coefficients 
\be\label{zeta}
d\zeta(u) = \zeta(u) \Gamma(P(u))du + \zeta(u) \Lambda(P(u)) d W_\Q(u),\ \ u\in [t,T], \quad \zeta(t)=1, 
\ee
with $\Gamma(p) := \frac{d}{dp}\big(\eta(p)-\beta(p) \frac{\theta(p)-r}{\sigma}\big)$ and $\Lambda(p) := \frac{d}{dp}\beta(p)$, and the process $P$ in \eqref{c_p=E} and \eqref{zeta} is the unique strong solution to \eqref{P''}.  
\end{itemize}
\end{lemma}

\begin{remark}\label{rem:Girsanov}
In Lemma~\ref{lem:Cauchy}, ``$P(u)\in(0,1)$ for all $u\ge t$'' ensures that $\frac{\theta(P(u))-r}{\sigma}$ is a bounded process, so that $Z$ in \eqref{Z} is a strictly positive martingale by Novikov's condition. Hence, $\Q$ in \eqref{Q} is a well-defined probability measure equivalent to $\P$. By Girsanov's theorem (\cite[Theorem 3.5.1]{Karatzas2000}), $W_\Q$ in \eqref{W_Q} is a Brownian motion, adapted to the same filtration $\{\mathcal F_u\}_{t\le u\le T}$, under $\Q$. 
\end{remark}

\begin{remark}\label{rem:P'}
By \cite[Theorem 5.3]{Friedman-book-75} and the definition above it, $\zeta$ in \eqref{zeta} is the ``derivative'' of $P$ in \eqref{P''} w.r.t.\ its initial value $P(t)=p$. %Specifically, 
If we write $P^{t,p}$ for $P$ to stress the initial condition in \eqref{P''}, then $\zeta(u)$ is the limit of $(P^{t,p+h}(u)-P^{t,p}(u))/h$ in $L^2(\Omega,\mathcal F_T,\Q)$ as $h\to 0$, for all $u\ge t$.
\end{remark}

As detailed in Section~\ref{subsec:extended HJB}, by plugging the ansatz \eqref{sol. form} and  the solution $c$ to \eqref{Cauchy} for every $i=1,...,N$, denoted by $\{c_i\}_{i=1}^N$, into \eqref{V_i} and matching the coefficients on both sides, one can solve for $A_i$ and $B_i$ explicitly and the remaining terms in the equation form the Cauchy problem
\begin{equation}\label{Cauchy'}
\begin{cases}
\partial_t C+\eta(p)\partial_pC+\frac{\beta(p)^2}{2}\partial_{pp}C+R_i\big(t,p,\partial_p c_1(t,p),\cdots, \partial_p c_N(t,p)\big)=0&\hbox{for}\ (t,p)\in Q,\\
C(T,p)=0,&\hbox{for}\ p\in (0,1),%\label{terminal} 
\end{cases}
\end{equation}
where $R_i$ is given by
\begin{align}
\nonumber R_i&\big(t,p,\partial_p c_1(t,p),\cdots,\partial_p c_N(t,p)\big)\\
\nonumber&:= (\theta(p)-r)\left\{ \left(\kappa_i\frac{\theta(p)-r}{\sigma^2}-\frac{\beta(p)}{\sigma}\partial_pc_i\right)+\frac{\lambda^V_i-\lambda_i^M}{1-\overline\lambda^V}\bigg(\overline  \kappa\frac{\theta(p)-r}{\sigma^2}- \frac{\beta(p)}{\sigma}\overline{\partial_pc}\bigg)\right\}\\
\notag &\hspace{0.4in}-\frac{\gamma_i\sigma^2}{2} \left(\kappa_i\frac{\theta(p)-r}{\sigma^2}-\frac{\beta(p)}{\sigma}\partial_pc_i\right)^2   \\
\nonumber &\hspace{0.4in}-\frac{\gamma_i\beta(p)^2}{2}(\partial_pc_i)^2-\gamma_i\sigma\beta(p)\partial_pc_i\left(\kappa_i \frac{\theta(p)-r}{\sigma^2}-\frac{\beta(p)}{\sigma}\partial_pc_i\right)\\
 &=\left(\frac{\theta(p)-r}{\sigma}\right)^2 \bigg\{\left(\kappa_i+\frac{\lambda^V_i-\lambda_i^M}{1-\overline\lambda^V}\overline  \kappa\right)-\frac{\gamma_i\kappa_i^2}{2} \bigg\}-\beta(p)\left(\frac{\theta(p)-r}{\sigma}\right) \bigg\{\left(\partial_pc_i+\frac{\lambda^V_i-\lambda_i^M}{1-\overline\lambda^V}\overline{\partial_pc}\right) \bigg\}.\label{Q_i}
\end{align}
Despite the complexity of $R_i$, it is {\it bounded} thanks to the boundedness of $\partial_p c$ in Lemma~\ref{lem:Cauchy}. This allows us to establish in the next result the existence of a unique solution to \eqref{Cauchy'}, following similar arguments in the proof of Lemma~\ref{lem:Cauchy}; see Section~\ref{subsec:proof of lem:Cauchy'} for details. 

\begin{corollary}\label{lem:Cauchy'}
Suppose that the conditions of Lemma~\ref{lem:Cauchy} hold. For any $i=1,...,N$, let $c_i\in C^{1,2}([0,T)\times (0,1))$ denote the unique solution to \eqref{Cauchy} obtained in Lemma~\ref{lem:Cauchy}. Then, the Cauchy problem \eqref{Cauchy'} has a unique solution $C\in C^{1,2}([0,T)\times (0,1))$ that is continuous up to the boundary $\{T\}\times (0,1)$. The solution $C$ is bounded and admits the stochastic representation 
\be\label{C=E}
C(t,p) = \EE^{t,p}\left[\int_{t}^T R_i\big(u,P(u),\partial_p c_1(u,P(u)),\cdots, \partial_p c_N(u,P(u))\big)du\right],\quad \forall (t,p)\in[0,T]\times (0,1), 
\ee
where $P$ is the unique strong solution to \eqref{P'''}.
%\be\label{P'''}
%dP(u)=\eta(P(u))du+\beta(P(u))d W(u),\ \ u\ge t, \quad P(t)=p.
%\ee
%Moreover, the first derivative $\partial_p c$ is also bounded on $[0,T)\times (0,1)$. 
\end{corollary}

Based on the solutions to the Cauchy problems \eqref{Cauchy} and \eqref{Cauchy'}, the next result presents a Nash equilibrium as in Definition~\ref{def:E'}. The proof is relegated to Section~\ref{subsec:proof of thm:E mu unknown}. %of the $N$-player game \eqref{MV_model P}. 

\begin{theorem}\label{thm:E mu unknown}
Recall \eqref{kappa}, \eqref{kappa bar}, and \eqref{theta}. A Nash equilibrium $\bm \pi^*=(\pi^*_1,...,\pi^*_N)\in\mathcal A^N_S$ for the $N$-player game \eqref{MV_model P}, subject to the dynamics $\bm X=(X_1,...,X_N)$ in \eqref{wealth full'} and $P$ in \eqref{P}, is given by
\be\label{NE-1}
\pi_i^*(t,p) = e^{-r(T-t)}\bigg\{\frac{\theta(p)-r}{\sigma^2}\left(\kappa_i+\frac{\lambda^V_i}{1-\overline\lambda^V}\overline  \kappa\right)-\frac{\beta(p)}{\sigma}\left(\partial_p c_i +\frac{\lambda^V_i}{1-\overline\lambda^V}\overline{\partial_p c}\right)\bigg\},\ \ i=1,...,N,
\en
where $c_i(t,p)$ is the unique solution to the Cauchy problem \eqref{Cauchy} with $\eta\equiv 0$ (Lemma~\ref{lem:Cauchy})  %among functions in $C^{1,2}([0,T)\times (0,1))$ that are continuous up to the boundary $\{T\}\times (0,1)$, 
and we use the notation $\overline{\partial_p c } := \frac{1}{N}\sum^N_{i=1}\partial_p c_i $. 
Moreover, the value function under $\bm \pi^*$ is
\be\label{value}
V_i(t,x,p) = e^{r(T-t)}\left(x_i-{\lambda^M_i}\bar x\right)+C_i(t,p),\ \ i=1,...,N,
\en
where $C_i(t,p)$ is the unique solution to the Cauchy problem \eqref{Cauchy'} with $\eta\equiv 0$ (Corollary~\ref{lem:Cauchy'}). % among functions in $C^{1,2}([0,T)\times (0,1))$ that are continuous up to the boundary $\{T\}\times (0,1)$. 
\end{theorem}

\begin{remark}\label{rem:NE-1'}
By \eqref{c_p=E}, $\pi^*_i$ in \eqref{NE-1} can be expressed as 
\begin{equation}\label{NE-1 with g}
\pi_i^*(t,p) = e^{-r(T-t)} g(t,p) \left(\kappa_i+\frac{\lambda^V_i}{1-\overline\lambda^V}\overline  \kappa\right),
\end{equation}
with 
\begin{equation}\label{g}
g(t,p) := \frac{\theta(p)-r}{\sigma^2} -\frac{2 \beta(p)}{\sigma}(\mu_1-\mu_2) \E_\Q^{t,p}\bigg[\int_t^T \zeta(u) \frac{\theta\left(P(u)\right) -r}{\sigma^2} du \bigg].
\end{equation}
\end{remark}

%In the first term, the crucial value 
%$$\kappa_i+\frac{\lambda^V_i}{1-\overline\lambda^V}\overline  \kappa$$
%is increasing in $\lambda^V$ but decreasing in $\lambda^M$ and $\gamma$ implying that investors consider investing more risky assets in small risk aversion and weaker relative consideration in mean but stronger consideration in variance. In this case, investors intend to maximize their own expected return rather than to beat the ensemble average with smaller $\lambda^M$ and to control the variance of the difference between their own wealth and the ensemble average with larger $\lambda^V$.  

The first term in \eqref{NE-1} is of the exact form of \eqref{NE_constant}. This means that under partial information, investors still wish to trade according to \eqref{NE_constant}---they simply replace the unknown $\mu$ by the estimate $\theta(p)=p\mu_1+(1-p)\mu_2$ in \eqref{theta}, the average of $\mu_1$ and $\mu_2$ (the two possible values of $\mu$) based on the current ``judgement'' $p = P(t) = % \widehat{\mathfrak p}_1(t) =
\P(\mu=\mu_1\mid \mathcal F^S_t)$ in \eqref{p_j} at time $t$. 
%As the judgement may change over time, 
The second term in \eqref{NE-1}, on the other hand, serves to {\it hedge against} the fluctuations in $P(\cdot)$ over time. 

This is reminiscent of the results in \cite{BC10}. To see this, first consider the case $\lambda_i^M=\lambda_i^V=0$ (i.e., investor $i$ disregards relative performance). As $\kappa_i$ now reduces to $1/\gamma_i$ by \eqref{kappa}, $c_i(t,p)$ as in \eqref{c=E} coincides with the {\it anticipated portfolio gains} in \cite[(23)]{BC10}, with the instantaneous stock mean return $\mu_s$ therein replaced by the estimated return $\theta(P(s))$. Moreover, $\pi_i^*(t,p)$ in \eqref{NE-1} becomes 
\begin{equation}\label{NE-1 lambda=0}
\pi_i^*(t,p) = e^{-r(T-t)}\bigg\{\frac{\theta(p)-r}{\gamma_i\sigma^2}-\frac{\beta(p)}{\sigma}\partial_p c_i \bigg\}, 
\end{equation}
which aligns with the investment policy in \cite[(26)]{BC10}. Indeed, by viewing the judgement process $P(\cdot)$ as an underlying market factor (i.e., in the role of $X$ in \cite{BC10}), the term $-\frac{\beta(p)}{\sigma}\partial_p c_i$ corresponds exactly to the last term in \cite[(26)]{BC10}. This allows the interpretation of myopic trading and intertemporal hedging in \cite{BC10} to be applied to our setting of partial information. The term ${(\theta(p)-r)}/{\gamma_i\sigma^2}$ in \eqref{NE-1 lambda=0} represents investor $i$'s myopic demand, which optimizes trading based on the currently estimated stock return $\theta(p)$ under partial information, %; the term $-\frac{\beta(p)}{\sigma}\partial_p c_i$ stands for investor $i$'s hedging demand, which accounts for future changes of investment opportunities due to fluctuations in $P(\cdot)$, 
without accounting for future changes of investment opportunities due to fluctuations in $P(\cdot)$. 
The term $-\frac{\beta(p)}{\sigma}\partial_p c_i$ in \eqref{NE-1 lambda=0}, in response, is investor $i$'s hedging demand---as explained below \cite[Proposition 1]{BC10}, the sensitivity of anticipated portfolio gains (i.e., $c_i(t,p)$ in our case) to the underlying market factor (i.e., $P(\cdot)$ in our case) identifies the investor's demand to hedge against future fluctuations in the underlying factor.  

The above interpretation extends to the general case $\lambda_i^M, \lambda_i^V\ge 0$. By \eqref{c_p=E}, the trading strategy under $\lambda_i^M=\lambda_i^V = 0$ (i.e., \eqref{NE-1 lambda=0}) can be expressed as $\pi_i^*(t,p) = e^{-r(T-t)} g(t,p)/\gamma_i$, with $g(t,p)$ given by \eqref{g}. Comparing this with \eqref{NE-1 with g}, an equivalent expression of the trading strategy \eqref{NE-1}, we find that the interpretation of myopic trading and intertemporal hedging still holds generally under $\lambda_i^M, \lambda_i^V\ge 0$, and the only change is that $1/\gamma_i$ %(the reciprocal of investor $i$'s risk aversion) 
in the $\lambda_i^M=\lambda_i^V=0$ case now becomes $\kappa_i+({\lambda^V_i}/{(1-\overline\lambda^V)})\overline  \kappa$, which is a linear combination of $\{1/\gamma_j\}_{j=1}^N$ depending on the relative performance coefficients $\{\lambda_j^M\}_{j=1}^N$ and $\{\lambda_j^V\}_{j=1}^N$ (see Remark~\ref{rem:1/gamma to kappa}). 

That is, Theorem~\ref{thm:E mu unknown} spells out how partial information and relative performance criteria jointly influence the results in \cite{BC10}, where mean-variance portfolio selection is studied for a single investor with full information. Simply put, partial information is dealt with by introducing an (artificial) market factor, i.e., the judgement process $P(\cdot)$, for estimating the unobserved stock mean return, while relative performance criteria alter investors' risk appetite---each investor now considers not only her own risk aversion coefficient but all other investors'. %, in an explicit (but nonlinear) manner. 
%by considering not only her own risk aversion coefficient but all other investors', the investor comes up with a weighted coefficient in a nonlinear (but explicit) manner.

\begin{remark}[Increased risk appetite]\label{re_1}
If $\lambda_i^M=\lambda^V_i=\lambda_i$ for all $i=1,...,N$, \eqref{NE-1 with g} simplifies to  
\begin{equation}\label{NE-2}
\pi_i^*(t,p) = e^{-r(T-t)} g(t,p) \left(\gamma_i^{-1}+\frac{\lambda_i}{1-\overline\lambda}\overline\gamma^{(-1)}_N\right),
\end{equation}
%\ba\label{NE-2}
%\nonumber \pi_i^*(t,p)&=&e^{-r(T-t)}\bigg\{\frac{\theta(p)-r}{\sigma^2}\left(\gamma_i^{-1}+\frac{\lambda_i}{1-\overline\lambda}\overline\gamma^{(-1)}\right)-\frac{\beta(p)}{\sigma}\left(\partial_p c_i +\frac{\lambda_i}{1-\overline\lambda}\overline{\partial_p c}\right)\bigg\}
%\ea
with $\overline\lambda:=\frac{1}{N}\sum_{j=1}^N\lambda_j$ and $\overline\gamma^{(-1)}_N:=\frac{1}{N}\sum_{j=1}^N\gamma_j^{-1}$.  
%Similar to the general case, $\left(\gamma_i^{-1}+\frac{\lambda_i}{1-\overline\lambda}\overline\gamma^{(-1)}\right)$ is increasing in $\lambda_i$ but decreasing in $\gamma_i$. The investors with risk preference with smaller $\gamma$ and stronger relative consideration intend to have more risky assets. It is interesting to observe that compared to the previous case with the varied $\lambda^M$ and $\lambda^V$, the objective function for the variance dominates the one for the mean. 
We see that investor $i$'s risky position is larger in magnitude when relative performance is considered (i.e., $\lambda_i>0$) than when it is disregarded (i.e., $\lambda_i=0$). Such increased risk appetite can be ascribed to the intent to enlarge $\E^{t,\bm x, p}\big[X_i(T)-\lambda_i\overline X(T)\big]$ in \eqref{MV_model P}, 
%the mean-variance objective \eqref{MV_model P}, particularly the mean term $\E^{t,\bm x, p}\big[X_i(T)-\lambda_i\overline X(T)\big]$
for which $X_i(T)$ needs to constantly exceed a random variable (i.e., $\lambda_i\overline X(T)$) if $\lambda_i>0$ or a constant (i.e., zero) if $\lambda_i=0$. Hence, the wealth process $X_i$ has to be more volatile (corresponding to enlarged magnitude of the risky position) in the former case than in the latter. 
\end{remark}

\begin{remark}\label{re_2}
If $\lambda_i^V=0$ for all $i=1,...,N$, 
%\begin{align*}
%\nonumber J_i(t, \bm x, p, {\overline \pi}_{(-i)}, \pi_i)&:=\EE^{t,\bm x, p}\left[\left(1-\frac{\lambda^M_i}{N}\right)X_i(T)-\lambda^M_i\overline X_{(-i)}(T)\right] -\frac{\gamma_i}{2}\mathrm{Var}^{t,\bm x, p}\left[X_i(T)\right], %\ \bigg |\ \bm X(t)= \bm x,\ P(t)=p\right].
%\end{align*}
investors consider a relative performance criterion for expected wealth, but not for the variance of wealth; recall \eqref{MV_model P}. Note that \eqref{NE-1 with g} now reduces to $\pi_i^*(t,p)=e^{-r(T-t)}g(t,p) \kappa_i$, 
%\be\label{NE-3}
%$\pi_i^*(t,p)=e^{-r(T-t)}\bigg\{\frac{\theta(p)-r}{\sigma^2} \kappa_i-\frac{\beta(p)}{\sigma} \partial_p c_i\bigg\}. $
%\en
which no longer depends on other investors' parameters. % (as $\kappa_i$ in \eqref{kappa} depends on only the $i$-th investor's parameters $\gamma_i$, $\lambda^M_i$, and $\lambda^V_i$).  
%The value $\kappa_i$ is reduced by $\left(1-\frac{\lambda^M_i}{N}\right)/\gamma_i$ implying that investors consider more risky assets in the case of weaker risk aversion and relative consideration through smaller $\gamma_i$ and $\lambda_i^M$. 
\end{remark}

Interestingly, our mathematical results can further explain the size of intertemporal hedging, by closely examining the {\it precision} of the judgement process $P(\cdot)$ over time.\footnote{We use ``precision'' as in the usual context of ``accuracy versus precision'' in taking measurements. Specifically, ``precision'' means how close the measurements (i.e., posterior probabilities at different times) are to each other.} %(under an equivalent measure $\Q$). 
%adjust the first term, based on (i) the current judgement {\it per se} and (ii) the {\it precision} of judgements over time\footnote{We use ``precision'' as in the usual context of ``accuracy versus precision'' in taking measurements. Specifically, ``precision'' means how close the measurements (i.e., posterior probabilities at different times) are to each other.} (under an equivalent measure $\Q$). 
%Specifically, $\beta(p)=\frac{\mu_1-\mu_2}{\sigma}p(1-p)$ in \eqref{NE-1} controls the level of adjustment through our current judgement $p=\widehat{\mathfrak p}_1(t)$. When $p$ is close to 1 (resp.\ 0), we are quite confident of $\mu=\mu_1$ (resp.\ $\mu=\mu_2$) at time $t$. In addition, thanks to \eqref{theta}, $\theta(p)$ is close to $\mu_1$ (resp.\ $\mu_2$), the likely true value of $\mu$. As $\theta(p)$ is likely close to $\mu$, the first term in \eqref{NE-1} should readily well approximate \eqref{NE_constant}. Hence, little adjustment is needed, corresponding to a small $\beta(p)$. When $p$ is far away from 1 and 0 (i.e., near $1/2$), as $\theta(p)$ now stands halfway between $\mu_1$ and $\mu_2$, the first term in \eqref{NE-1} clearly differs from \eqref{NE_constant}. Stronger adjustment is then required, corresponding to a larger $\beta(p)$. 
By Lemma~\ref{lem:hW}, starting with $p\in (0,1)$ at time $t\in[0,T)$, $P^{t,p}(\cdot)={\mathfrak p}_1(\cdot)$ evolves continuously as the SDE \eqref{P}. Intuitively, if our judgements over time are ``{imprecise},'' in the sense that $P^{t,p}(\cdot)$ will oscillate wildly on $[t,T]$, the myopic trading based on $p$ (i.e., the first term in \eqref{NE-1}) should be hedged against significantly. If our judgements over time are ``{precise},'' in the sense that $P^{t,p}(\cdot)$ will stay near $p$ on $[t,T]$, %the myopic trading based on $p$ could be good enough and there should be 
the hedging demand should be much less. Such nuances, in fact, are readily encoded in $\{\partial_p c_i\}_{i=1}^N$. 
%materialized through $\{\partial_p c_i\}_{i=1}^N$ in the second term of \eqref{NE-1}. 

By \eqref{c_p=E}, $\partial_p c_i(t,p)$ is determined by $\zeta(\cdot)$ in \eqref{zeta} and $\theta(P^{t,p}(\cdot))-r$ on $[t,T]$, under a measure $\Q$ equivalent to $\P$. As $\zeta(\cdot)$ is the stochastic exponential of $\int_t^\cdot\Gamma(P^{t,p}(u))du +\int_t^\cdot \Lambda(P^{t,p}(u)) d W_\Q(u)$, it takes values on the entire half line $(0,\infty)$. By contrast, $\theta(P^{t,p}(\cdot))-r$ lies within $[\mu_2-r,\mu_1-r]$ and can take negative values if $\mu_2<r$. %It then follows from \eqref{c_p=E} that $\zeta(\cdot)$ mainly affects the magnitude of $\partial_p c_i$, while $\theta(\widehat{\mathfrak p}_1(u))-r$ influences the sign of $\partial_p c_i$. 
Now, given $h\neq 0$, consider $\tau(h):= \inf\{t'\ge 0: P^{0,p}(t') = p+h\}$ and observe that for any $u\in [t,T]$,
\begin{equation}\label{zeta=P'}
\zeta(u) = \lim_{h\to 0}\frac{P^{t,p+h}(u)-P^{t,p}(u)}{h} = \lim_{h\to 0}\frac{P^{t,p}(u+\tau(h))-P^{t,p}(u)}{h} \quad \hbox{in $L^2(\Omega,\mathcal F_T,\Q)$},%\quad \forall u\in [t,T],
\end{equation}
where the first equality is due to Remark~\ref{rem:P'} and the second equality stems from the time-homogeneity and uniqueness of strong solutions of $P$ in \eqref{P}. Since $\tau(h)\downarrow 0$ as $h\to 0$, $\zeta(u)$ measures the rate of change of $P^{t,p}(\cdot)$ at time $u$ (under $\Q$), capturing how intensely $P^{t,p}(\cdot)$ will move away from its current value $P^{t,p}(u)$ (under $\Q$). Hence, if $P^{t,p}(\cdot)$ is ``precise'' under $\Q$ (i.e., $\zeta(u)$ is small for all $u\in[t,T]$), $|\partial_p c_i|$ will be small for all $i=1,...,N$ by \eqref{c_p=E}. This shrinks the second term of \eqref{NE-1} (i.e., the hedging demand), consistently with our intuition.  
%leaving the first term to take command. 
Similarly, if $P^{t,p}(\cdot)$ is ``imprecise'' under $\Q$ (i.e., $\zeta(u)$ is large for some $u\in[t,T]$), $|\partial_p c_i|$ can be large for all $i=1,...,N$. In some cases, this significantly inflates the second term of \eqref{NE-1} (i.e., the hedging demand), which alters the myopic trading drastically; see Section~\ref{subsec:discuss} for concrete illustrations.  

%By Lemma~\ref{lem:p to 1} and $\Q$ being equivalent to $\P$, we have  $P^{t,p}(u)=\widehat{\mathfrak p}_1(u)\to 1$ (or 0) under $\Q$ when $\mu=\mu_1$ (or $\mu=\mu_2$). If the current judgement $p=P^{t,p}(t)$ (computed under $\P$) is near $1/2$, it must be ``unreliable'' under $\Q$, as $P^{t,p}(\cdot)$ under $\Q$ will eventually converge to 1 or 0. Hence, by the arguments below \eqref{zeta=P'}, the second term of \eqref{NE-1} tends to be large. If the current judgement $p$ is near 1, as investors do not know the true value of $\mu$, it is not immediately clear whether $P^{t,p}(\cdot)$ will remain close to 1 (and thus converges to 1 eventually) or move a long way towards the other end (and converges to 0 eventually).  

%Note that we have the same results in Remarks \ref{re_1}-\ref{re_2} for the Nash equilibrium \eqref{NE_Markov}. 

%%%%%%%%%%%%%%%%%%%%%%%%%%%%%%%%%%%%%

%%%%%%%%%%%%%%%%%%%%%%%%%%%%%%%%%%%%%
%%%%%%%%%%%%%%%%%%%%%%%%%%%%%%%%%%%%%

\section{The Case of Alternating Investment Opportunities}\label{sec:alternating mu}
%In Section~\ref{sec:constant mu}, the stock $S$ is assumed to have a {\it constant} expected return $\mu\in\R$, yet investors might only know its two possible values $\mu_1$ and $\mu_2$ (with $\mu_1>\mu_2$) and are uncertain which is the true value. 
In this section, we take $\mu:\Omega\times [0,\infty)\to\R$ in \eqref{risky_0} to be a continuous-time Markov chain that alternates between two values $\mu_1$ and $\mu_2$ (with $\mu_1>\mu_2$). Specifically,
\begin{equation}\label{mu process}
\mu(u) := 
\begin{cases}
\mu_1,\quad \hbox{if}\ M(u)=1,\\
\mu_2,\quad \hbox{if}\ M(u)=2,
\end{cases}
\qquad \forall u\ge 0, 
\end{equation}
where $M:\Omega\times [0,\infty)\to\{1,2\}$ is a continuous-time Markov chain whose generator is given by
\begin{equation}\label{generator}
    G= \left(
       \begin{array}{cc}
         -q_1& q_1 \\
       q_2 & -q_2 \\
       \end{array}
     \right),
     \end{equation}
where $q_1, q_2>0$ are given constants. The stock dynamics \eqref{risky_0} then becomes the regime-switching model in Dai et al.\ \cite{DZZ2010}, which admits a concrete financial interpretation: ``$M(t)=1$'' represents the good state (i.e., a bull market) where $S$ is expected to grow at the larger rate $\mu_1$, while ``$M(t)=2$'' represents the bad state (i.e., a bear market) where $S$ is expected to grow at the smaller rate $\mu_2$. We further assume that the processes $M$ and $W$ are independent. This essentially stipulates that the transitions between a bull market and a bear market are determined by macro-economic factors and thus distinct from the day-to-day micro-level randomness (encoded through $W$) that drives the asset price volatility. 

%We then modify the stock dynamics \eqref{risky_1} to 
%\begin{align}
%\label{risky} dS(u)&=\mu(M(u)) S(u)du+\sigma S(u) dW(u),\quad S(t)=s,
%%\label{riskless}dS_0(u)&=&rS_0(u)du,\; S_0(t)=s_0,
%\end{align}
%where $\mu:\{1,2\}\to\R$ is defined by $\mu(1) = \mu_1$ and $\mu(2) =\mu_2$, with $\mu_1>\mu_2$. That is, the expected return $\mu$ of the stock $S$ {\it alternates} between $\mu_1$ and $\mu_2$ (with $\mu_1>\mu_2$). 

%%%%%%%%%%%%%%%%%%%%%%%%%%%%%%%%

\subsection{Analysis under Full Information $\F^{\mu,S}$}\label{subsec:full alternating mu}
Assume that investors have access to $\F^{\mu,S}$. The dynamics of $S$ in \eqref{risky_0}, including $\mu(u)$ as in \eqref{mu process} and $dW(u)$, is then fully known. %Equivalently speaking, investors always observe the present market state $M(u)$. 
%Assume that the dynamics of $S$ in \eqref{risky} is fully known; in particular, investors observe the process $M(t)$, i.e., know precisely the present market state. As a result, for any $i=1,...,N$ and $\pi_i\in\mathcal A$, the wealth process of investor $i$ at time $t\in [0,T)$ is
%\ba\label{wealth full''}
 %dX_i(u)= \big(rX_i(u)+\pi_i(u)\big(\mu(M(u))-r\big)\big)du+\pi_i(u)\sigma d W(u),\ \ u\in[t,T],\quad X_i(t)=x_i\in\R, %\\
%%\nonumber dp(u)&=&\eta(p(u))du+\beta(p(u))d\widehat W(u),\quad p(t)=p,
%\ea 
%which is simply \eqref{wealth full} with $\mu$ therein replaced by $\mu(M(t))$. 
%Similarly, the average wealth (${\overline X}$) and the average wealth excluding investor $i$ (${\overline X}_{(-i)}$) are given by \eqref{bc_ave} and \eqref{bc_ave_-i}, respectively, with $\mu$ therein replaced by $\mu(M(t))$. 

\begin{remark}\label{rem:Markov mu alternating known}
For any $t\ge 0$, because $\{\mu(t) = \mu_j\} = \{M(t) = j\}$ for $j=1,2$, we will substitute ``conditioning on $M(t) = m\in\{1,2\}$'' for ``conditioning on $\mu(t)=b\in\{\mu_1,\mu_2\}$,'' whenever it is convenient (e.g., in \eqref{integrability}). % (and correspondingly the variable $m\in\{1,2\}$ for $b\in\{\mu_1,\mu_2\}$)
Similarly, for a Markov $\bm \pi=(\pi_1,...,\pi_N)\in\mathcal A_{\mu,S}^N$, we will substitute $\xi_i (t,\bm X(t),M(t))$ for $\xi_i (t,\bm X(t),\mu(t))$ in Definition~\ref{def:Markov}. 
\end{remark}

For each $i=1,2,...,N$, at any current time $t\in[0,T]$, wealth levels $\bm x=(x_1,\cdots,x_N)\in\R^N$, and market state $m\in\{1,2\}$, given the trading strategies of the other $N-1$ investors (i.e., $\pi_j\in\mathcal A_{\mu,S}$ for all $j\neq i$), investor $i$, in line with \eqref{MV_model}, looks for a trading strategy $\pi_i\in\mathcal A_{\mu,S}$ that maximizes the mean-variance objective % $U_i(t)=U_i(t,\bm X(t) ,p(t))$ is 
%the $i$-th investor's mean-variance objective \eqref{MV_model} then becomes
\begin{align}
J_i\big(t, \bm x, m, \{\pi_j\}_{j\neq i}, \pi_i\big)&:=\EE^{t,\bm x,m}\left[X_i(T)-\lambda^M_i\overline X(T)\right]-\frac{\gamma_i}{2}\mathrm{Var}^{t,\bm x,m}\left[X_i(T)-\lambda^V_i\overline X(T)\right],\label{MV_model Z}
%:=\EE^{t,\bm x,m}\left[\left(1-\frac{\lambda^M_i}{N}\right)X_i(T)-\lambda^M_i\overline X_{(-i)}(T)\right]\\
%&\hspace{0.2in}-\frac{\gamma_i}{2}\mathrm{Var}^{t,\bm x,m}\left[\left(1-\frac{\lambda^V_i}{N}\right)X_i(T)-\lambda^V_i\overline X_{(-i)}(T)\right].\label{MV_model Z}
\end{align}
where the superscript ``${t,\bm x,m}$'' denotes conditioning on $\bm X(t)=\bm x$ and $M(t)=m$. 

%Due to the additional state process $M$, Markov trading strategies (Definition \ref{def:Markov}) need to be modified accordingly.

%\begin{definition}\label{def:Markov'}
%We say $\bm \pi=(\pi_1,...,\pi_N)\in\mathcal A^N$ is Markov for the $N$-player game \eqref{MV_model Z}, if for any $i=1,...,N$, there exists a Borel measurable $\xi_i:[0,T]\times\R^d\times \{1,2\}\to \R$ such that $\pi_i(t) = \xi_i(t,\bm X(t), M(t))$ for a.e.\ $t\in[0,T]$ a.s.  We will write $\pi_i$ and $\xi_i$ interchangeably. 
%\end{definition}

%Similarly to \eqref{concatenate}, given two Markov $\bm\pi, \bm\theta\in\mathcal A^N$ for the game \eqref{MV_model Z} and $s\in (0,T)$, we can define the concatenation of $\bm \theta$ and $\bm \pi$ at time $s$, denoted by $\bm\theta\otimes_s\bm\pi = (\theta_1\otimes_s\pi_1,...,\theta_N\otimes_s\pi_N)$, as
%\[
%(\theta_i\otimes_s\pi_i)(t,\bm x,m) := 
%\begin{cases}
%\theta_i(t,\bm x,m),\quad &\hbox{for}\ 0\le t<s,\ \bm x\in\R^N,\ m\in\{1,2\},\\
%\pi_i(t,\bm x,m),\quad &\hbox{for}\ s\leq t\le T,\ \bm x\in\R^N,\ m\in\{1,2\}, 
%\end{cases} 
%\quad \forall i=1,...,N. 
%\]
%Note that $\bm\theta\otimes_s\bm\pi\in\mathcal A^N$ is Markov for the game \eqref{MV_model Z} by construction. We now define a Nash equilibrium for \eqref{MV_model Z} in line with Definition~\ref{def:E}.

\begin{definition}\label{def:E''}
We say $\bm \pi^*=(\pi^*_1,...,\pi^*_N)\in\mathcal A^N_{\mu,S}$ is a Nash equilibrium for the $N$-player game \eqref{MV_model Z}, subject to the dynamics $\bm X=(X_1,...,X_N)$ in \eqref{wealth full}  with $\mu(u)$ as in \eqref{mu process}, if $\bm \pi^*$ is Markov %w.r.t.\ $(\bm X,M)$ %for \eqref{MV_model Z} 
and for any $(t,\bm x,m)\in[0,T)\times\R^N\times \{1,2\}$ and Markov $\bm\pi=(\pi_1,...,\pi_N)\in\mathcal A_{\mu,S}^N$, % w.r.t.\ $(\bm X,M)$, %for \eqref{MV_model Z}, 
\be\label{intra E''}
\liminf_{h\downarrow 0}\frac{J_i\big(t,\bm{x},m, \{\pi^*_j\}_{j\neq i},\pi_i^*\big)-J_i\big(t,\bm{x},m, \{\pi^*_j\}_{j\neq i},{\pi_i\otimes_{t+h}\pi^*_i}\big)}{h}\geq 0,\quad \forall i=1,...,N.
\en
\end{definition}

%The $i$-th investor, for all $i=1,...,N$, looks for an investment strategy $\pi_i\in\mathcal A$ that maximizes the mean-variance objective \eqref{MV_model}. Our goal is to find a Nash equilibrium for this $N$-player game as formulated in Definition~\ref{def:E}. 

%\be\label{X_i_MV}
%dX_i(u)=rX_i(t)+\pi_i(u)(\mu(\alpha(u))-r)du+\pi_i(u)\sigma d W(u),\; X_i(t)=x_i. 
%\en 

Such a Nash equilibrium is presented in the next result, whose proof is relegated to Section~\ref{sec:proof of main'}.
 
\begin{theorem}\label{thm:E M known} 
Recall \eqref{kappa}, \eqref{kappa bar}, and \eqref{theta}. A Nash equilibrium $\bm \pi^*=(\pi^*_1,...,\pi^*_N)\in\mathcal A_{\mu,S}^N$ for the $N$-player game \eqref{MV_model Z}, subject to the wealth dynamics \eqref{wealth full} with $\mu(u)$ as in \eqref{mu process}, is given by
\be\label{NE_Markov}
\pi_i^*(t,m)=e^{-r(T-t)}\bigg\{\frac{\mu_m-r}{\sigma^2}\left(\kappa_i+\frac{\lambda^V_i}{1-\overline\lambda^V}\overline  \kappa\right)\bigg\},\ \ i=1,...,N.
\en
Moreover, the value function under the Nash equilibrium $\bm \pi^*$ is
\be\label{value'}
V_i(t,x,m)=e^{r(T-t)}\left(x_i- {\lambda^M_i}\overline x\right)+C_i(t,m),\ \ i=1,...,N. 
\en
where $C_i(t,m)$, $m\in\{1,2\}$, is defined as
\ba
\label{C_1_M}C_i(t,1)&:=&\frac{q_2\widetilde Q_i^1+q_1\widetilde Q_i^2}{q_1+q_2}(T-t)+\frac{q_1}{(q_1+q_2)^2}\left(\widetilde Q_i^1-\widetilde Q_i^2\right)\left(1-e^{(q_1+q_2)(T-t)}\right)\\
\label{C_2_M} C_i(t,2)&:=&\frac{q_2\widetilde Q_i^1+q_1\widetilde Q_i^2}{q_1+q_2}(T-t)-\frac{q_2}{(q_1+q_2)^2}\left(\widetilde Q_i^1-\widetilde Q_i^2\right)\left(1-e^{(q_1+q_2)(T-t)}\right)
\ea
with $\widetilde Q_i^m$, $m=1,2$, given by
\ban
\nonumber \widetilde Q_i^m
&:=&\left(\frac{\mu_m-r}{\sigma}\right)^2 \bigg\{\left(\kappa_i-\frac{\lambda^V_i-\lambda_i^M}{1-\overline\lambda^V}\overline  \kappa\right)-\frac{\gamma_i\kappa_i^2}{2} \bigg\}.
\ean
\end{theorem}

%%%%%%%%%%%%%%%%%%%%%%%%%%%%%%%%%%%%%%%

\subsection{Analysis under Partial Information $\F^S$}\label{subsec:partial alternating mu}
Assume that investors have access to only $\F^{S}$, i.e., they see the stock $S$ in \eqref{risky_0} evolve over time, but the alternating expected return $\mu(u)$, defined as in \eqref{mu process}, remains unknown. Equivalently speaking, investors observe only the current stock price $S(u)$, but not the current market state $M(u)$.
As a result, the Nash equilibrium formula in Theorem~\ref{thm:E M known} is no longer of use and we will rely on the posterior probability \eqref{p_j_0}, which now takes the form: given a (subjective) prior probability $p_0\in(0,1)$ of the event $\{M(0) = 1\}$, 
%For any $t\ge 0$, given that the prior distribution of $M(t)$ (before observations of $S$ are made) is $M(t)=1$ with some probability $p\in (0,1)$, we consider the posterior probability
\be\label{p_j'}
{\mathfrak p}_j(u) =\PP(M(u)=j\mid \mathcal F^S_u),\quad j=1,2,\quad \forall u> 0.
\ee 
Similarly to Lemma~\ref{lem:hW}, ${\mathfrak p}_j(\cdot)$ above can be characterized as the unique strong solution to an SDE, as shown in the next result (whose proof is relegated to Section~\ref{subsec:proof of lem:tW}). 
 
\begin{lemma}\label{lem:tW}
Fix any $t\ge 0$. 
\begin{itemize}
\item [(i)] Let $B$ be any standard Brownian motion. For any $p \in (0,1)$, the SDE 
\be\label{P eta'}
dP(u)=\Big(-(q_1+q_2)P(u)+q_2\Big)du + \frac{\mu_1-\mu_2}{\sigma}P(u)(1-P(u))  dB(u),\ \ u\ge t,\quad P(t) =p,
\ee
has a unique strong solution, which satisfies $P(u)\in(0,1)$ for all $u\ge t$ a.s.
\item [(ii)] Given $S$ in \eqref{risky_0} (with $\mu(u)$ as in \eqref{mu process}) and $\mathfrak{p}_1$ in \eqref{p_j'}, the process $\widehat W$ in \eqref{Inno_W}
%\be\label{Inno_W'}
%\widetilde W(u) := \frac1\sigma \bigg[\log\bigg(\frac{S(u)}{S(t)}\bigg)-(\mu_1-\mu_2)\int_t^u\widetilde{\mathfrak p}_1(s)ds-\left(\mu_2-\frac{\sigma^2}{2}\right)(u-t)\bigg],\quad t\ge 0,
%\en
is a standard Brownian motion adapted to $\{\mathcal F^S_u\}_{u\ge t}$. Moreover, ${\mathfrak p}_1$ % in \eqref{p_j} 
is the unique strong solution to 
\be\label{P'}
dP(u)=\Big(-(q_1+q_2)P(u)+q_2\Big)du +\frac{\mu_1-\mu_2}{\sigma}P(u)(1-P(u))  d\widehat W(u),\ \ u\ge t\quad P(t) = {\mathfrak p}_1(t) \in (0,1). 
\ee
Hence, $S$ in \eqref{risky_0} can be expressed equivalently as
\ba 
\label{risky_1''} dS(u)= \Big((\mu_1-\mu_2)P(u)+\mu_2\Big) S(u)du+\sigma S(u) d\widehat W(u),\ \ u\ge t\quad S(t)=s>0,%\\
%\label{riskless_1}dS_0(u)&=&rS_0(u)du,\; S_0(t)=s_0,
\ea
where $P$ is the unique strong solution to \eqref{P'}.
\end{itemize}
\end{lemma}
The importance of \eqref{risky_1''} is that it rewrites $S$ in \eqref{risky_0}, which depends on the unknown $M(u)$ (through $\mu(u)$ defined in \eqref{mu process}), in terms of the known constants $\mu_1$, $\mu_2$, $\sigma$, and the observable process $P(\cdot) = {\mathfrak p}_1(\cdot)$. When investors view the stock $S$ as \eqref{risky_1''}, their wealth processes can also be expressed in terms of $P$ in \eqref{P'} and $\widehat W$ in \eqref{Inno_W}, such that the dynamics of every process involved is fully observable. Specifically, for any $i=1,...,N$ and $\pi_i\in\mathcal A_S$, the wealth process \eqref{wealth full} of investor $i$ can be equivalently expressed as \eqref{wealth full'}, 
%\be\label{wealth full'''}
 %dX_i(u) = \Big[ rX_i(u)+\pi_i(u)\big((\mu_1-\mu_2)P(u)+\mu_2-r\big)\Big]du+\pi_i(u)\sigma d\widetilde W(u),\ \ u\in [t,T],\ \ X_i(t)=x_i,
%\ee
where $P$ is now the unique strong solution to \eqref{P'}. %This is simply \eqref{wealth full'}, with $P$ and $\widehat W$ therein replaced by the unique strong solution to \eqref{P'} and $\widetilde W$, respectively. %Similarly, the average wealth ${\overline X}$  %and the average wealth excluding the $i$-th investor ${\overline X}_{(-i)}$ are 
%is given by \eqref{bc_ave P}, %and \eqref{bc_ave_-i P}, 
%with $P$ and $\widehat W$ therein replaced by the unique strong solution to \eqref{P'} and $\widetilde W$. 
For each $i=1,...,N$, given the current time $t\in[0,T]$, wealth levels $\bm x=(x_1,\cdots,x_N)\in\R^N$, and posterior probability $p\in(0,1)$ of the event $\{M(t)=1\}$, as well as the trading strategies of the other $N-1$ investors (i.e., $\pi_j\in\mathcal A_S$ for all $j\neq i$), investor $i$ looks for a trading strategy $\pi_i\in\mathcal A_S$ that maximizes the mean-variance objective \eqref{MV_model P}, with $P$ therein being the unique strong solution to \eqref{P'}. Our goal is to find a Nash equilibrium for this $N$-player game defined as below. %in Definition~\ref{def:E'}. 

\begin{definition}\label{def:E''}
We say $\bm \pi^*=(\pi^*_1,...,\pi^*_N)\in\mathcal A^N_S$ is a Nash equilibrium for the $N$-player game \eqref{MV_model P}, subject to the dynamics $\bm X=(X_1,...,X_N)$ in \eqref{wealth full'} and $P$ in \eqref{P'}, if $\bm\pi^*$ is Markov %w.r.t.\ $(\bm X, P)$ % for \eqref{MV_model P} 
and for any $(t,\bm x,p)\in[0,T)\times\R^N\times (0,1)$ and Markov $\bm\pi=(\pi_1,...,\pi_N)\in\mathcal A_S^N$, %w.r.t.\ $(\bm X, P)$, %for \eqref{MV_model P}, 
\eqref{intra E'} holds. 
\end{definition}

\begin{theorem}\label{THM_1}
Recall \eqref{kappa}, \eqref{kappa bar}, and \eqref{theta}. A Nash equilibrium $\bm \pi^*=(\pi^*_1,...,\pi^*_N)\in\mathcal A_S^N$ for the $N$-player game \eqref{MV_model P}, subject to the dynamics $\bm X=(X_1,...,X_N)$ in \eqref{wealth full'} and $P$ in \eqref{P'}, is given by \eqref{NE-1}, 
%\be\label{NE-1}
%\pi_i^*(t,p) = e^{-r(T-t)}\bigg\{\frac{\theta(p)-r}{\sigma^2}\left(\kappa_i+\frac{\lambda^V_i}{1-\overline\lambda^V}\overline  \kappa\right)-\frac{\beta(p)}{\sigma}\left(\partial_p c_i +\frac{\lambda^V_i}{1-\overline\lambda^V}\overline{\partial_p c}\right)\bigg\},
%\en
where $c_i$ is the unique solution to the Cauchy problem \eqref{Cauchy} obtained in Lemma~\ref{lem:Cauchy}, with $\eta$ therein taken to be
\be\label{eta}
\eta(p):= -(q_1+q_2)p+q_2\quad p\in[0,1].
\ee
%among functions in $C^{1,2}([0,T)\times (0,1))$ that are continuous up to the boundary $\{T\}\times (0,1)$. %, and we use the notation $\overline{\partial_p c } := \frac{1}{N}\sum^N_{i=1}\partial_p c_i $. 
%\be\label{c_i_T}
%c_i(t,p)=\widetilde\EE\left[\int_{t}^T \kappa_i\left(\frac{\mu(p(u))-r}{\sigma}\right)^2du\bigg |{\mathcal{F}}(t)\right],
%\en
%where $\widetilde\EE$ denotes the expected value under the risk neutral measure $\widetilde\PP$ subject to 
%\be\label{p_1}
%dp(u)=\left(\alpha(p(u))-\beta(p(u)) \left(\frac{\mu(p(u))-r}{\sigma}\right)\right)du+\beta(p(u))d\widetilde W(u),\; p(t)=p,
%\en
%where $\widetilde W$ is a standard correlated Brownian motion under the new measure $\widetilde\PP$ and 
%\be\label{C_i_T}
%C_i(t,p)=\EE\left[\int_{t}^T Q_i(u,p(u),\partial_p c_1(u,p(u)),\cdots\partial_p c_N(u,p(u)))du\bigg |{\mathcal{F}}(t)\right]
%\en
%where $Q$ is given by \eqref{Q_i} with 
%\[
%dp(u)=\alpha(p(u))du+\beta(p(u))d\widehat W(u), \quad p(t)=p,
%\]
%where $\widehat W$ is a standard Brownian motion given by \eqref{Inno_W}. 
Also, the value function under the Nash equilibrium $\bm \pi^*$ is given by \eqref{value}, 
%\be\label{value}
%V_i(t,x,p) = e^{r(T-t)}\left(1-\frac{\lambda^M}{N}\right)x_i-e^{r(T-t)}\frac{\lambda^M}{N}\overline x_{(-i)}+C_i(t,p),
%\en
where $C_i$ is the unique solution to the Cauchy problem \eqref{Cauchy'} obtained in Corollary~\ref{lem:Cauchy'}, with $\eta$ therein taken to be \eqref{eta}. %, among functions in $C^{1,2}([0,T)\times (0,1))$ that are continuous up to the boundary $\{T\}\times (0,1)$. 
\end{theorem}

The proof of Theorem~\ref{THM_1} is relegated to Section~\ref{subsec:proof of THM_1}. Let us stress that the interpretations of the first and second terms of \eqref{NE-1}, discussed in detail below Remark~\ref{rem:NE-1'}, still hold in the present setting, once we replace the SDE \eqref{P} therein by \eqref{P'}. Similarly, Remarks \ref{re_1} and \ref{re_2} also hold in the current setting of Theorem~\ref{THM_1}.

%%%%%%%%%%%%%%%%%%%%%
%%%%%%%%%%%%%%%%%%%%%

\section{Discussion and Numerical Illustration}\label{sec:discuss}%: How Partial Information Breeds Systemic Risk}\label{sec:discuss%%%%%%%%%%%%%%%%%%%%%%

\subsection{Myopic Trading versus Intertemporal Hedging}\label{subsec:discuss}
%{The Case of Constant Investment Opportunities}\label{subsec:discuss}
As explained below Remark~\ref{rem:NE-1'}, under partial information, the full-information trading strategy \eqref{NE_constant} changes into \eqref{NE-1}, consisting of myopic trading and intertemporal hedging. In particular, myopic trading (i.e., the first term of \eqref{NE-1}) is in the exact form of \eqref{NE_constant}, with $\mu$ in \eqref{NE_constant} (now unknown) approximated by $\theta(p)$ in \eqref{NE-1}. It is of interest to see how investors' wealth is affected by partial information and whether myopic trading or intertemporal hedging is the main driver. 
%As shown in Figures \ref{fig:mu=mu_1_relative} and \ref{fig:mu alternative_relative}, investors' wealth under \eqref{NE-1} is significantly reduced from that under \eqref{NE_constant}. It is then of interest to investigate whether myopic trading or intertemporal hedging is the main culprit. 

\begin{figure}[htbp]
\centering
\includegraphics[width=6cm,height=5.55cm]{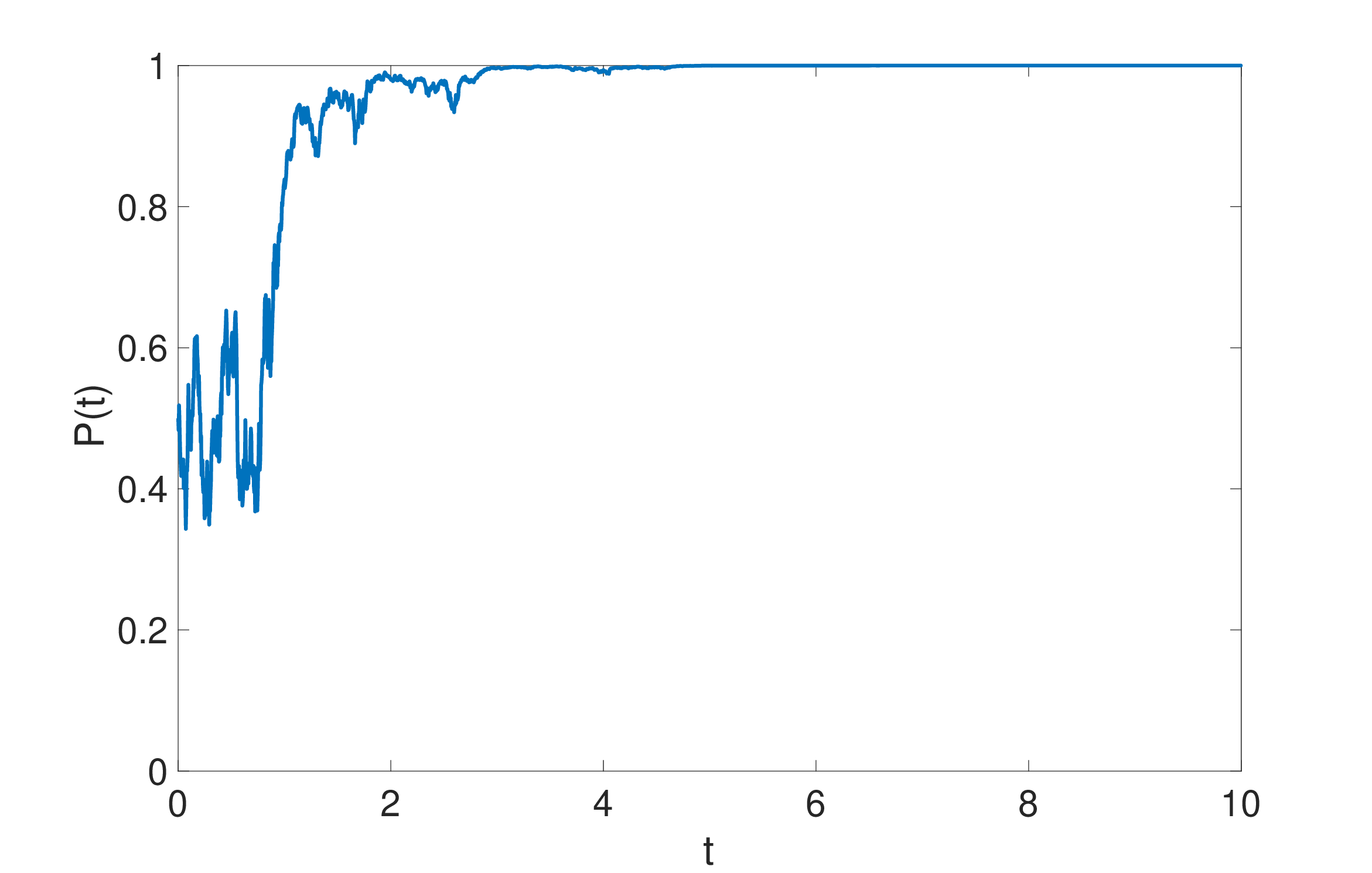}
\includegraphics[width=6cm,height=5.55cm]{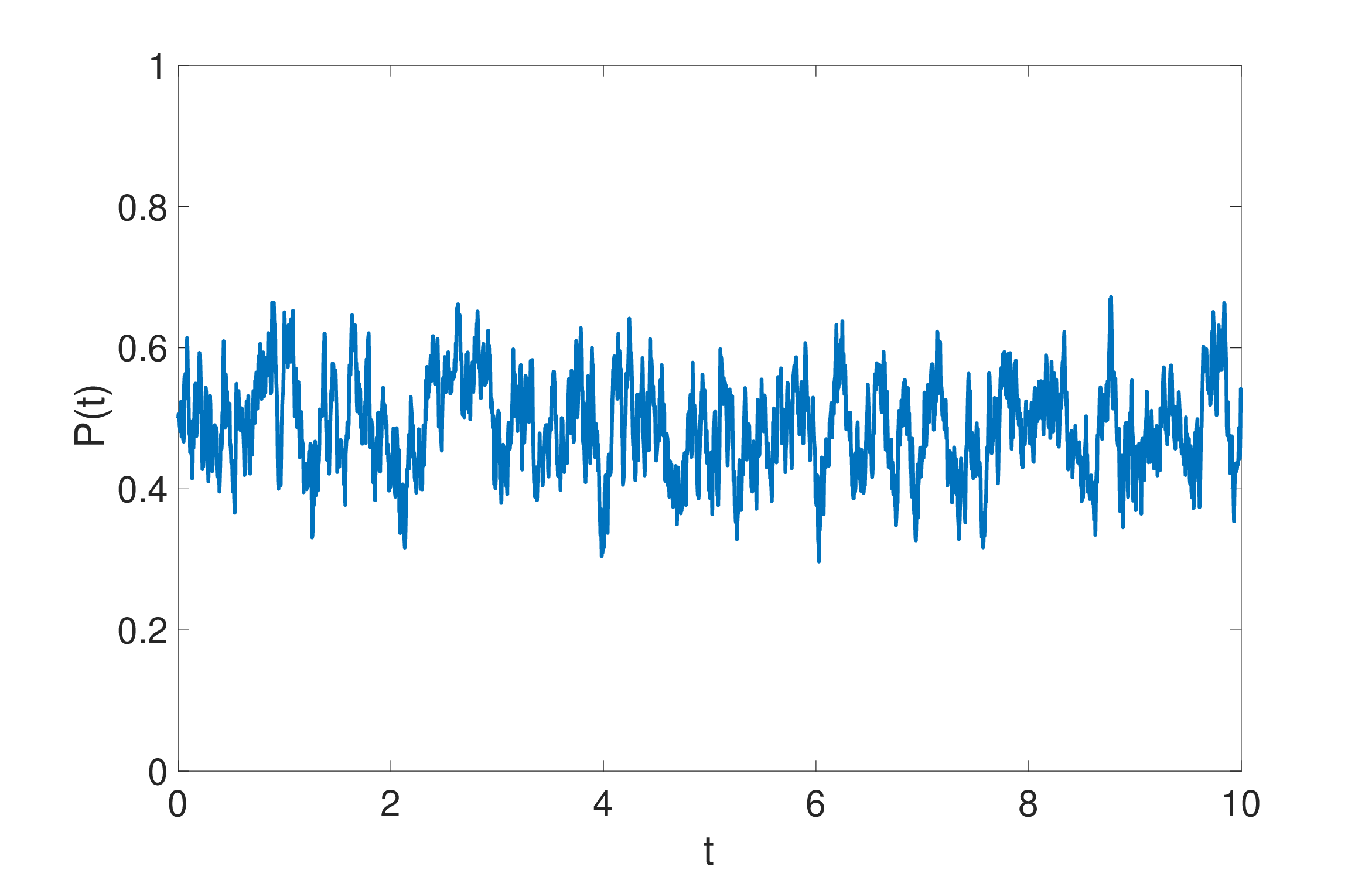}
\caption{One realization of $P(\cdot)$ in \eqref{P} (left) and $P(\cdot)$ in \eqref{P'} (right), under $\mu_1=0.2$, $\mu_2=0.02$, $\sigma=0.1$, and $q_1=q_2=10$.}
\label{fig:P}
\end{figure}

\begin{figure}[htbp]
\centering
\begin{subfigure}[b] {1.025\textwidth}
        \centering
	\includegraphics[width=5.55cm,height=5.55cm]{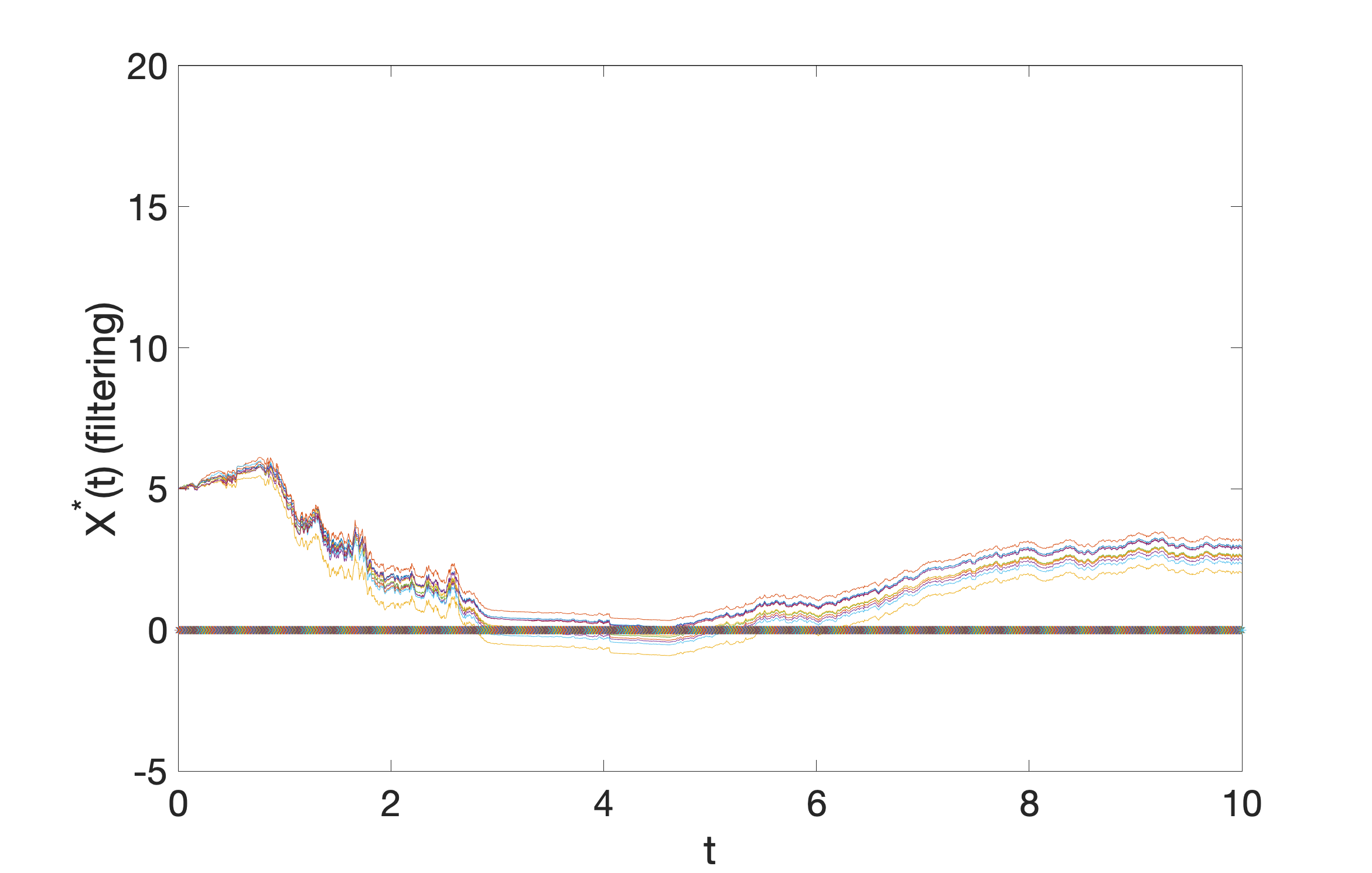} 
	\includegraphics[width=5.55cm,height=5.55cm]{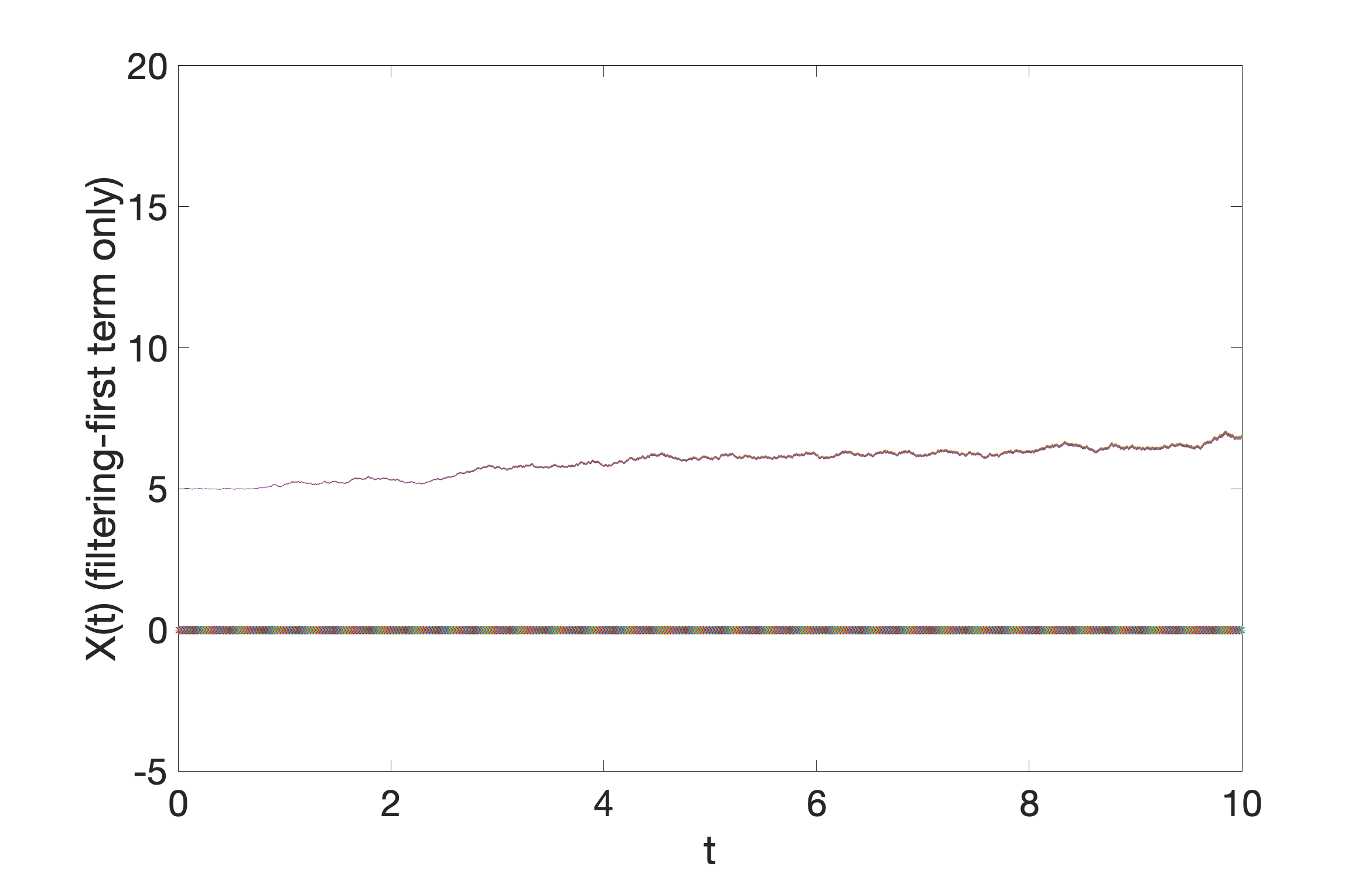}
	\includegraphics[width=5.55cm,height=5.55cm]{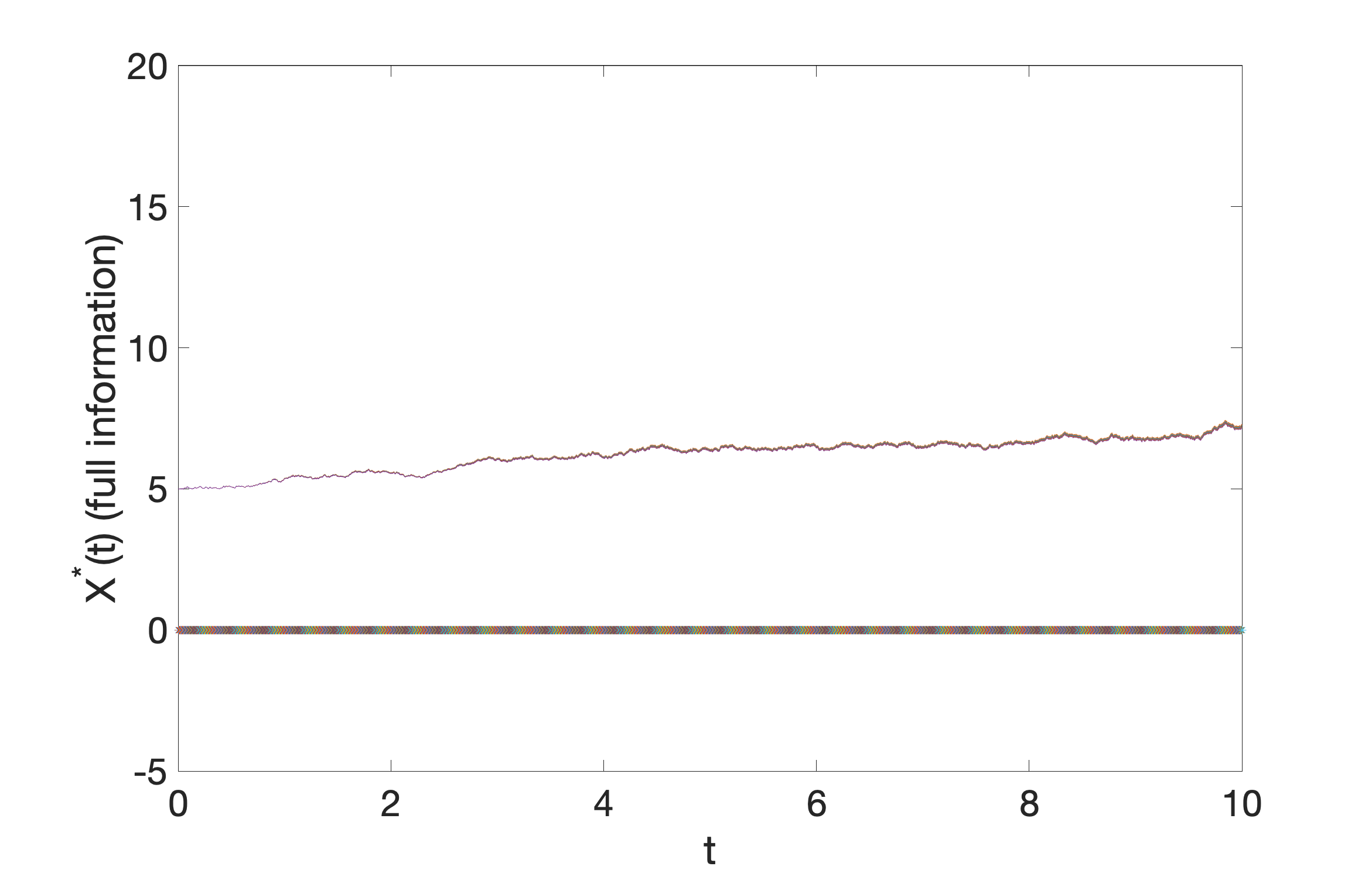}
        \caption{Wealth processes $X_i$ for $i=1,...10$}
        \label{subfig:X_i}
    \end{subfigure}
\begin{subfigure}[b]{1.025\textwidth}
        \centering
\includegraphics[width=5.55cm,height=5.55cm]{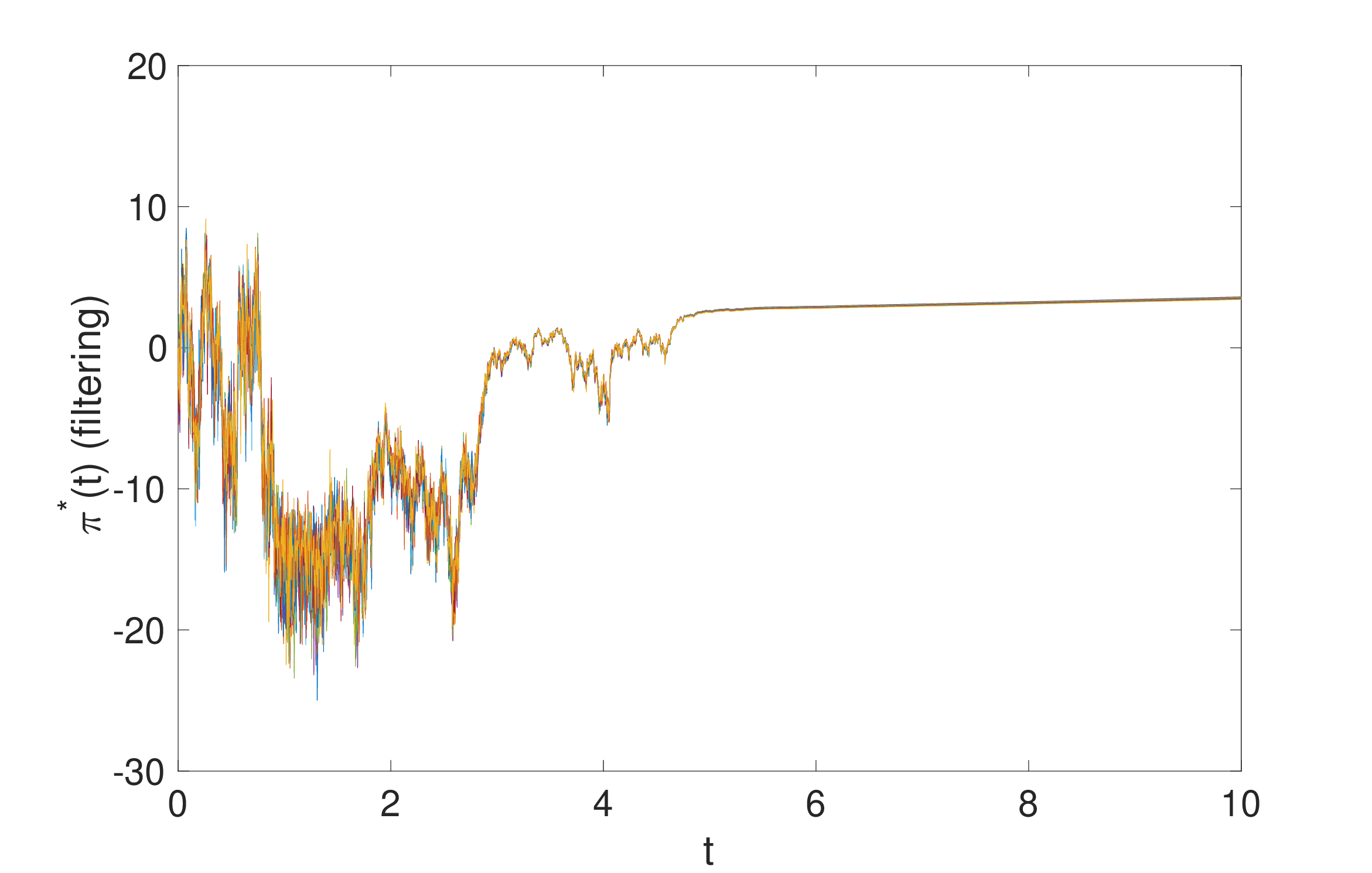} 
\includegraphics[width=5.55cm,height=5.55cm]{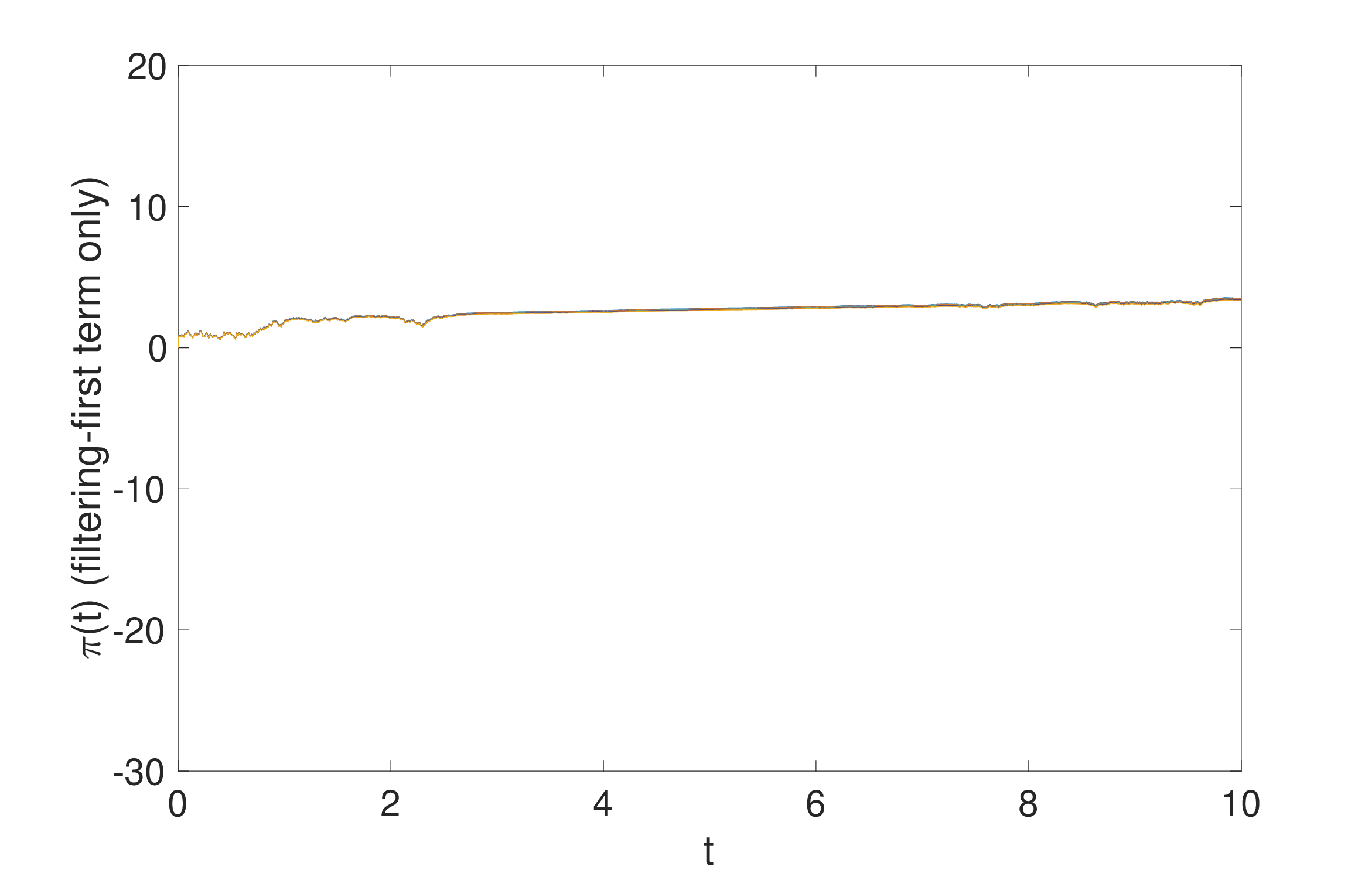}
\includegraphics[width=5.55cm,height=5.55cm]{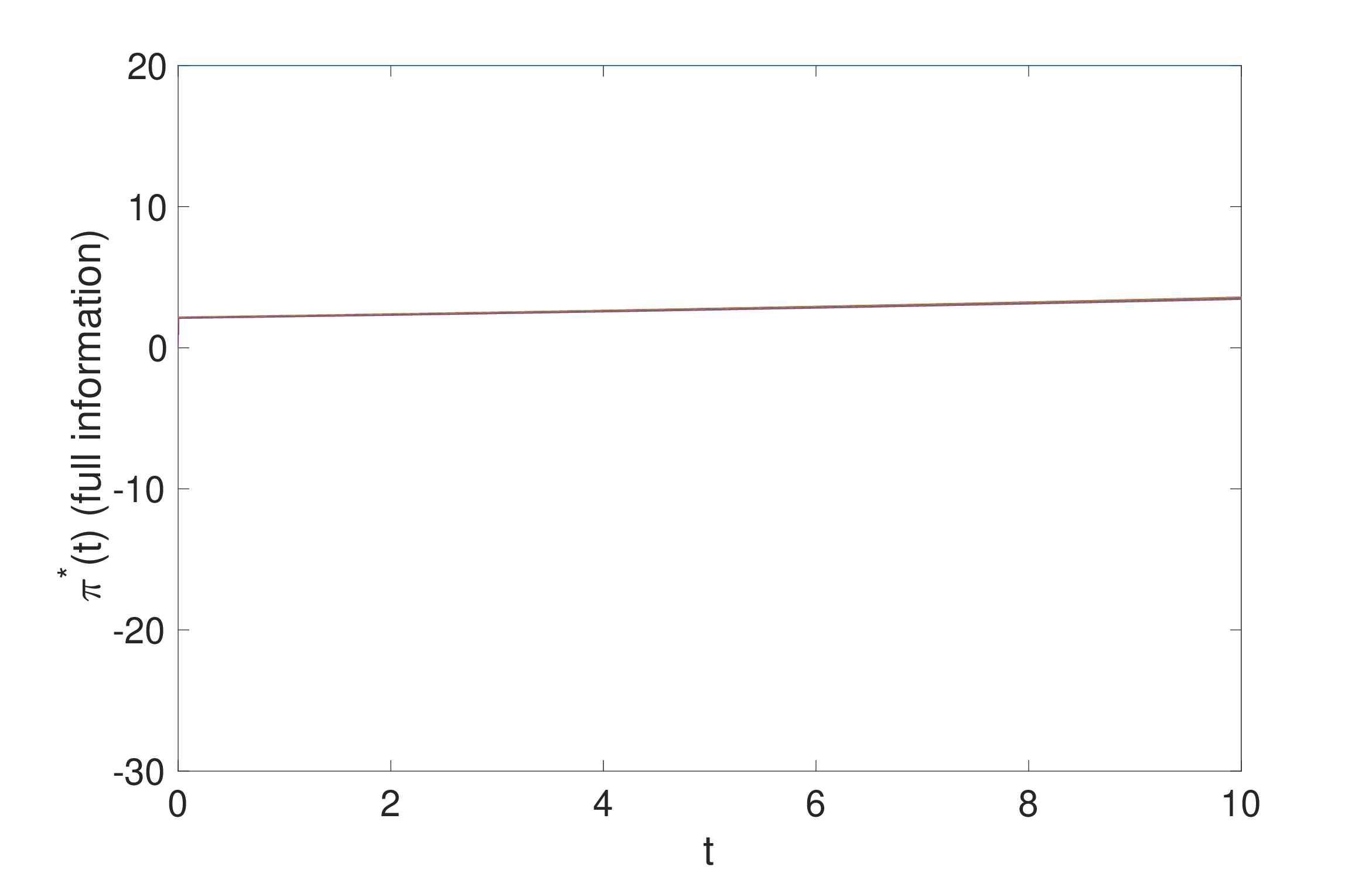}
        \caption{Trading strategies $\pi_i$ for $i=1,...10$}
        \label{subfig:pi_i}
    \end{subfigure}
%\begin{subfigure}[b]{1.025\textwidth}
   %     \centering
%\includegraphics[width=5.55cm,height=5.55cm]{q0g8_r_p.eps}
%\includegraphics[width=5.55cm,height=5.55cm]{q0g8_r_l5.eps}
 %\caption{Posterior probability $P$ and empirical loss distributions}
 %\label{subfig:P}
    %\end{subfigure}
\caption{Under $T=10$, $N = 10$, $r = 0.05$, $\mu=\mu_1=0.2$, $\mu_2=0.02$, $\sigma=0.1$,  $\lambda^M_i=\lambda^V_i=0.5$ and $\gamma_i=8+0.1i$ for $i=1,...,10$, Panel (b) computes one realization of trading strategies $\pi_i$, $i=1,...,10$, with $\pi_i$ taken to be \eqref{NE-1}, the first term of \eqref{NE-1}, and \eqref{NE_constant}, respectively (from left to right). Panel (a) presents one realization of wealth processes $X_i(t)$, $i=1,...,10$, computed under the corresponding trading strategies in Panel (b). 
%By simulating 100 realizations and recording for each realization how many of $X_i(t)$, $i=1,...,10$, fall below the default level 0 by time $T$, the second plot in Panel (c) presents the empirical loss distributions under full information (dashed line) and partial information (solid line).
}
\label{fig:mu=mu_1}
%\end{center}
\end{figure}

For the case of a constant $\mu$, by Lemmas~\ref{lem:p to 1} and \ref{lem:hW}, $P(\cdot)={\mathfrak p}_1(\cdot)$ in \eqref{p_j} satisfies SDE \eqref{P}, which might oscillate wildly between 0 and 1 before converging to one of them. It is then possible that $\theta(P(\cdot))$ drifts far away from the true value of $\mu$, making myopic trading distinct from the ideal level \eqref{NE_constant}, for quite some time. %This can potentially curtail, or even reverse, the growth of wealth. 
%Take the case $\mu=\mu_1>r$ for example. As it takes time for $\theta(P(\cdot))$ to move near $\mu=\mu_1$, the first term of \eqref{NE-1} specifies markedly less stock holding than \eqref{NE_constant} for some while. With a positive risk premium $\mu-r$, when one takes $\pi_i$ in \eqref{wealth full} to be the first term of \eqref{NE-1}, rather than \eqref{NE_constant}, the momentum for wealth accumulation (resulting from $\pi_i(u)(\mu-r)du$ in \eqref{wealth full}) weakens. The situation gets worse if we also have $\mu_2<r$. When $\theta(P(\cdot))$ drifts near $\mu_2$, the first term in \eqref{NE-1} becomes negative, such that $\pi_i(u)(\mu-r)du$ in \eqref{wealth full} actually reduces wealth.
The first plot in Figure~\ref{fig:P} shows that $P(\cdot)={\mathfrak p}_1(\cdot)$ in \eqref{p_j} does oscillate wildly before converging to 1 after $t=4$. Accordingly, myopic trading %the first term of \eqref{NE-1} 
(the second plot, Figure~\ref{subfig:pi_i}) deviates from \eqref{NE_constant} (the last plot, Figure~\ref{subfig:pi_i}) for $t\in [0,4]$. This deviation, while discernible, is generally small. The resulting wealth processes (the last two plots, Figure~\ref{subfig:X_i}) are thus similar. Indeed, while $P$ oscillates forcefully, it generally moves in the right direction: it rises above 0.8 quickly after time 1 and stay within $[0.8,1]$ afterwards. This allows $\theta(P(\cdot))$ to stay near $\mu=\mu_1$ after $t=1$, such that myopic trading %the first term of \eqref{NE-1} 
and \eqref{NE_constant} remain close for most of the time. 

It is intertemporal hedging (i.e., the second term of \eqref{NE-1}) that drives %the trading strategy 
\eqref{NE-1} significantly away from \eqref{NE_constant}. Recall that under the measure $\Q$ in \eqref{Q}, $P(\cdot)$ in \eqref{P} becomes \eqref{P''} with $\eta\equiv0$, which is the original driftless $\P$-dynamics plus a new drift term. As the added drift term is bounded, the strong oscillation (or ``impreciseness'') of \eqref{P} under $\P$ is largely inherited by \eqref{P''} under $\Q$. The arguments below \eqref{zeta=P'} then suggest an inflated second term of \eqref{NE-1}, or intertemporal hedging. As shown in the first two plots of Figure~\ref{subfig:pi_i}, intertemporal hedging drastically alters the trading strategy and thus severely impacts investors' wealth (the first plot, Figure~\ref{subfig:X_i}). 

\begin{figure}[htbp]
\centering
\begin{subfigure}[b] {1.025\textwidth}
        \centering
\includegraphics[width=5.55cm,height=5.55cm]{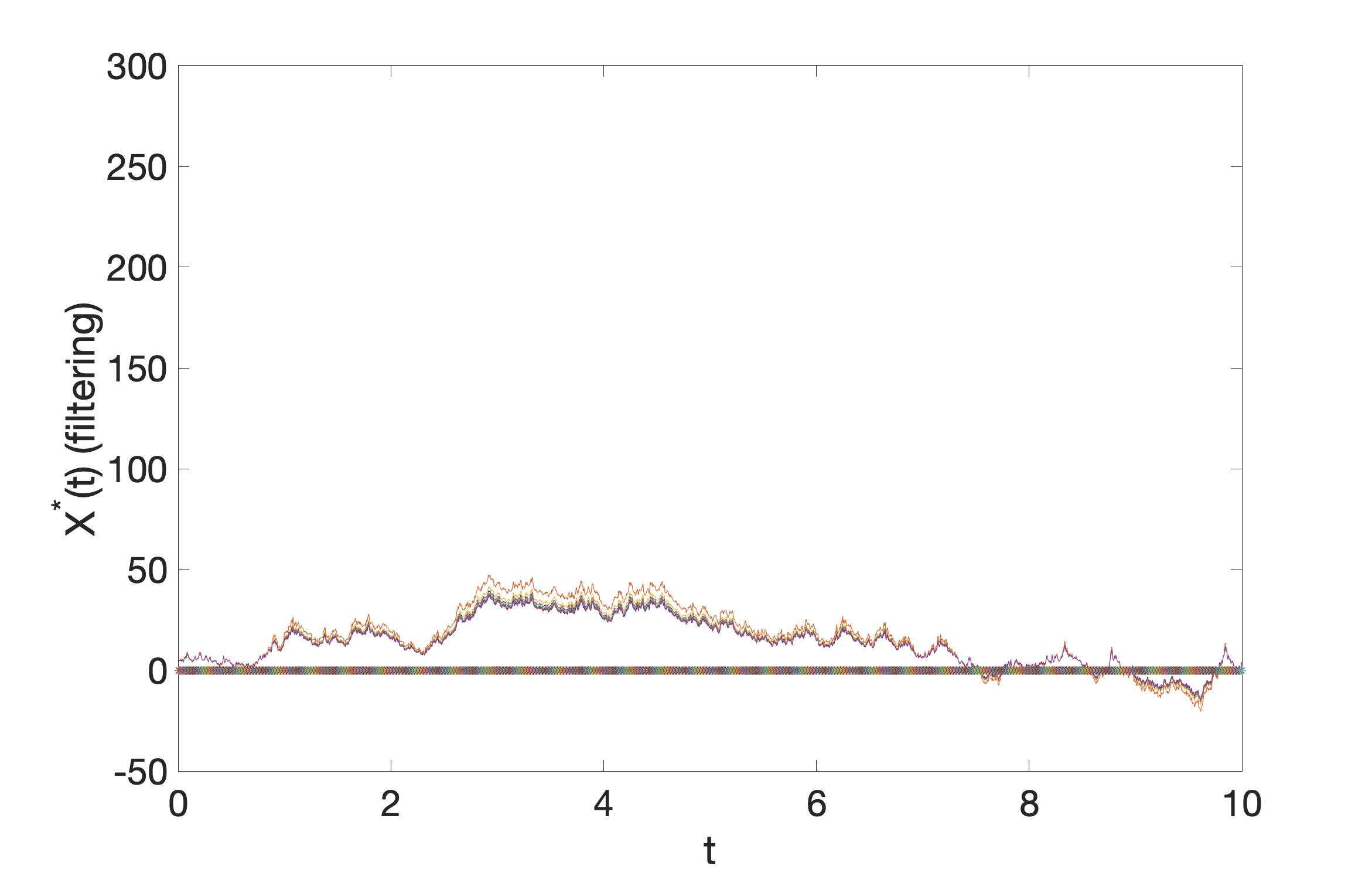}
\includegraphics[width=5.55cm,height=5.55cm]{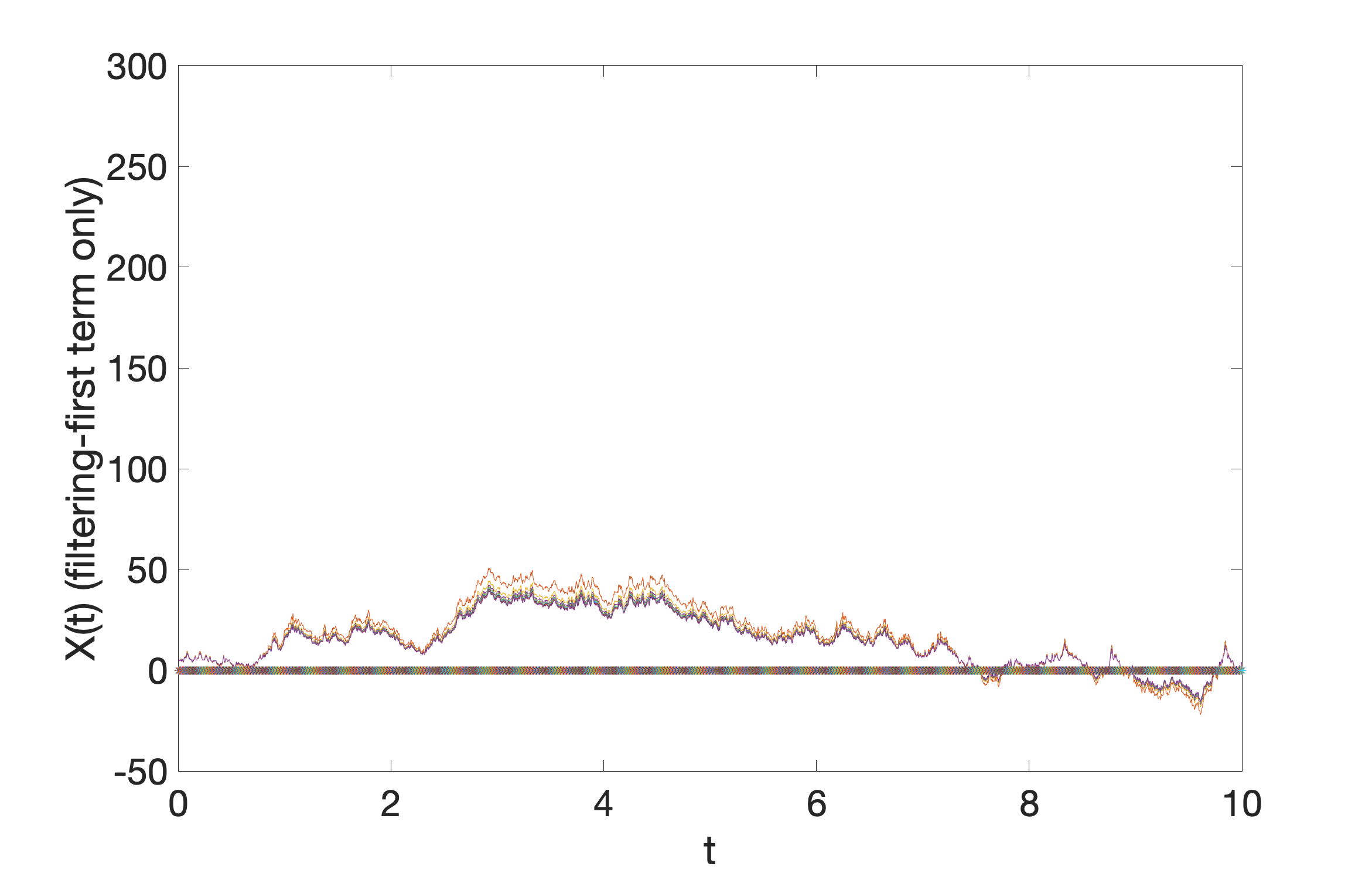}
\includegraphics[width=5.55cm,height=5.55cm]{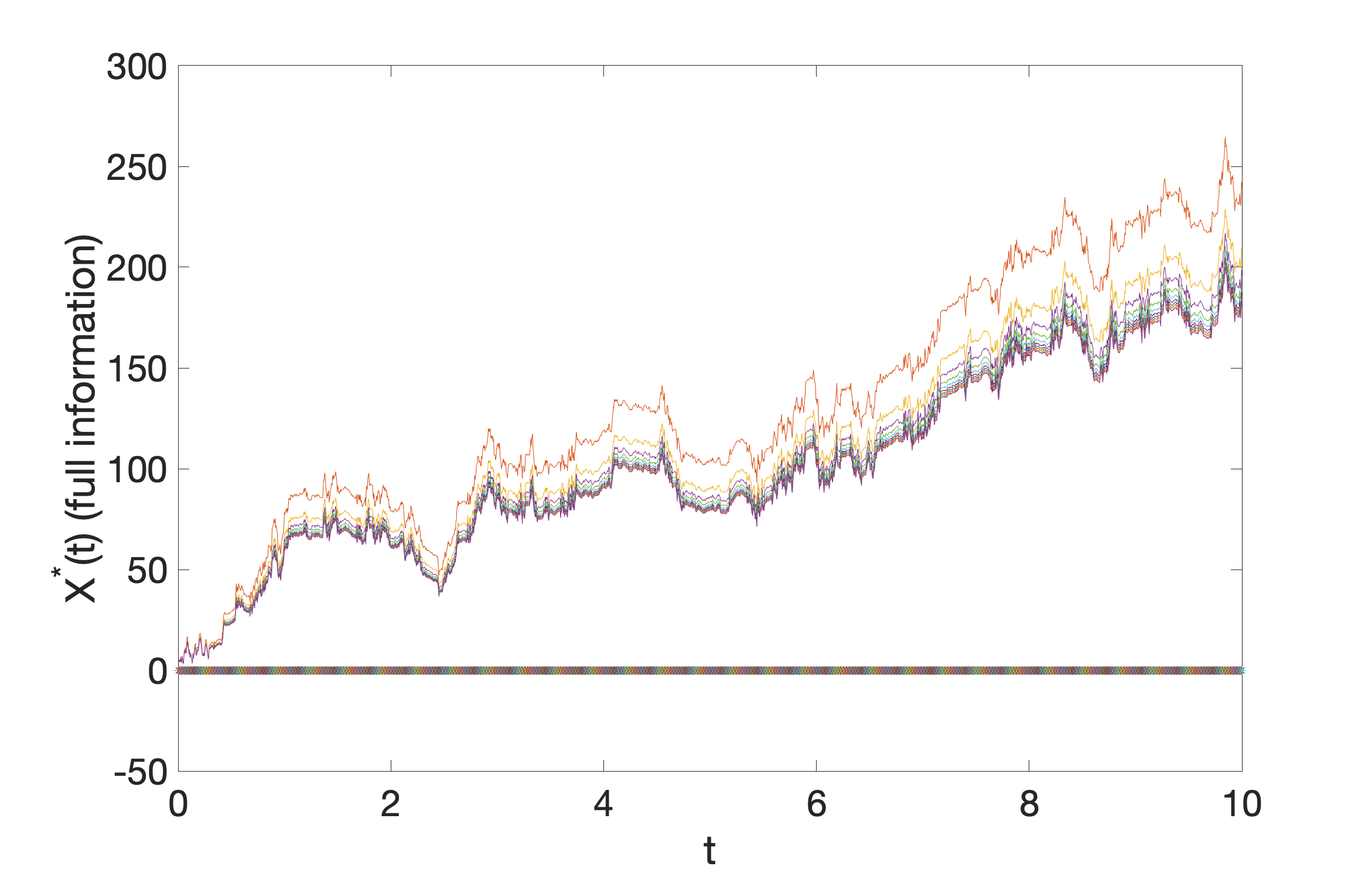}
\caption{Wealth processes $X_i$ for $i=1,...10$}
        \label{subfig:X_i'}
    \end{subfigure}
\begin{subfigure}[b]{1.025\textwidth}
        \centering
	\includegraphics[width=5.55cm,height=5.55cm]{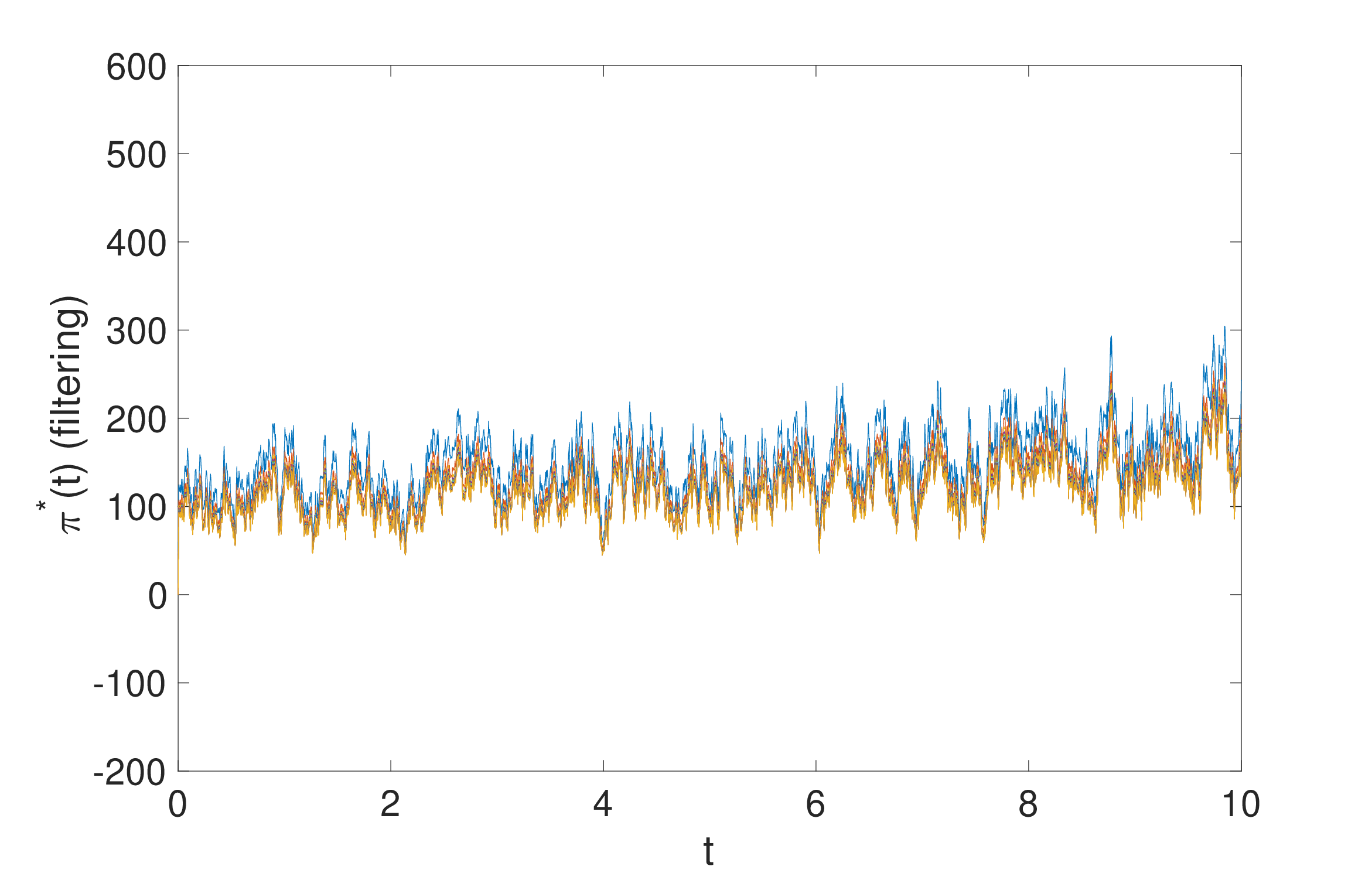}
	\includegraphics[width=5.55cm,height=5.55cm]{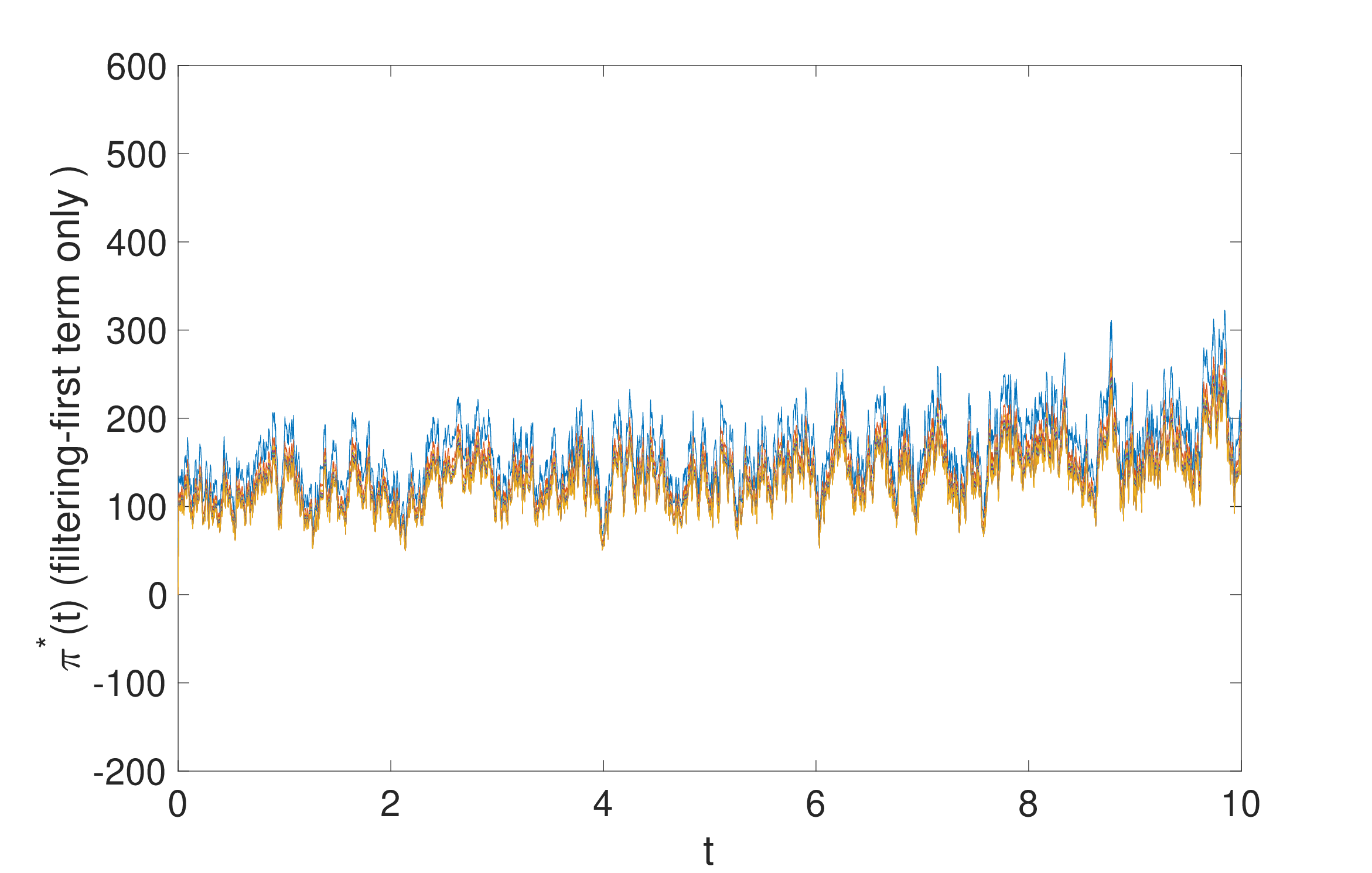}
	\includegraphics[width=5.55cm,height=5.55cm]{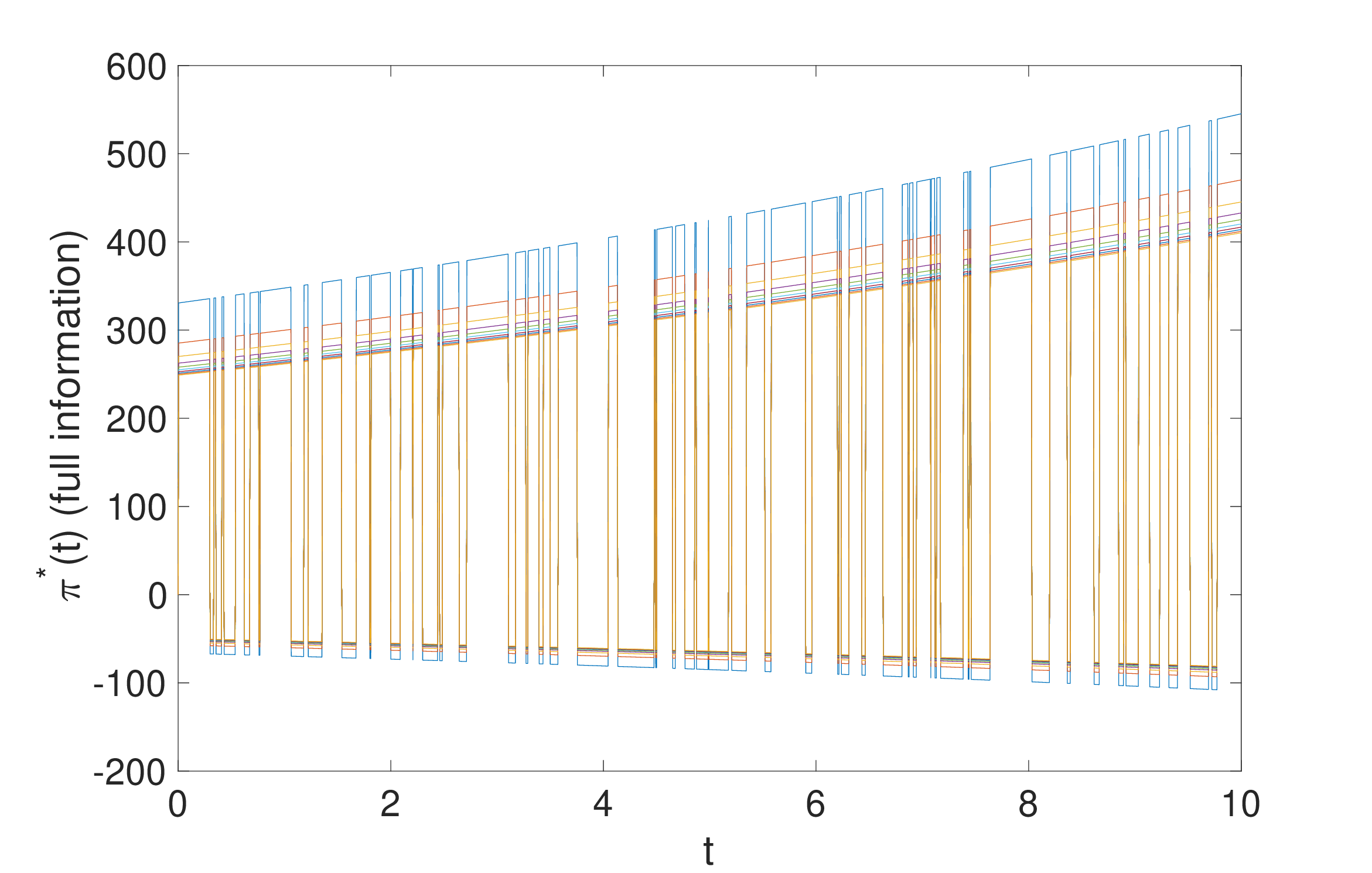}
\caption{Trading strategies $\pi_i$ for $i=1,...10$}
        \label{subfig:pi_i'}
    \end{subfigure}
%\begin{subfigure}[b]{1.025\textwidth}
   %     \centering
	%\includegraphics[width=5.55cm,height=5.55cm]{p_q10l9.eps}
	%\includegraphics[width=5.55cm,height=5.55cm]{q10l9.eps}
	 %\caption{Posterior probability $P$ and empirical loss distributions}
 	%\label{subfig:P'}
        %\end{subfigure}
\caption{Under $T=10$, $N = 10$, $r = 0.05$, $\mu$ alternating between $\mu_1=0.2$ and $\mu_2=0.02$ with $q_1=q_2=10$ (recall \eqref{mu process}), $\sigma=0.1$, $\lambda^M_i=\lambda^V_i=0.9$ and $\gamma_i=0.1i$ for $i=1,...,10$, Panel (b) computes one realization of trading strategies $\pi_i$, $i=1,...,10$, with $\pi_i$ taken to be \eqref{NE-1}, the first term of \eqref{NE-1}, and \eqref{NE_Markov}, respectively (from left to right), where ``$\eta\equiv 0$'' is replaced by \eqref{eta}. Panel (a) presents one realization of wealth processes $X_i(t)$, $i=1,...,10$, computed under the corresponding trading strategies in Panel (b). 
%By simulating 100 realizations and recording for each realization how many of $X_i(t)$, $i=1,...,10$, fall below the default level 0 by time $T$, the second plot in Panel (c) presents the empirical loss distributions under full information (dashed line) and partial information (solid line). 
}
\label{fig:mu alternates}
\end{figure}

%\subsection{The Case of Alternating Investment Opportunities}\label{subsec:discuss'}
For the case of an alternating $\mu$, 
%Theorem~\ref{THM_1} stipulates that the equilibrium trading strategy still takes the form \eqref{NE-1}, with ``$\eta\equiv0$'' therein replaced by \eqref{eta}. While the same analysis for the case of a constant $\mu$ still applies, a very different story unfolds.
by Lemma~\ref{lem:tW}, $P(\cdot)={\mathfrak p}_1(\cdot)$ in \eqref{p_j'} satisfies SDE \eqref{P'}, which behaves quite differently from \eqref{P} (the SDE under a fixed $\mu$). As \eqref{P} involves only a Brownian motion term, it can oscillate wildly between 0 and 1, despite its ultimate convergence to 0 or 1 (by Lemma~\ref{lem:p to 1}). By contrast, the drift term of \eqref{P'} makes the process mean-reverting. As \eqref{P'} always tends to return to the mean level $\frac{q_2}{q_1+q_2}$, its oscillation between 0 and 1 is dampened. In other words, with \eqref{P}, investors' judgment of $\mu$ changes widely over time, but a firm clear opinion will take shape eventually; with \eqref{P'}, investors' judgment is stable over time but remains vague throughout.   
This lasting vagueness keeps myopic trading (i.e., the first term of \eqref{NE-1}) away from the full-information trading strategy \eqref{NE_Markov} {\it perpetually}. As $P(\cdot)$ in \eqref{P'} never stays near 0 or 1 for long, $\theta(P(\cdot))$ is constantly far away from $\mu_1$ and $\mu_2$ (and thus from $\mu(\cdot)$ in \eqref{mu process}). In the second plot of Figure~\ref{fig:P}, $P(\cdot)$ in \eqref{P'} %=\widetilde{\mathfrak p}_1(\cdot)$ in \eqref{p_j'}
oscillates around $\frac{q_2}{q_1+q_2}=0.5$, taking values mostly within $[0.4,0.6]$. Accordingly, $\theta(P(\cdot))$ never gets close to $\mu(\cdot)$ in \eqref{mu process}, leaving myopic trading distinct from \eqref{NE_Markov} perpetually (the last two plots, Figure~\ref{subfig:pi_i'}). The resulting wealth processes (the last two plots, Figure~\ref{subfig:X_i'}) then differ substantially. %: trading according to the first term of \eqref{NE-1} severely reduces wealth, causing many defaults. % (i.e., their wealth to fall below 0). 

Interestingly, intertemporal trading (i.e., the second term of \eqref{NE-1}) does not play a significant role here, as opposed to the case of a constant $\mu$. Under the measure $\Q$ in \eqref{Q}, $P(\cdot)$ in \eqref{P'} becomes \eqref{P''}, with $\eta$ given by \eqref{eta}. By the definitions of $\eta$ and $\beta$ in \eqref{eta} and \eqref{theta}, the mean-reverting effect remains dominant whenever $P(\cdot)$ moves near 0 or 1 under $\Q$. As $P(\cdot)$ is made more stable (or,  more ``precise'') on $[t,T]$ under $\Q$ by this mean-reverting effect, the arguments below \eqref{zeta=P'} imply %that $\partial_p c_i(t,p)$ tend to be closer to 0, which 
a diminished second term of \eqref{NE-1}, or intertemporal hedging. As shown in the first two plots of Figure~\ref{subfig:pi_i'}, adding the second term of \eqref{NE-1} does not change the trading strategy much and %(except reducing stock holding very slightly). 
the resulting wealth processes (the first two plots, Figure~\ref{subfig:X_i'}) are quite similar. %, with wealth reduced only slightly when the second term of \eqref{NE-1}  is added. 

In summary, when $\mu$ is constant, due to the strong oscillation (or ``impreciseness'') of the posterior probability $P(\cdot)$, intertemporal hedging (against such uncertainty) 
%the second term  of \eqref{NE-1} (i.e., intertemporal hedging) 
is dominant in driving the trading strategy \eqref{NE-1} away from the ideal level \eqref{NE_constant}. % {\it for quite some time}. %This results from the strong oscillation (or ``impreciseness'') of the posterior probability $P(\cdot)$, from which investors cannot quickly read the true value of $\mu$. 
When $\mu$ is alternating, due to the mean-reverting feature of $P(\cdot)$, %from which investors {\it never} learn the true state of $\mu$.
myopic trading is dominant in driving \eqref{NE-1} away from the ideal level \eqref{NE_Markov}. % {\it perpetually}. 
That is, the main culprit of the deviation from ideal trading (and the resulting wealth reduction) can vary, depending crucially on the behavior of the posterior probability $P(\cdot)$. 

\subsection{Flocking Behavior and Self-Reinforcement}\label{subsec:DSR}
The intent to reduce $\mathrm{Var}^{t,\bm x, p}\big[X_i(T)-\lambda^V_i\overline X(T)\big]$ in \eqref{MV_model P} motivates investor $i$ to keep her wealth process $X_i$ at a constant distance from $\lambda^V_i\overline X$ (whether the distance is small or large), which encourages comovement of $X_i$ and $\overline X$. As a result, the joint evolution of $\{X_i\}_{i=1}^N$ likely forms a ``flocking behavior'' around $\overline X$, where each $X_i$ tends to move in the same direction as $\overline X$. This suggests two consequences of relative performance criteria for investors' wealth evolution, i.e., (i) {\it contagion:} if $X_i$ falls (resp.\ rises) significantly so that $\overline X$ starts to decrease (resp.\ increase), all other $X_j$'s will tend to follow suit; (ii) {\it self-reinforcement:} if all wealth processes $\{X_i\}_{i=1}^N$ simultaneously fall (resp.\ rise), such that so does $\overline X$, all of $\{X_i\}_{i=1}^N$ will tend to fall (resp.\ rise) further.
%When $\overline X$ decreases, caused by the decrease of some or all investors' wealth, all investors' wealth tend to decrease. That is, a decrease in $X_i$ can be contagious to $X_j$ if $X_j$ is not currently decreasing, and reinforcing to $X_j$ if $X_j$ is currently decreasing. 

\begin{figure}[htbp]
\centering
\begin{subfigure}[b] {1.025\textwidth}
        \centering
	\includegraphics[width=5.55cm,height=5.55cm]{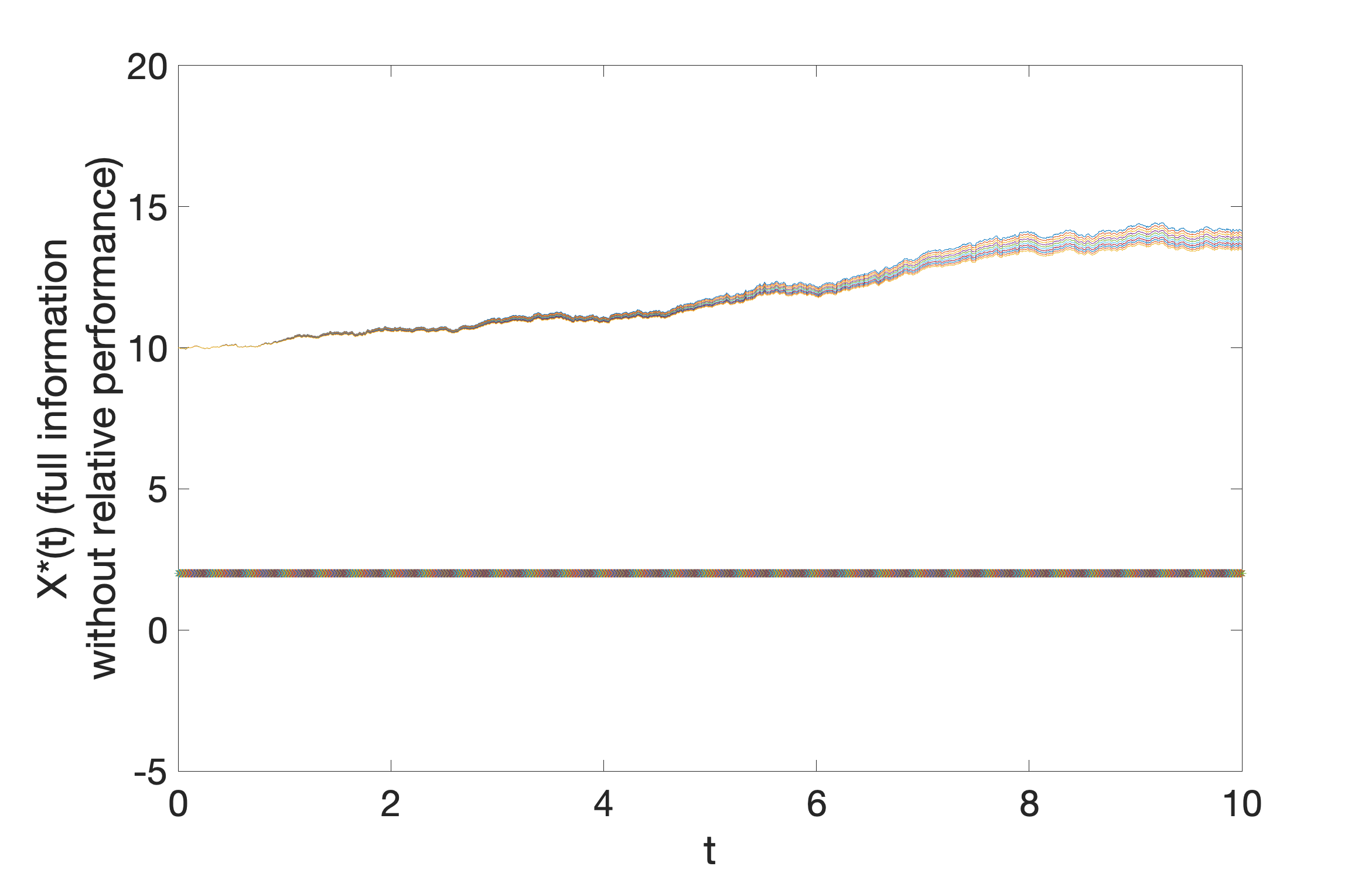} 
	\includegraphics[width=5.55cm,height=5.55cm]{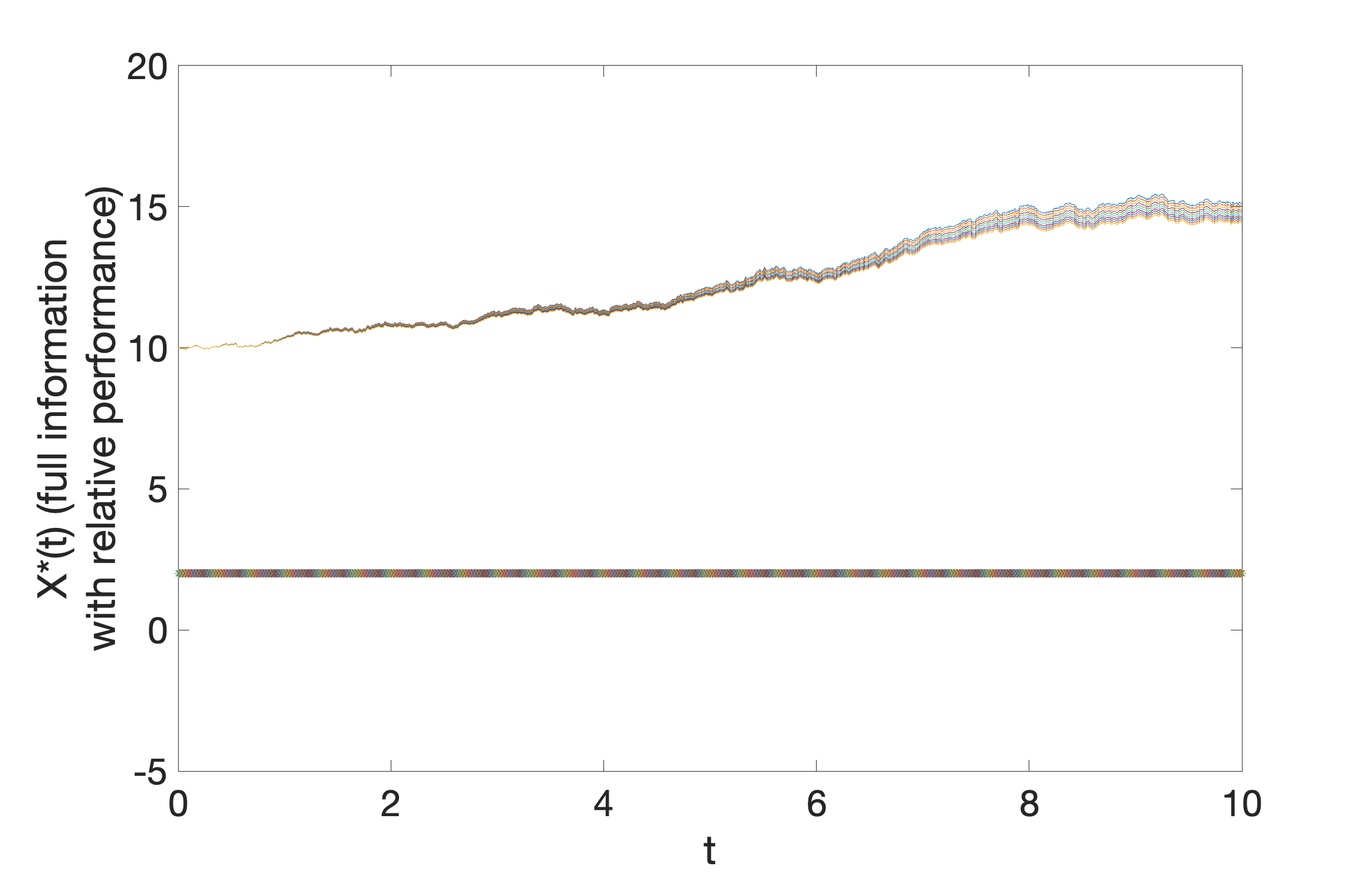}
	\includegraphics[width=5.55cm,height=5.55cm]{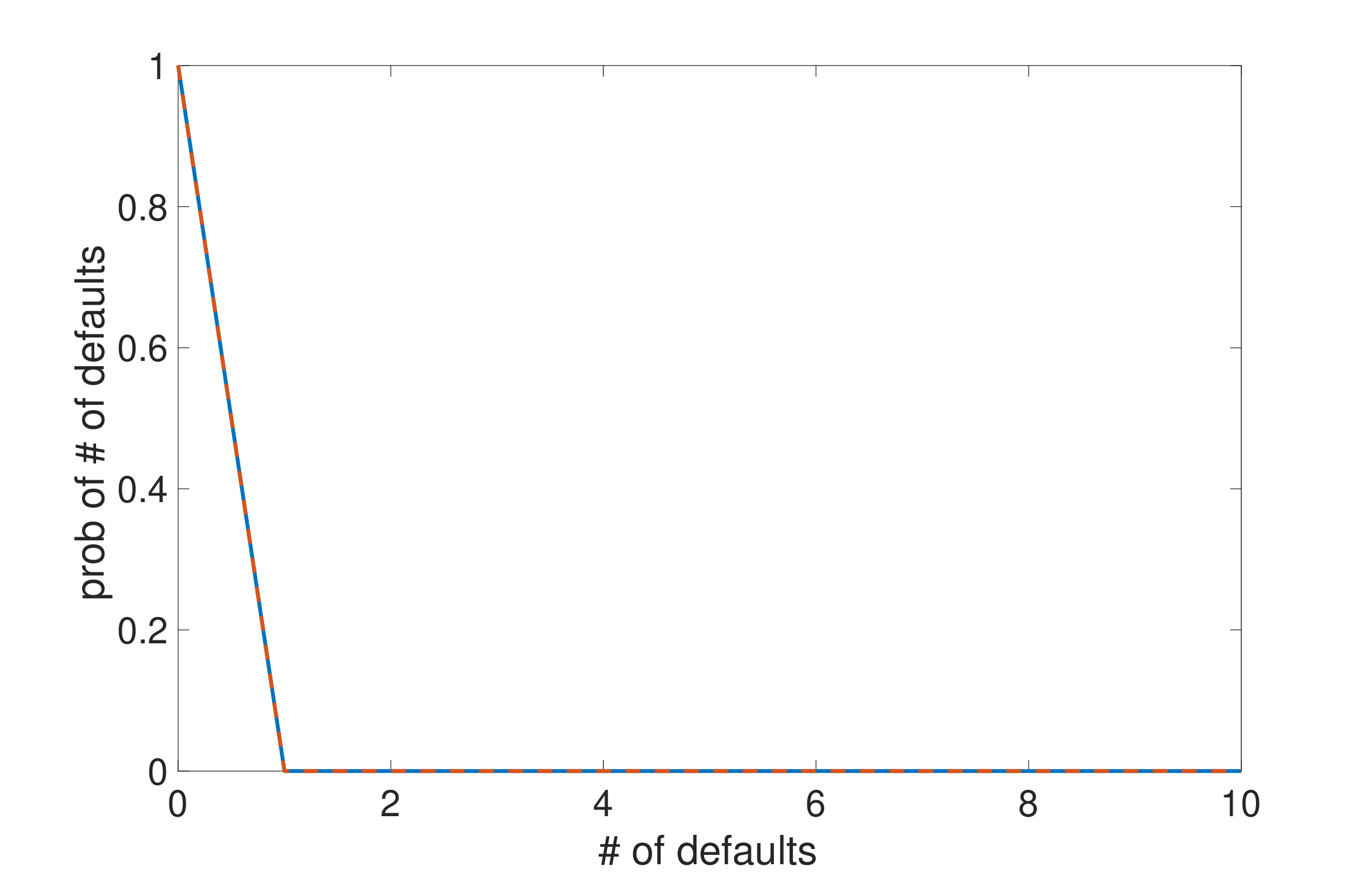}
        \caption{full information}
        \label{subfig:full}
    \end{subfigure}
\begin{subfigure}[b]{1.025\textwidth}
        \centering
\includegraphics[width=5.55cm,height=5.55cm]{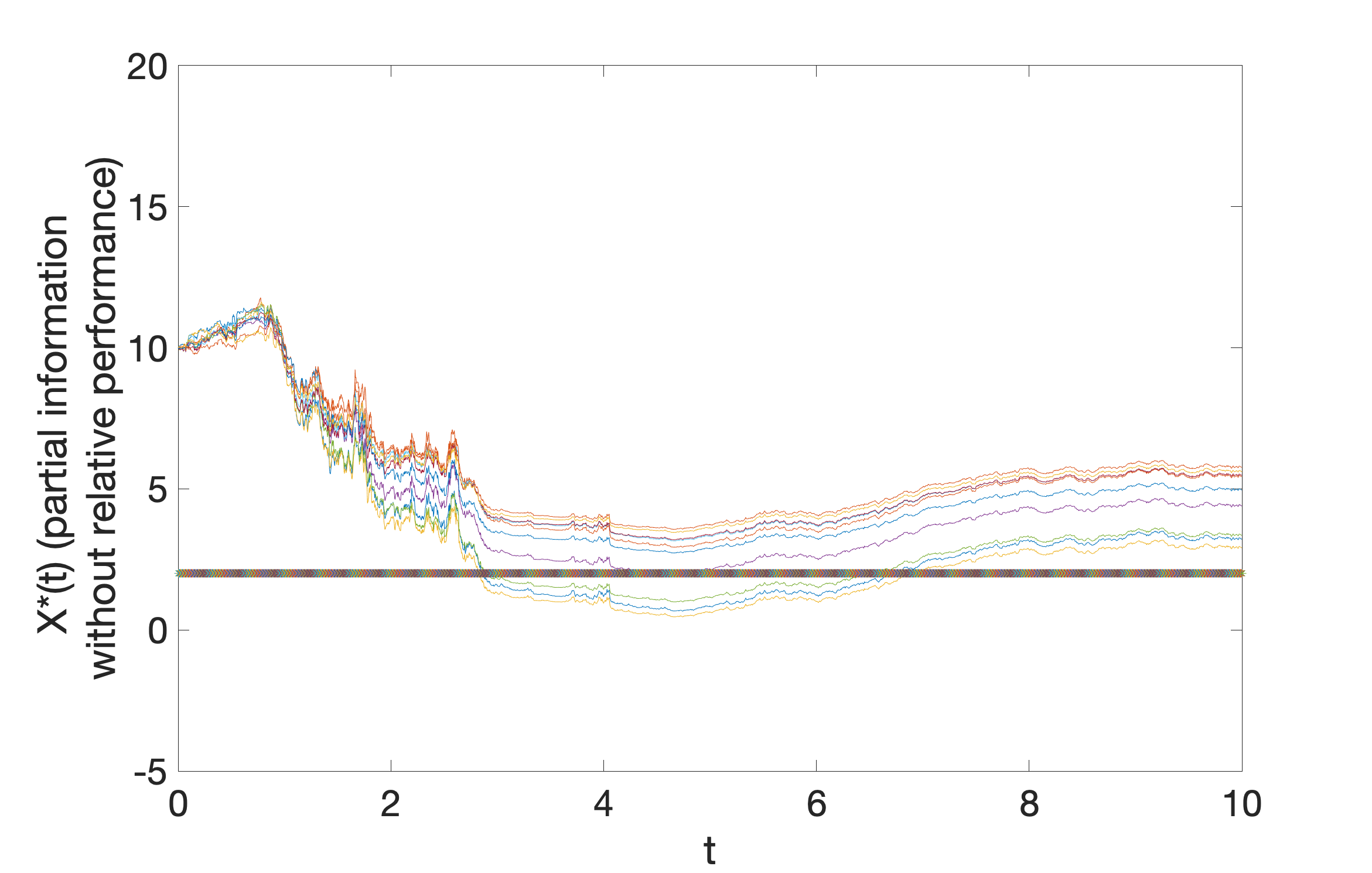} 
	\includegraphics[width=5.55cm,height=5.55cm]{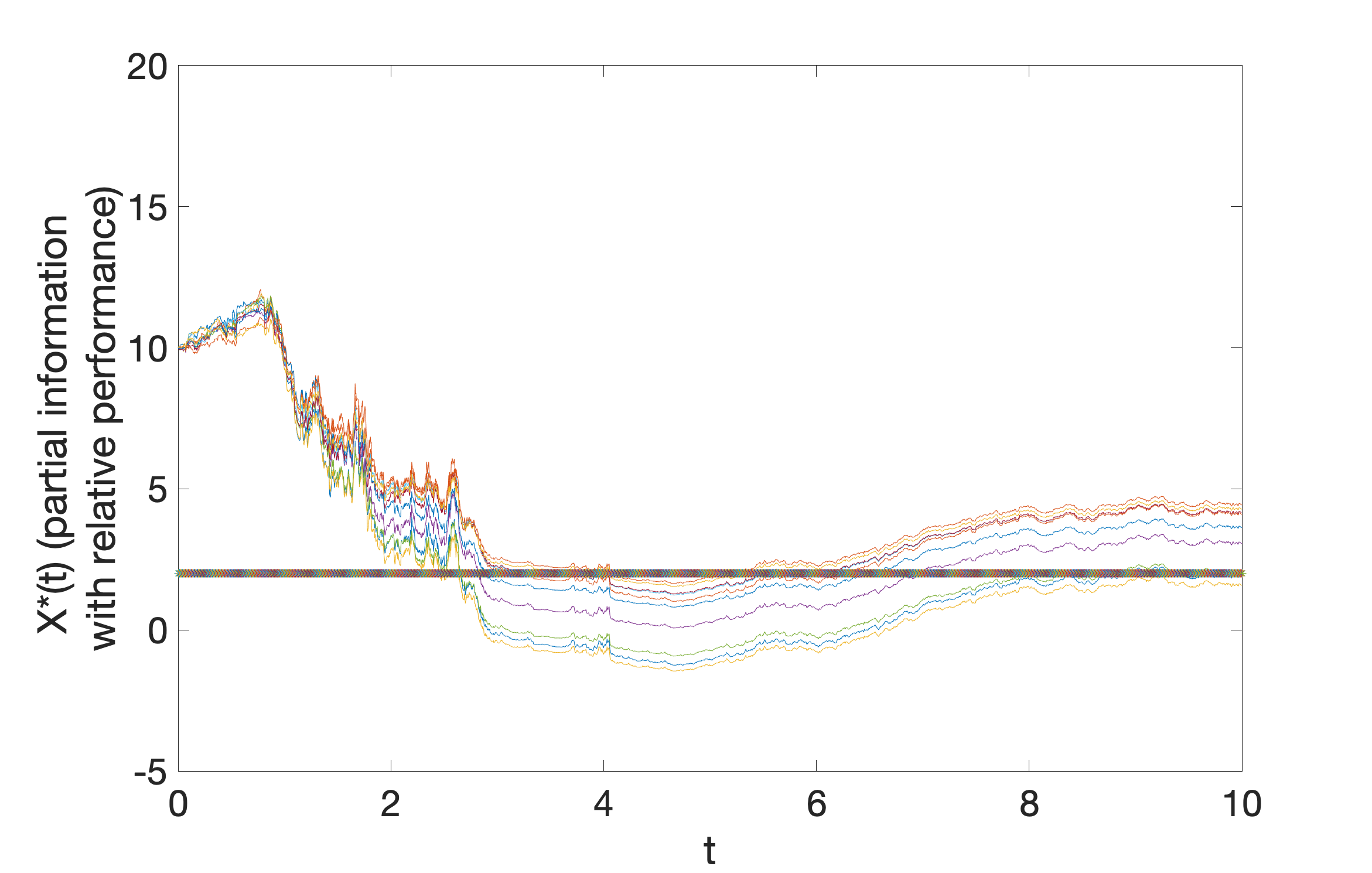}
	\includegraphics[width=5.55cm,height=5.55cm]{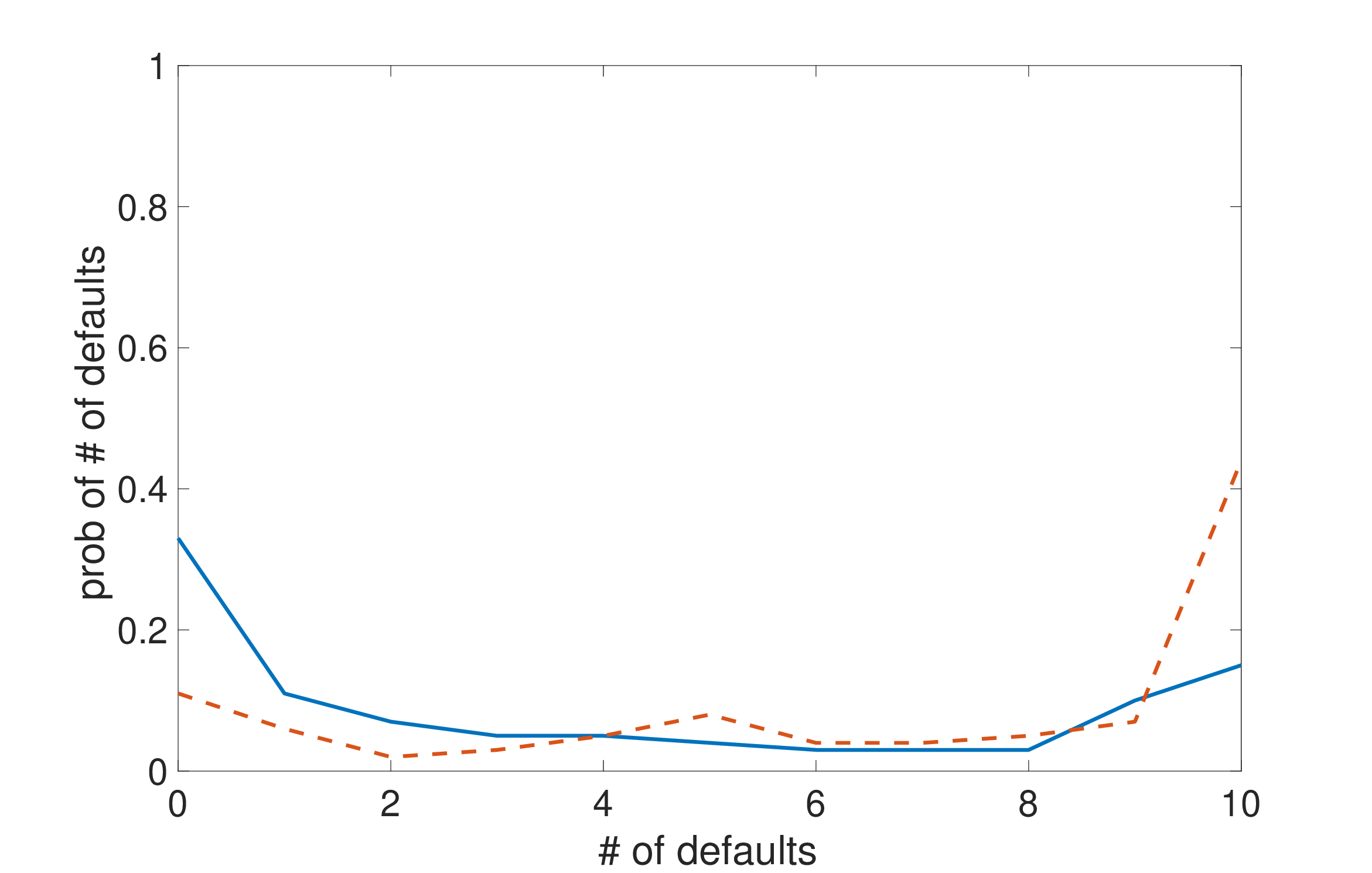}
        \caption{partial information}
        \label{subfig:partial}
    \end{subfigure}
\caption{Under $T=10$, $N = 10$, $r = 0.05$, $\mu=\mu_1=0.2$, $\mu_2=0.02$, $\sigma=0.1$ and $\gamma_i=4.5+0.1i$ for $i=1,...,10$, the first column presents one realization of wealth processes $X_i(t)$, $i=1,...,10$ when relative performance is disregarded ($\lambda^M_i=\lambda^V_i=0$) under full and partial information; the second column presents one realization of wealth processes $X_i(t)$, $i=1,...,10$ when relative performance is considered ($\lambda^M_i=\lambda^V_i=0.2$) under full and partial information. 
By simulating 100 realizations and recording for each realization how many of $X_i(t)$, $i=1,...,10$, fall below the warning threshold 2 by time $T$, the third column presents the empirical warning distributions with relative performance considered (dashed line) and disregarded (solid line), under full and partial information.}
\label{fig:mu=mu_1_relative}
%\end{center}
\end{figure}

This hints at an interesting role of partial information---to amplify  {\it downward} self-reinforcement. When there is a misbelief of the $\mu$ value among investors, $\{X_i\}_{i=1}^N$ are more likely to fall sharply simultaneously. We therefore expect {downward self-reinforcement} to be more pronounced under partial information than under full information.
%For the case of a constant $\mu$, 
%The first two plots in Figure~\ref{subfig:full} shows that, under full information, whether relative performance is considered does not quite affect investors' wealth. 
Figure~\ref{subfig:partial} shows that investors' wealth can fall sharply simultaneously under partial information (the first plot) and the consideration of relative performance further reduces all investors' wealth significantly (the second plot). Under full information, such sharp simultaneous declines of wealth do not exist (the first plot, Figure~\ref{subfig:full}) and considering relative performance has an negligible effect (the second plot, Figure~\ref{subfig:full}). %consistently with our intuition that downward self-reinforcement (caused by relative performance) is amplified under partial information. 
%For the case of an alternating $\mu$, the same phenomenon can be observed in Figure 2.    

It is of interest as future research to investigate if the ``downward amplification'' effect of partial information on investors' wealth can be theoretically proved and characterized, beyond the motivating numerical illustration in Figure~\ref{fig:mu=mu_1_relative}.

%%%%%%%%%%%%%%%%%%%%%%

\subsection{Implications for Systemic Risk}\label{subsec:systemic risk}
As spelled out in Shleifer and Vishny \cite{SV11} and Duarte and Eisenbach \cite{DE21}, systemic risk commonly unfolds as fire-sale spillovers. A distressed financial institution quickly sells an asset at a discount in exchange for cash. Such a {fire sale} slashes the market price of the asset, posing new financial distresses to other slightly healthier financial institutions. This triggers a new wave of fire sales, which further diminishes the asset price and drags down even healthier financial institutions. The procedure continues indefinitely, affecting even institutions considered financially robust. 

In particular, a mutual fund or a hedge fund resorts to fire sales (specifically, quickly selling securities held in its portfolio) when it faces significant capital withdrawals by its shareholders, i.e., a run on the fund; see Coval and Stafford \cite{CS07}. The discussion %``downward self-reinforcement'' effect (
in Section~\ref{subsec:DSR} then suggests that fund managers' consideration of relative performance may exacerbate fire-sale spillovers. Indeed, as trading decisions are practically made under partial information, considering relative performance may markedly reduce fund values due to downward self-reinforcement. %, as explained and numerically illustrated in Section~\ref{subsec:DSR}. 
With lower fund values, an initial fire sale can occur more easily, when shareholders of a fund rush to redeem their investment once they see the fund value fall below a certain threshold. Moreover, the ensuing spillovers are likely more devastating, as other funds are more vulnerable now due to smaller balances. 

A regulator's attempt to intervene can also be made more costly. Suppose that whenever a fund's value falls below a warning threshold (set higher than shareholders' threshold for a run on the fund), the regulator gets prepared to intervene by collecting sufficient monetary support. The third plot in Figure~\ref{subfig:full} shows that, under full information, the empirical warning distribution (i.e., how likely a certain number of funds among $\{X_i\}_{i=1}^{N}$ will trigger a warning by time $T$, computed from 100 simulations of  $\{X_i\}_{i=1}^{N}$) stays unchanged whether fund managers consider relative performance or not. However, in the realistic case of partial information, fund managers' consideration of relative performance significantly impacts the warning distribution (the third plot, Figure~\ref{subfig:partial}): the no-warning probability drops by more than 20\%, while the probability that every fund triggers a warning (and thus requires monetary support) climbs by about 25\%. 

To understand the precise impact on systemic risk, we need to characterize the probability of warning (or default). For $i=1,...,N$, let $x_i\in\R$ be the initial wealth of investor $i$, $D_i<x_i$ be the warning threshold (or the threshold of a run on the fund), and $\tau_i : =\inf\{t\ge 0 : X_i(t) < D_i\}$ be the warning (or default) time. As detailed in Appendix~\ref{sec:default prob.}, while Girsanov's theorem yields an explicit formula for $\PP(\tau_i<T)$ in the case of a constant $\mu$ under full information (as in  Section~\ref{subsec:full constant mu}), the same method does not work under partial information or an alternating $\mu$ (as in Sections~\ref{subsec:partial constant mu}, \ref{subsec:full alternating mu}, and \ref{subsec:partial alternating mu}). This suggests two important directions for future research: (i) the general characterization of $\P(\tau_i<T)$; (ii) solving the mean-variance problem \eqref{MV_model}, with $T$ replaced by the random horizon $\tau_i\wedge T$. The study of (ii) can reveal how investor $i$'s knowledge of a potential default changes her trading behavior in this paper (where we implicitly assume that investors disregard defaults).

%%%%%%%%%%%%%%%%%%%%%
%%%%%%%%%%%%%%%%%%%%%

\appendix
\section{Proofs of Auxiliary Results}
\subsection{Proof of Lemma~\ref{lem:p to 1}}\label{subsec:proof of lem:p to 1}
%\begin{proof}
By Wonham \cite[Section 2]{Wonham1965}, ${\mathfrak p}_j(u)$ can be computed explicitly as %, using the known constants $\mu_1$, $\mu_2$, $\sigma$ and observations of $S$ in \eqref{risky_1}. Specifically,
\be\label{p_j formula}
{\mathfrak p}_j(u)=\frac{{\mathfrak p}_j(t)\exp\left\{\frac{\mu_j-\sigma^2/2}{\sigma^2}\big(\log (S(u)/S(t))\big)-\frac{(\mu_j-\sigma^2/2)^2}{2\sigma^2}(u-t)\right\}}{\sum_{k=1}^2 {\mathfrak p}_k(t)\exp\left\{\frac{\mu_k-\sigma^2/2}{\sigma^2}\big(\log (S(u)/S(t))\big)-\frac{(\mu_k-\sigma^2/2)^2}{2\sigma^2}(u-t)\right\}}.
\en
If $\mu=\mu_1$, then $\log(S(u)/S(t)) = (\mu_1-\sigma^2/2)(u-t)+\sigma (W(u)-W(t))$. Hence, \eqref{p_j formula} becomes
\begin{align}
\nonumber {\mathfrak p}_1(u)&=\frac{{\mathfrak p}_1(t)\exp\left\{\frac{\mu_1-\sigma^2/2}{\sigma^2}\left((\mu_1-\sigma^2/2)(u-t)+\sigma (W(u)-W(t))\right)-\frac{(\mu_1-\sigma^2/2)^2}{2\sigma^2}(u-t)\right\}}{\sum_{k=1}^2 {\mathfrak p}_k(t)\exp\left\{\frac{\mu_k-\sigma^2/2}{\sigma^2}\left((\mu_1-\sigma^2/2)(u-t)+\sigma (W(u)-W(t))\right)-\frac{(\mu_k-\sigma^2/2)^2}{2\sigma^2}(u-t)\right\}}\\ 
%\nonumber &=&\frac{1}{1+\frac{p_2(0)\exp\left\{\frac{\mu_2-\sigma^2/2}{\sigma^2}\bigg((\mu_1-\sigma^2/2)u+\sigma W(u)\bigg)-\frac{\mu_2-\sigma^2/2}{2\sigma^2}u\right\}}{p_1(0)\exp\left\{\frac{\mu_1-\sigma^2/2}{\sigma^2}\bigg((\mu_1-\sigma^2/2)u+\sigma W(u)\bigg)-\frac{\mu_1-\sigma^2/2}{2\sigma^2}u\right\}}}\\
&=\left({1+\frac{{\mathfrak p}_2(t)}{{\mathfrak p}_1(t)}\exp\left\{-\frac{(\mu_1-\mu_2)^2}{2\sigma^2}(u-t)+\frac{\mu_2-\mu_1}{\sigma}(W(u)-W(t))\right\}}\right)^{-1}.\label{p_1 calculation}
\end{align}
As $\exp\big\{-\frac{(\mu_1-\mu_2)^2}{2\sigma^2}(u-t)+\frac{\mu_2-\mu_1}{\sigma}(W(u)-W(t))\big\}\to 0$ as $u\to\infty$, we conclude ${\mathfrak  p}_1(u)\to 1$ as $u\to\infty$. If $\mu=\mu_2$, a calculation similar to \eqref{p_1 calculation} yields 
\ba
\nonumber {\mathfrak p}_1(u)&=&\left({1+\frac{{\mathfrak p}_1(t)}{{\mathfrak p}_2(t)}\exp\left\{\frac{(\mu_1-\mu_2)^2}{2\sigma^2}(u-t)+\frac{\mu_2-\mu_1}{\sigma}(W(u)-W(t))\right\}}\right)^{-1}.
\ea
As $\exp\big\{\frac{(\mu_1-\mu_2)^2}{2\sigma^2}(u-t)+\frac{\mu_2-\mu_1}{\sigma}(W(u)-W(t))\big\}\rightarrow \infty$ as $u\to\infty$, we conclude ${\mathfrak  p}_1(u)\to 0$ as $u\to\infty$. 
%\end{proof}

%%%%%%%%%%%%%%%%%%%%%%

\subsection{Proof of Lemma~\ref{lem:hW}}\label{subsec:proof of lem:hW}
%\begin{proof}
(i) By \cite[Theorem 5.5.15]{Karatzas2000}, there exists a weak solution $P$ to \eqref{P eta} up to a possibly finite explosion time beyond the interval $(0,1)$, which is defined by
\be\label{tau}
\tau := \lim_{n\to\infty} \tau_n,\quad \hbox{with}\ \tau_n:=\inf\{u\ge t : P(u)\notin (1/n,1-1/n)\}\ \ \forall n\in\N.
\ee
%and the solution is unique in the sense of probability law. 
With no drift term in \eqref{P eta}, the corresponding scale function (see e.g., \cite[eqn. (5.5.42)]{Karatzas2000}) simplifies to $p(x) = x-c$ for some $c\in(0,1)$. Hence, the function $v(x)$ in \cite[eqn. (5.5.65)]{Karatzas2000} takes the form 
\[
v(x) = \frac{2\sigma^2}{(\mu_1-\mu_2)^2}\int_c^x \frac{x-y}{y^2(1-y)^2} dy,\quad x\in(0,1). 
\]
For any $x\in(0,c)$, observe that
\begin{align*}
\int_c^x \frac{x-y}{y^2(1-y)^2} dy &= \int_x^c \frac{y-x}{y^2(1-y)^2} dy \ge  \int_x^c \frac{y-x}{y^2} dy = \int_x^c \bigg(\frac{1}{y}-\frac{x}{y^2}\bigg)dy\\
&= \log(c)-\log(x)+x/c-1\to \infty,\quad \hbox{as}\ x\downarrow 0,
\end{align*}
which implies $v(0+)=\infty$. For any $x\in(c,1)$, note that
\begin{align*}
\int_c^x \frac{x-y}{y^2(1-y)^2} dy &\ge \int_c^x \frac{x-y}{(1-y)^2} dy  = \int_c^x \bigg(\frac{1}{1-y}-\frac{1-x}{(1-y)^2}\bigg)dy\\
&= -\log(1-x)+\log(1-c)+\frac{1-x}{1-c}-1\to \infty,\quad \hbox{as}\ x\uparrow 1,  
\end{align*}
which implies $v(1-)=\infty$. Hence, by Feller's test for explosion (see \cite[Proposition 5.5.32]{Karatzas2000} and \cite[Theorem 5.1.5]{Pinsky-book-95}), $\P[\tau=\infty]=1$. That is, $\{P(u)\}_{u\ge t}$ never leaves the interval $(0,1)$ a.s. and is thus a genuine weak solution to \eqref{P eta}. Now, as $p\mapsto p(1-p)$ is locally Lipschitz on $\R$, the uniqueness of strong solutions to \eqref{P eta} holds (\cite[Theorem 5.2.5]{Karatzas2000}), which implies pathwise uniqueness for weak solutions %to \eqref{P eta} 
(\cite[Remark 5.3.3]{Karatzas2000}). As a weak solution exists and pathwise uniquessness holds, \cite[Corollary 5.3.23] {Karatzas2000} asserts that there exists a unique strong solution to \eqref{P eta}.\footnote{As an alternative to the arguments here based on \cite[Chapter 5]{Karatzas2000}, one may apply \cite[Theorem IX.2.1]{RY-book-99} with the choice $f(t,\omega):= \omega_t(1-\omega_t) 1_{(0,1)}(\omega_t)$ for all $t\ge0$ and continuous paths $\omega:[0,\infty)\to \R$.} As argued above, this solution must lie within $(0,1)$ a.s.  

%the weak solution $P$ is in fact the unique strong solution to \eqref{P eta}. 

(ii) Let us write $S$ in \eqref{risky_1} as $S(u) = S(t) e^{Y(u)}$, with $Y(u) := (\mu-\frac{\sigma^2}{2}) (u-t) +\sigma (W(u)-W(t))$ for $u\ge t$. As $Y(u)=\log\big(\frac{S(u)}{S(t)}\big)$ and $\E[\mu\mid\mathcal \cF^S_u] = \mu_1 {\mathfrak p}_1(u)+\mu_2(1-{\mathfrak p}_1(u))$, we deduce from \eqref{Inno_W} that 
\[
\widehat W(u) = \frac1\sigma\bigg[Y(u)-\int_t^u \bigg(\E[\mu\mid\mathcal F^S_v]-\frac{\sigma^2}{2} \bigg) dv\bigg],\quad t\ge 0,
\]
which is a standard Brownian motion w.r.t.\ $\{\mathcal F^S_u\}_{u\ge t}$ thanks to \cite[Lemma 11.3]{LS2013}. By Wonham \cite[eqn. (12)]{Wonham1965}, ${\mathfrak p}_1$ satisfies 
\begin{align}
d{\mathfrak p}_1(u) &= -\frac{1}{\sigma^2}\left((\mu_1-\mu_2){\mathfrak p}_1(u)+ \mu_2-\frac{\sigma^2}{2}\right)(\mu_1-\mu_2)(1-{\mathfrak p}_1(u)){\mathfrak p}_1(u)du \notag\\
&\hspace{2in} + \frac{1}{\sigma^2} (\mu_1-\mu_2)(1-{\mathfrak p}_1(u)){\mathfrak p}_1(u) dY(u).\label{p_1 dynamics}
%\\
%&= -\frac{1}{\sigma^2}\left((\mu_1-\mu_2)\mathfrak p_1(t)+ \mu_2-\frac{\sigma^2}{2}\right)(\mu_1-\mu_2)(1-\mathfrak p_1(t))\mathfrak p_1(t)dt \\
%&\hspace{2in} + \frac{1}{\sigma^2} (\mu_1-\mu_2)(1-\mathfrak p_1(t))\mathfrak p_1(t) 
\end{align}
To simplify this, we apply It\^{o}'s formula to \eqref{Inno_W} and get
\begin{align}
\sigma d \widehat W(u) &= \frac{dS(u)}{S(u)} -\frac{1}{2 S^2(u)} \sigma^2 S^2(u) du - \left((\mu_1-\mu_2){\mathfrak p}_1(u)+\mu_2-\frac{\sigma^2}{2}\right)du. \label{Ito to hW}
%\\
%& = \frac{dS(t)}{S(t)} - \left((\mu_1-\mu_2)p_1(t)+\mu_2\right)dt,
\end{align}
As $Y(u) = \log(S(u)/S(t))$, we have 
\be\label{dY}
dY(u) = \frac{dS(u)}{S(u)} -\frac{1}{2 S^2(u)} \sigma^2 S^2(u) du= \left((\mu_1-\mu_2){\mathfrak p}_1(u)+\mu_2-\frac{\sigma^2}{2}\right)du + \sigma d\widehat W(u),
\ee
where the second equality stems from \eqref{Ito to hW}. Plugging this into \eqref{p_1 dynamics} yields $d{\mathfrak p}_1(u)= \frac{\mu_1-\mu_2}{\sigma}(1-{\mathfrak p}_1(u)){\mathfrak p}_1(u) d\widehat W(u)$. Taking $B=\widehat W$ in (i), we conclude that ${\mathfrak p}_1$ is the unique strong solution to \eqref{P}. As \eqref{Ito to hW} can be rearranged as $dS(u)= \big((\mu_1-\mu_2){\mathfrak p}_1(u)+\mu_2\big) S(u)du+\sigma S(u) d\widehat W(u)$, we obtain \eqref{risky_1'}. 
%\end{proof}

%%%%%%%%%%%%%%%%%%%%%%%

\subsection{Proof of Lemma~\ref{lem:Cauchy}}\label{subsec:proof of lem:Cauchy}
%\begin{proof}[Proof of Lemma~\ref{lem:Cauchy}]
(i) Recall from Remark~\ref{rem:Girsanov} that the measure $\Q$ in \eqref{Q} is equivalent to $\P$ and $W_\Q$ in \eqref{W_Q} is a Brownian motion (adapted to the same filtration $\{\mathcal F_u\}_{t\le u\le T}$) under $\Q$. It follows that the unique strong solution $P$ to \eqref{P'''} (under $\P$) becomes a strong solution to \eqref{P''} (under $\Q$), which satisfies $P(u)\in (0,1)$ for $t\le u\le T$ $\Q$-a.s.\  (as $\Q$ is equivalent to $\P$). Because the coefficients of \eqref{P''}, i.e., $p\mapsto \eta(p)- \beta(p)\big(\frac{\theta(p)-r}{\sigma}\big)$ and $p\mapsto \beta(p)$, are Lipschitz on $(0,1)$, 
conditions (A1) and (A2) in Heath and Schweizer \cite{HS00} are satisfied. As $p\mapsto \theta(p)$ is bounded and locally Lipschitz on (0,1), \cite[Lemma 2]{HS00} asserts that $f(t,p) := \kappa_i \E_\Q^{t,p}\big[\int_t^{T}\big(\frac{\theta(P(u))-r}{\sigma}\big)^2 du\big]$ is bounded and continuous on $[0,T]\times(0,1)$. Moreover, as $p\mapsto \beta(p)^2$ is Lipschitz on $(0,1)$ and bounded away from 0 on $D_n:=(1/n,1-1/n)$ for all $n\in\N$, conditions (A3') in \cite{HS00} is also fulfilled. Hence, by \cite[Theorem 1]{HS00} and the remark below it, $f(t,p)$ is the unique solution in $C^{1,2}([0,T)\times(0,1))$ to the Cauchy problem \eqref{Cauchy}. 

(ii) As the coefficients of the SDE \eqref{zeta} is linear in the state variable, there exists a unique strong solution $\zeta$ to \eqref{zeta} for any $(t,p)\in[0,T)\times (0,1)$. Note that $p\mapsto \eta(p)-\beta(p) (\frac{\theta(p)-r}{\sigma})$ and $p\mapsto\beta(p)$ are Lipschitz on $[0,1]$ and we denote by $L>0$ their Lipschitz constant. By definition, $\Gamma$ and $\Lambda$ are bounded by $L$ on $[0,1]$, such that \eqref{zeta} fulfills \cite[conditions (1.14)-(1.16)]{Pham-book-09}.
%For any $(t,p)\in[0,T)\times (0,1)$, thanks to the boundedness of $\Gamma$ and $\Lambda$, the SDE with random coefficients \eqref{zeta} satisfies \cite[conditions (1.14)-(1.16)]{Pham-book-09}. 
Hence, by \cite[Theorem 1.3.15]{Pham-book-09}, there exists $C_1>0$, depending only on $T$ and $L$, such that %the unique strong solution $\zeta$ satisfies
\be\label{E[sup]}
\E_\Q\bigg[\sup_{t\le u\le T} \zeta(u)^2\bigg]\le C_1 (1+\zeta(t)^2) = 2 C_1. 
\ee

Now, we write $P^{t,p}$ for $P$ to stress the initial condition in \eqref{P''}. By \cite[Theorem 5.3]{Friedman-book-75}, for any $t\le u\le T$, $(P^{t,p+h}(u)-P^{t,p}(u))/h \to \zeta(u)$ in $L^2(\Omega,\cF,\Q)$ as $h\to 0$. 
%Hence, \cite[Theorem 5.1]{Friedman-book-75} asserts that $\zeta$ is the unique strong solution to \eqref{zeta} such that $\E[\int_t^T \zeta(u)^2 du]<\infty$. Now, 
By \eqref{c=E}, %and the definition of $\theta$ in \eqref{theta} that 
\begin{align}
\notag\lim_{h\to 0}&\frac{c(t,p+h)- c(t,p)}{h} = \lim_{h\to 0} \frac{\kappa_i}{h} \E_\Q\bigg[\int_t^T \left(\frac{\theta(P^{t,p+h}(u))-r}{\sigma}\right)^2-\left(\frac{\theta(P^{t,p}(u))-r}{\sigma}\right)^2 du\bigg]\\
&= \frac{\kappa_i}{\sigma^2}(\mu_1-\mu_2) \bigg(\lim_{h\to 0}\int_t^T \E_\Q\bigg[\frac{P^{t,p+h}(u)-P^{t,p}(u)}{h} \notag \\
&\hspace{1.85in}\cdot\left((\mu_1-\mu_2) (P^{t,p+h}(u)+P^{t,p}(u))+2(\mu_2-r)\right) \bigg] du\bigg),\label{c_p'}
\end{align}
where the second equality stems from the definition of $\theta$ in \eqref{theta} and Fubini's theorem. By \cite[Lemma 3.1]{Pham98}, there exists $C_2>0$, depending only on $T$ and $L$, such that $\E_\Q[|P^{t,p+h}(u)-P^{t,p}(u)|^2]\le C_2 h^2$ for all $h>0$ and $u\in [t,T]$. Thus, $\E_\Q[|P^{t,p+h}(u)-P^{t,p}(u)|]\le (\E_\Q[|P^{t,p+h}(u)-P^{t,p}(u)|^2])^{1/2}\le C_2^{1/2} h$  for all $h>0$ and $u\in [t,T]$. This allows the use of the dominated convergence theorem to exchange the limit and the integral in \eqref{c_p'}. By doing so and recalling $(P^{t,p+h}(u)-P^{t,p}(u))/h\to\zeta(u)$ in $L^2(\Omega,\cF,\Q)$ (and thus in $L^1(\Omega,\cF,\Q)$), we conclude from \eqref{c_p'} that 
\begin{align}\label{c_p}
\partial_p c(t,p) &=  \frac{\kappa_i}{\sigma^2} (\mu_1-\mu_2) \int_t^T \E_\Q\Big[\zeta(u) \left((\mu_1-\mu_2) 2 P^{t,p}(u) +2(\mu_2-r)\right)\Big] du\notag\\
&= \frac{2\kappa_i}{\sigma^2} (\mu_1-\mu_2) \int_t^T \E_\Q \Big[\zeta(u) \Big( \theta\left(P^{t,p}(u)\right) -r\Big)\Big] du\notag\\
&= \frac{2\kappa_i}{\sigma^2} (\mu_1-\mu_2) \E_\Q \bigg[\int_t^T \zeta(u) \Big( \theta\left(P^{t,p}(u)\right) -r\Big) du \bigg],
\end{align}
where the last line follows from Fubini's theorem. As $P(u)\in (0,1)$ for $t\le u\le T$ $\Q$-a.s., $\theta\left(P^{t,p}(u)\right)$ is a bounded process by the definition of $\theta$ in \eqref{theta}. Also, in view of \eqref{zeta}, $\zeta(u)$ is a nonnegative process. Hence, we conclude from \eqref{c_p} and \eqref{E[sup]} that $\partial_p c(t,p)$ is bounded on $[0,T]\times (0,1)$. 
%\begin{align}
%\notag\bigg|\lim_{h\to 0}\frac{c(t,p+h)- c(t,p)}{h}\bigg| &\le \lim_{h\to 0} \frac{\kappa_i}{h} \E\bigg[\int_t^T \bigg|\left(\frac{\theta(P^{t,p+h}(u))-r}{\sigma}\right)^2-\left(\frac{\theta(P^{t,p}(u))-r}{\sigma}\right)^2 \bigg|du\bigg]\\
%\notag&\le \kappa_i\lim_{h\to 0}\int_t^T \E\bigg[ \left|\frac{P^{t,p+h}(u)-P^{t,p}(u)}{h} \right|\bigg]du\\
%&\le \kappa_i\lim_{h\to 0}\int_t^T \E\bigg[ \left|\frac{P^{t,p+h}(u)-P^{t,p}(u)}{h} \right|^2 \bigg]du, \label{c_p}
%%&= \kappa_i \int_t^T \E \left[\left|\zeta(u)\right|^2\right] du \le 2 T C_T,
%\end{align}
%where the second inequality follows from the Lipschitz continuity of $p\mapsto (\frac{\theta(p)-r}{\sigma})^2$ on $[0,1]$ and Fubini's theorem. By \cite[Lemma 3.1]{Pham98}, there exists $C_2>0$, depending only on $T$ and $L$, such that $\E[|P^{t,p+h}(u)-P^{t,p}(u)|^2]\le C_2 h^2$ for all $h>0$ and $u\in [t,T]$. This ensures that we can apply the dominated convergence theorem to exchange the limit and the integral in \eqref{c_p}. By doing so and recalling that $(P^{t,p+h}(u)-P^{t,p}(u))/h\to\zeta(u)$ in $L^2(\Omega)$, we conclude from \eqref{c_p} that 
%\[
%\bigg|\lim_{h\to 0}\frac{c(t,p+h)- c(t,p)}{h}\bigg| \le  \kappa_i \int_t^T \E \left[\left|\zeta(u)\right|^2\right] du \le 2\kappa_i T C_1, 
%\]
%where the last inequality stems from \eqref{E[sup]}. That is, we have $|\partial_p c(t,p)|\le 2T C_T$ on $[0,T)\times (0,1)$, as desired.  
%\end{proof} 

\begin{remark}
In the proof of Lemma~\ref{lem:Cauchy} (i) above, instead of using the approximation result in \cite{HS00}, one can construct a different approximation following Chen et al.\ \cite[Lemma 4.1]{CHSZ17}. The latter requires much less regularity than \cite{HS00} and can be useful for more general models. 
%the argument above \eqref{c_n bdd} is motivated by Chen et al.\ \cite[Lemma 4.1]{CHSZ17}, which studies a degenerate parabolic equation associated with an SDE without a drift. 
\end{remark}

%%%%%%%%%%%%%%%%%%%%%%%

\subsection{Proof of Corollary~\ref{lem:Cauchy'}}\label{subsec:proof of lem:Cauchy'}
%\begin{proof}
The desired results follow from the same arguments in the proof of Lemma~\ref{lem:Cauchy} (i) (see Section~\ref{subsec:proof of lem:Cauchy}), with $\eta(p)- \beta(p)\big(\frac{\theta(p)-r}{\sigma}\big)$ and $\kappa_i\big(\frac{\theta(p)-r}{\sigma}\big)^2$ replaced by $\eta(p)$ and $R_i\big(t,p,\partial_p c_1(t,p),\cdots, \partial_p c_N(t,p)\big)$, and the unique strong solution $P$ to \eqref{P''} (under $\Q$) replaced by that to \eqref{P'''} (under $\P$). In particular, for any $i=1,...,N$, as $\partial_p c_i$ is bounded and continuous by Lemma~\ref{lem:Cauchy} (ii), we see that $R_i\big(t,p,\partial_p c_1(t,p),\cdots, \partial_p c_N(t,p)\big)$ by definition is bounded and locally Lipschitz in $p$, uniformly in $t$. Hence, results in \cite{HS00} can be used as in the proof of Lemma~\ref{lem:Cauchy} (i) to get our desired results.
%when $\kappa_i\big(\frac{\theta(p)-r}{\sigma}\big)^2$ and $\E_\Q$ in \eqref{c=c_n} is replaced by $R_i\big(t,p,\partial_p c_1(t,p),\cdots, \partial_p c_N(t,p)\big)$ and $\E$, respecticely, the arguments below \eqref{c=c_n} still hold.
%\end{proof}

%%%%%%%%%%%%%%%%%%%%%%%

\subsection{Proof of Lemma~\ref{lem:tW}}\label{subsec:proof of lem:tW}
%\begin{proof}
(i) By \cite[Theorem 5.5.15]{Karatzas2000}, there exists a weak solution $P$ to \eqref{P eta'} up to a possibly finite explosion time beyond the interval $(0,1)$, defined as in \eqref{tau}. %, and the solution is unique in the sense of probability law. 
In view of the dynamics in \eqref{P eta'}, for any $c\in (0,1)$, the corresponding scale function (see e.g., \cite[eqn. (5.5.42)]{Karatzas2000}) takes the form
\begin{align}
\notag p(x) &= \int_c^x \exp\bigg(\frac{2\sigma^2}{(\mu_1-\mu_2)^2}\int_c^y \frac{(q_1+q_2)z-q_2}{z^2(1-z)^2} dz\bigg) dy\\
&= \int_c^x \exp\bigg(\frac{2\sigma^2}{(\mu_1-\mu_2)^2}\bigg[ \frac{q_2}{y}+\frac{q_1}{1-y}+(q_1-q_2) \log\bigg(\frac{y}{1-y}\bigg) -C \bigg] \bigg) dy,\label{p calc}
\end{align} 
where $C\in\R$ denotes a generic constant that depends only on $q_1$, $q_2$, and $c$ and may change from line to line. 
For $x\in (c,1)$, we deduce from the above and $e^z \ge z$ for all $z\in\R$ that 
\begin{align}\label{p calc'}
\notag p(x) &\ge  \frac{2\sigma^2}{(\mu_1-\mu_2)^2} \int_c^x \bigg[ \frac{q_2}{y}+\frac{q_1}{1-y}+(q_1-q_2) \log\bigg(\frac{y}{1-y}\bigg) -C \bigg] dy\\
&= \frac{2\sigma^2}{(\mu_1-\mu_2)^2} \bigg[ q_2 \log(x)-q_1 \log({1-x}) + (q_1-q_2)(x\log x+(1-x)\log(1-x))-C\bigg]. 
\end{align}
Hence, we get $p(1-)=\infty$. For $x\in (0,c)$, a similar calculation shows that the above inequality ``$\ge$'' turns into ``$\le$'', which then implies $p(0+)=-\infty$. 
%\begin{align*}
%p(x) &\le % \frac{-2\sigma^2}{(\mu_1-\mu_2)^2} \int_x^c \bigg[ \frac{q_2}{y}+\frac{q_1}{1-y}+(q_1-q_2) \log\bigg(\frac{y}{1-y}\bigg) -C \bigg] dy\\%&= 
%\frac{2\sigma^2}{(\mu_1-\mu_2)^2} \bigg[ q_2 \log(x)-q_1 \log({1-x}) + (q_1-q_2)(x\log x+(1-x)\log(1-x))-C\bigg].
%\end{align*}
%Hence, we obtain $p(0+)=-\infty$. 
We therefore conclude from \cite[Problem 5.5.27]{Karatzas2000} and Feller's test for explosion (see e.g., \cite[Theorem 5.5.29]{Karatzas2000}) that $\P[\tau=\infty]=1$. That is, $\{P(u)\}_{u\ge t}$ never leaves the interval $(0,1)$ a.s. and is thus a genuine weak solution to \eqref{P eta}. Now, as the maps $p\mapsto -(q_1+q_2)p$ and $p\mapsto p(1-p)$ are locally Lipschitz, we can argue as in the last five lines of the proof of Lemma~\ref{lem:hW} (i) (see Section~\ref{subsec:proof of lem:Cauchy}) that a unique strong solution to \eqref{P eta} exists and it lies in $(0,1)$ a.s.

%the weak solution $P$ is in fact the unique strong solution to \eqref{P eta'}. 

(ii) Let us write $S$ in \eqref{risky_0} as $S(u) = S(t) e^{Y(u)}$, with $Y(u) := \int_t^u \big(\mu(\alpha(v))-\frac{\sigma^2}{2}\big) dv +\sigma (W(u)-W(t))$ for $u\ge t$. As $Y(u)=\log\big(\frac{S(u)}{S(t)}\big)$ and $\E[\mu(u)\mid\mathcal F^S_u]=\mu_1{\mathfrak p}_1(u)+\mu_2(1-{\mathfrak p}_1(u))$, \eqref{Inno_W} becomes 
\[
\widehat W(u) = \frac1\sigma\bigg[Y(u)-\int_t^u \bigg(\E[\mu(v)\mid\mathcal F^S_v]-\frac{\sigma^2}{2}\bigg) dv\bigg],\quad u\ge t,
\]
which is a standard Brownian motion w.r.t.\ $\{\mathcal F^S_u\}_{u\ge t}$ thanks to \cite[Lemma 11.3]{LS2013}.\footnote{While \cite[Lemma 11.3]{LS2013} is stated for diffusion processes, the arguments in its proof still hold when applied to $\mu(t)$ in \eqref{mu process}, a continuous-time Markov chain.} By Wonham \cite[eqn. (21)]{Wonham1965}, ${\mathfrak p}_1$ satisfies 
\begin{align}
d{\mathfrak p}_1(u) &= \Big(-(q_1+q_2){\mathfrak p}_1(u)+q_2\Big)du -\frac{1}{\sigma^2}\left((\mu_1-\mu_2){\mathfrak p}_1(u)+ \mu_2-\frac{\sigma^2}{2}\right)(\mu_1-\mu_2)(1-{\mathfrak p}_1(u)){\mathfrak p}_1(u)du \notag\\
&\hspace{2in} + \frac{1}{\sigma^2} (\mu_1-\mu_2)(1-{\mathfrak p}_1(u)){\mathfrak p}_1(u) dY(u).\label{p_1 dynamics'}
%\\
%&= -\frac{1}{\sigma^2}\left((\mu_1-\mu_2)\mathfrak p_1(t)+ \mu_2-\frac{\sigma^2}{2}\right)(\mu_1-\mu_2)(1-\mathfrak p_1(t))\mathfrak p_1(t)dt \\
%&\hspace{2in} + \frac{1}{\sigma^2} (\mu_1-\mu_2)(1-\mathfrak p_1(t))\mathfrak p_1(t) 
\end{align}
To simplify this, we apply It\^{o}'s formula to \eqref{Inno_W} and get \eqref{Ito to hW}, so that the dynamics of $Y$ in \eqref{dY} still holds. Plugging this into \eqref{p_1 dynamics'} yields $d{\mathfrak p}_1(u)= \big(-(q_1+q_2){\mathfrak p}_1(u)+q_2\big)du+\frac{\mu_1-\mu_2}{\sigma}(1-{\mathfrak p}_1(u)){\mathfrak p}_1(u) d\widehat W(u)$. By taking $B=\widehat W$ in part (i), we conclude that ${\mathfrak p}_1$ is the unique strong solution to \eqref{P'}. Finally, since rearranging \eqref{Ito to hW} gives $dS(u)= \big((\mu_1-\mu_2)
{\mathfrak p}_1(u)+\mu_2\big) S(u)du+\sigma S(u) d\widehat W(u)$, we obtain \eqref{risky_1''}. 
%\end{proof}

%%%%%%%%%%%%%%%%%%%%%
%%%%%%%%%%%%%%%%%%%%%

\section{Proofs of Theorems~\ref{thm:E mu known}, \ref{thm:E mu unknown}, and \ref{THM_1}}\label{sec:proof of main} % of Theorems~\ref{thm:E mu unknown} and \ref{THM_1}
As we will see in this section, the proofs of Theorems~\ref{thm:E mu known}, \ref{thm:E mu unknown}, and \ref{THM_1} share a similar structure. This is because the state processes involved, i.e., $\bm X$ and $P$, always change continuously as solutions to SDEs. By contrast, the proof of Theorem~\ref{thm:E M known} (postponed to Section~\ref{sec:proof of main'}) requires a different derivation to accommodate the state process $M$, which jumps between two distinct states. 
%we will first solve the two Cauchy problems \eqref{Cauchy}-\eqref{Cauchy'}. The main challenge here is twofold: they are both degenerate and one depends on the other through first derivatives. We will perform elliptic regularization and derive a uniform bound for first derivatives through probabilistic estimates; see Section~\ref{subsec:Cauchy}. Next, 

Section~\ref{subsec:extended HJB} below solves the extended HJB equations for the problem \eqref{MV_model P}, i.e., the coupled system \eqref{V_i}-\eqref{g_i} for $i=1,...,N$. 
%of $2N$ partial differential equations (PDEs)---two PDEs for each investor. 
Based on the semi-explicit form of the solution (in terms of the Cauchy problems \eqref{Cauchy} and \eqref{Cauchy'}), 
%By solving the extended HJB equation semi-explicitly (in terms of the solutions to the Cauchy problems \eqref{Cauchy} and \eqref{Cauchy'}), 
we will construct the Nash equilibria stated in Theorems~\ref{thm:E mu unknown} and \ref{THM_1}; see Sections~\ref{subsec:proof of thm:E mu unknown} and \ref{subsec:proof of THM_1}. Theorem~\ref{thm:E mu known} will follow as a special case; see Section~\ref{subsec:proof of thm:E mu known}. 

%%%%%%%%%%%%%%%%%%%

\subsection{Solving the Extended HJB Equations}\label{subsec:extended HJB}
Let $B$ be a standard Brownian motion. Consider the problem \eqref{MV_model P}, subject to the wealth dynamics 
\be\label{wealth full'''-eta}
 dX_i(u) = rX_i(u)+\pi_i(u)\Big((\mu_1-\mu_2)P(u)+\mu_2-r\Big)du+\pi_i(u)\sigma d B(u),\ \ u\in[t,T],\quad X_i(t)=x_i,
\ee
where $P$ is the unique strong solution to 
\be\label{P'-eta}
dP(u)=\eta\big(P(u)\big)dt +\frac{\mu_1-\mu_2}{\sigma}P(u)(1-P(u))  d B(u),\ \ u\in[t,T],\quad P(t) = p \in (0,1). 
\ee
For the case $\mu_1=\mu_2=\mu$, \eqref{wealth full'''-eta} corresponds to the wealth dynamics \eqref{wealth full} in Section~\ref{subsec:full constant mu}; moreover, as there is no longer $P(u)$-dependence in the wealth dynamics, the problem \eqref{MV_model P} reduces to \eqref{MV_model}. For the case $\mu_1>\mu_2$, by taking $\eta\equiv 0$ in \eqref{P'-eta}, \eqref{wealth full'''-eta} corresponds to the wealth dynamics in Section~\ref{subsec:partial constant mu}; by taking $\eta:[0,1]\to\R$ to be \eqref{eta} in \eqref{P'-eta}, \eqref{wealth full'''-eta}  represents the wealth dynamics in Section~\ref{subsec:partial alternating mu}. 

From the detailed explanations between \eqref{theta} and \eqref{Cauchy}, we know that the extended HJB equations are \eqref{V_i}-\eqref{g_i} for $i=1,...,N$. By plugging the ansatz  \eqref{sol. form} into \eqref{V_i}-\eqref{g_i} and using the notation
\[
{\overline \pi}^*_{(-i)}(u) := \frac{1}{N}\sum_{j=1,...,N,j\neq i} \pi^*_j(u) \qquad \forall i=1,...,N,
\]
we get 
\begin{align}
\nonumber\partial_t A_i x_i&+rA_ix_i+\partial_t B_i \overline x_{(-i)}+rB_i\overline x_{(-i)}+\partial_t C_i+\eta(p)\partial_pC_i+\frac{\beta(p)^2}{2}\partial_{pp}C_i\\
\nonumber &+B_i(\theta(p)-r)\overline \pi^*_{(-i)}-\frac{\gamma_i\sigma^2}{2}\left(\frac{b_i}{N}\right)^2\sum_{j\neq i}\sum_{k\neq i}\pi_j^*\pi^*_k-\frac{\gamma_i\beta(p)}{2}(\partial_pc_i)^2-\gamma\sigma\beta(p)b_i\partial_pc_i\overline \pi^*_{(-i)}\\
&+\sup_{\pi_i}\bigg\{A_i(\theta(p)-r)\pi_i-\frac{\gamma_i\sigma^2}{2}\pi_i^2 a_i^2-\gamma_i\sigma^2a_ib_i\pi_i\overline\pi^*_{(-i)}-\gamma_i\sigma\pi_i\beta(p)a_i\partial_pc_i\bigg\}=0\label{V_i_2}
\end{align}
with the terminal conditions $A_i(T)=1-\frac{\lambda^M_i}{N}$, $B_i(T)=-\lambda^M_i$, and $C_i(T,y)=0$, as well as
\ba
\nonumber &&\partial_t a_i x_i+\partial_t b_i \overline x_{(-i)}+ra_ix_i+a_i(\theta(p)-r)\pi_i^*+rb_i\overline x_{(-i)}+b_i(\theta(p)-r)\overline \pi^*_{(-i)}\\
&&+\partial_t c_i+\eta(p)\partial_pc_i+\frac{\beta(p)^2}{2}\partial_{pp}c_i=0\label{g_i_2}
\ea
with the terminal conditions $a_i(T)=1-{\lambda^V_i}/{N}$, $b_i(T)=-\lambda^V_i$, and $c_i(T,y)=0$. 
For \eqref{V_i_2} to hold for all $\bm x=(x_1,...,x_N)\in\R^N$, it is necessary that the coefficients of $x_i$ and the coefficients of $\overline{x}_{(-i)}$ sum up to 0, respectively. This leads to the ordinary differential equations (ODEs) 
$A_i'(t)+rA_i(t)=0$ with $A_i(T)=1-{\lambda^M_i}/{N}$ and $B_i'(t)+rB_i(t)=0$ with $B_i(T)=-\lambda^M_i$, which admit explicit solutions
\be\label{A_i}
A_i(t)=\left(1-\frac{\lambda^M_i}{N}\right)e^{r(T-t)},\qquad B_i(t)=-\lambda^M_ie^{r(T-t)}.
\en
Similarly, for \eqref{g_i_2} to hold for all $\bm x=(x_1,...,x_N)\in\R^N$, it is necessary that the coefficients of $x_i$ and the coefficients of $\overline{x}_{(-i)}$ sum up to 0, respectively. This leads to the ODEs $a_i'(t)+ra_i(t)=0$ with $a_i(T)=1-{\lambda^V_i}/{N}$ and $b_i'(t)+rb_i(t)=0$ with $b_i(T)=-\lambda^V_i$, which admit explicit solutions
\be\label{a_i}
a_i(t)=\left(1-\frac{\lambda^V_i}{N}\right)e^{r(T-t)},\qquad b_i(t)=-\lambda^V_ie^{r(T-t)}.
\en

By solving for the maximizer of the supremum in \eqref{V_i_2}, we find that a Nash equilibrium $\bm \pi^*=(\pi^*_1,...,\pi^*_N)\in\mathcal A^N$ needs to satisfy% candidate for the optimal strategy for the $i$-th investor
\be\label{pi_can}
\pi_i^*=\frac{A_i}{a_i^2}\frac{\theta(p)-r}{\gamma_i\sigma^2}-\frac{b_i}{a_i}\overline \pi^*_{(-i)}-\frac{\beta(p)}{\sigma}\frac{\partial_pc_i}{a_i},\quad \forall i=1,...,N.
\en
Inserting \eqref{A_i} and \eqref{a_i} into \eqref{pi_can} yields
\begin{align}
\nonumber\pi_i^*&=\left(1-\frac{\lambda^V_i}{N}\right)^{-2} \left(1-\frac{\lambda^M_i}{N}\right)\frac{\theta(p)-r}{\gamma_i\sigma^2}e^{-r(T-t)}\\
\nonumber&\hspace{0.5in}+\left(1-\frac{\lambda^V_i}{N}\right)^{-1}\bigg\{ \lambda^V_i\overline \pi^*_{(-i)}-e^{-r(T-t)}\frac{\beta(p)}{\sigma}\partial_pc_i\bigg\}\\
&=\left(1-\frac{\lambda^V_i}{N}\right)^{-1}\bigg\{\frac{\theta(p)-r}{\sigma^2} \kappa_ie^{-r(T-t)}+\lambda^V_i\overline \pi^*_{(-i)}-e^{-r(T-t)}\frac{\beta(p)}{\sigma}\partial_pc_i\bigg\},
\label{NE} 
\end{align} 
where the second equality follows from the definition of $\kappa_i$ in \eqref{kappa}.  %$ \kappa_i=\frac{1}{\gamma_i}\left(1-\frac{\lambda^V_i}{N}\right)^{-1} \left(1-\frac{\lambda^M_i}{N}\right)$. 
Recall that $\overline \pi^* = \frac{1}{N}\sum_{j=1}^N\pi^*_j = \pi^*_i/N + \frac{1}{N}\sum_{j\neq i}\pi^*_j= \pi_i^*/N + \overline \pi^*_{(-i)}$. It follows that
\begin{align}\label{xi-xbar}
\pi_i^*-\lambda^V_i{\overline\pi^*} &=  \bigg(1-  \frac{\lambda_i^V}{N} \bigg)\pi^*_i- \lambda_i^V \overline \pi^*_{(-i)}=e^{-r(T-t)}\bigg\{ \kappa_i\left(\frac{\theta(p)-r}{\sigma^2}\right)-\frac{\beta(p)}{\sigma}\partial_pc_i\bigg\},
\end{align}
where the second equality follows from \eqref{NE}. Summing this up over all $i=1,...,N$ then gives
\be\label{xbar}
{\overline\pi^*}=(1-\overline\lambda^V)^{-1}e^{-r(T-t)}\bigg\{\overline  \kappa\left(\frac{\theta(p)-r}{\sigma^2}\right)- \frac{\beta(p)}{\sigma}\overline{\partial_pc}\bigg\},
\ee
with $\overline  \kappa$ and $\overline\lambda^V$ defined as in \eqref{kappa bar} %=\frac{1}{N}\sum_{j=1}^N\lambda^V_j$, $\overline  \kappa=\frac{1}{N}\sum_{j=1}^N \kappa_j$, 
and $\overline{\partial_pc} := \frac{1}{N}\sum^N_{j=1}\partial_pc_j$. Plugging this back into \eqref{xi-xbar} leads to the formula of $\pi^*_i$ in \eqref{NE-1}. %Hence, a Nash equilibrium is 
%\[
%\nonumber \pi_i^*=e^{-r(T-t)}\bigg\{\frac{\theta(p)-r}{\sigma^2}\left(\kappa_i+\frac{\lambda^V_i}{1-\overline\lambda^V}\overline  \kappa\right)-\frac{\beta(p)}{\sigma}\left(\partial_pc_i+\frac{\lambda^V_i}{1-\overline\lambda^V}\overline{\partial_pc}\right)\bigg\}
%\]
%for $i=1,\cdots,N$. 

Now, we look back again at \eqref{g_i_2}. With the coefficients of $x_i$ and those of $\overline x_{(-1)}$ both summing up to 0, the remaining terms in \eqref{g_i_2} should also sum up to 0. By plugging \eqref{a_i} into the remaining terms and using the relation \eqref{xi-xbar}, we find that $c_i(t,p)$ needs to solve the Cauchy problem \eqref{Cauchy}, which indeed admits a unique solution in $C^{1,2}([0,T)\times (0,1))$ (by Lemma~\ref{lem:Cauchy}). Similarly, for \eqref{V_i_2}, as the coefficients of $x_i$ and those of $\overline x_{(-1)}$ both sum up to 0, the remaining terms in \eqref{V_i_2} should also sum up to 0. By plugging \eqref{A_i} and \eqref{a_i} into the remaining terms and using the relation \eqref{xi-xbar} and \eqref{xbar}, we find that $C_i(t,p)$ needs to solve the Cauchy problem \eqref{Cauchy'}, which indeed admits a unique solution in $C^{1,2}([0,T)\times (0,1))$ (by Corollary~\ref{lem:Cauchy'}). All the above proves the following.

\begin{proposition}\label{prop:sol. to HJB}
For any Lipschitz $\eta:[0,1]\to\R$, the extended HJB equations \eqref{V_i}-\eqref{g_i} for $i=1,...,N$ has a solution $\{(V_i, g_i)\}_{i=1}^N$ of the form \eqref{sol. form}, where $A_i$, $B_i$, $a_i$, and $b_i$ are given by \eqref{A_i} and \eqref{a_i}, $c_i$ is the unique classical solution to \eqref{Cauchy} in Lemma~\ref{lem:Cauchy}, and $C_i$ is the unique classical solution to \eqref{Cauchy'} in Corollary~\ref{lem:Cauchy'}. Moreover, $\pi^*_i$ in \eqref{NE-1} attains the supremum in \eqref{V_i}. 
\end{proposition}

%%%%%%%%%%%%%%%%%%%%%%%%%%%%%

\subsection{Proof of Theorem~\ref{thm:E mu unknown}}\label{subsec:proof of thm:E mu unknown} %, \ref{thm:E mu unknown}, and \ref{THM_1}
%By the preparations in Sections~\ref{subsec:Cauchy} and \ref{subsec:extended HJB}, we are now ready to prove Theorems~\ref{thm:E mu unknown} and \ref{THM_1}.
%\begin{proof}[Proof of Theorem~\ref{thm:E mu unknown}]%[Proof of Theorem~\ref{THM_1}]
As the wealth dynamics \eqref{wealth full'}, with $P$ therein the unique strong solution to \eqref{P}, corresponds to \eqref{wealth full'''-eta}-\eqref{P'-eta} with $\eta\equiv 0$, we take $\eta\equiv 0$ in Proposition~\ref{prop:sol. to HJB}. This readily gives $\{(V_i, g_i, \pi^*_i)\}_{i=1}^N$ that fulfill the following: for any fixed $i=1,...,N$, with $\{\pi^*_j\}_{j\neq i}$ all given, $(V_i, g_i)$ solves \eqref{V_i}-\eqref{g_i} (with $\eta\equiv 0$) and $\pi^*_i$, given by \eqref{NE-1}, attains the supremum in \eqref{V_i}. To prove that $(\pi^*_1,...,\pi^*_N)$ form a Nash equilibrium per Definition~\ref{def:E'}, we need to show that each $\pi^*_i$ satisfies \eqref{intra E'} and is admissible. Note that with $\{\pi^*_j\}_{j\neq i}$ given, this is essentially the mean-variance problem for one single agent (i.e., investor $i$) and one may use the verification result \cite[Theorem 5.2]{BKM2017} to show that $\pi^*_i$ satisfies \eqref{intra E'}. Indeed, by the construction of $(V_i,g_i,\pi^*_i)$ in Proposition~\ref{prop:sol. to HJB}, we see that (i) $V_i(t,\bm x,p)$ and $g_i(t,\bm x,p)$ belong to $C^{1,\infty,2}([0,\infty)\times\R^N\times (0,1))$ and their first derivatives in $x_i$, $i=1,...,N$, and $p$ are all bounded; (ii) $\pi^*_i$ is also bounded.  With such regularity and boundedness, the variational argument in \cite[pp. 342-343]{BKM2017} can be applied to $J_i(t,\bm x, p, \{\pi^*_j\}_{j\neq i},\pi^*_i)$, the value function of investor $i$ under $\pi_i^*$, and it yields the relation \eqref{intra E'}. 
%applying the same limit taking argument in \cite[pp. 342-343]{BKM2017} to the extended HJB equations \eqref{V_i}-\eqref{g_i}  (with $\pi^*_i$ in place of $\pi_i$, so that the supremum can be dropped) yields the relation \eqref{intra E'}. 
On the other hand, by \eqref{NE-1}, $\pi^*_1(t,p) = e^{-r(t-T)} K(t,p)$ with $K(t,p)$ bounded on $[0,T]\times [0,1]$, as $\theta(p)= p\mu_1+(1-p)\mu_2$, $\beta(p)=p(1-p)$, and $\partial_p c_i(t,p)$ therein are all bounded; recall Lemma~\ref{lem:Cauchy} (ii).  This, along with $P(u)={\mathfrak{p}}_1$ lies in $(0,1)$ a.s.\ and is $\F^S$-adapted (by Lemma~\ref{lem:hW} (ii) and \eqref{p_j}), implies that $\pi^*_i(t,P(t))$ is $\F^S$-progressively measurable and satisfies \eqref{integrability'}, i.e., $\pi^*_i\in\mathcal A_S$. As this holds for all $i=1,...,N$,  $\bm \pi^*=(\pi^*_1,...,\pi^*_N)\in\mathcal A_{S}^N$ is a Nash equilibrium for the $N$-player game \eqref{MV_model P}, subject to the dynamics \eqref{wealth full'} and \eqref{P}.

%%%%%%%%%%%%%%%%%%%%%%%%

\subsection{Proof of Theorem~\ref{THM_1}}\label{subsec:proof of THM_1} %, \ref{thm:E mu unknown}, and \ref{THM_1}
%\begin{proof}[Proof of Theorem~\ref{THM_1}]
As the wealth dynamics \eqref{wealth full'}, with $P$ therein the unique strong solution to \eqref{P'}, corresponds to \eqref{wealth full'''-eta}-\eqref{P'-eta} with $\eta:[0,1]\to\R$ given by \eqref{eta}, we take such an $\eta$ function in Proposition~\ref{prop:sol. to HJB}. The desired results then follow from exactly the same arguments in Section~\ref{subsec:proof of thm:E mu unknown}, %the proof of Theorem~\ref{thm:E mu unknown}, 
with ``$\eta\equiv 0$'' replaced by ``$\eta$ given by \eqref{eta}''. 
%\end{proof}

%%%%%%%%%%%%%%%%%%%%%%%%

\subsection{Proof of Theorem~\ref{thm:E mu known}}\label{subsec:proof of thm:E mu known} %, \ref{thm:E mu unknown}, and \ref{THM_1}
%\begin{proof}[Proof of Theorem~\ref{thm:E mu known}] %\label{subsec:proof of thm:E mu known}
When deriving the extended HJB equations (in line with the explanation between \eqref{theta} and \eqref{Cauchy}), as there is no $p$ dependence in the present case, we obtain a simplified version of \eqref{V_i}-\eqref{g_i}, where $\theta(p)$ is replaced by $\mu\in\R$ and every term with a derivative in $p$ disappears. By plugging the ansatz \eqref{sol. form} (yet without the $p$ variable) into this simplified extended HJB equation, we get $A_i$ and $B_i$ as in \eqref{A_i}, $a_i$ and $b_i$ as in \eqref{a_i}, and 
\ban
c_i(t) =\kappa_i\left(\frac{\mu-r}{\sigma}\right)^2(T-t),\quad C_i(t)=(T-t) N_i.
\ean
with $N_i$ given as in \eqref{N_i const}. This in turn implies that the maximizer of the supremum in the simplified \eqref{V_i} takes the form \eqref{NE_constant}. Relying on the explicit form of $(V_i, g_i, \pi^*_i)$, we may argue as in Section~\ref{subsec:proof of thm:E mu unknown} that the methods for \cite[Theorem 5.2]{BKM2017} can be applied here to show that $\pi^*_i$ satisfies \eqref{intra E}. Also, by \eqref{NE_constant}, $\pi^*_i$ is a constant multiple of $e^{-r(T-t)}$, where the constant depends on $\mu\in\R$ (but not on $\bm X$). It follows that $\pi^*_i(t)$ is $\F^{\mu,S}$-progressively measurable and fulfills \eqref{integrability mu constant known}. Hence, by Remark~\ref{rem:Markov mu constant known}, $\pi^*_i\in\mathcal A_{\mu,S}$. As this holds for all $i=1,...,N$, $\bm \pi^*=(\pi^*_1,...,\pi^*_N)\in\mathcal A_{\mu,S}^N$ is a Nash equilibrium for the $N$-player game \eqref{MV_model}, subject to the wealth dynamics \eqref{wealth full} with $\mu(u)\equiv \mu\in\R$. 
%\end{proof} 
 
 %%%%%%%%%%%%%%%%%%%%%%%%%%
 %%%%%%%%%%%%%%%%%%%%%%%%%%%
 
\section{Proof of Theorem~\ref{thm:E M known}}\label{sec:proof of main'}
In the following, we derive the extended HJB equation following Bj\"ork et al.\ \cite[Section 10]{BKM2016}, but in a regime-switching model similar to \cite{Zhou2003}. 
%We will extend the extended HJB equation in \cite{Zhou2003} for a regime-switching model to a multi-agent setting. 
For the problem \eqref{MV_model Z}, subject to the wealth dynamics \eqref{wealth full} with $\mu(u)$ as in \eqref{mu process}, the extended HJB equation for a Nash equilibrium $\bm \pi^*=(\pi^*_1,...,\pi^*_N)\in\mathcal A_{\mu,S}^N$ and the corresponding equilibrium value functions $\{V_i^{(m)}(t,\bm x)\}_{i=1}^N$, $m=1,2$, %$\{V_i(t,\bm x,m)\}_{i=1}^N$ 
takes the form: for any $i=1,...,N$,  
% and $\{g_i^{(\ell)}(t,\bm x)$ read 
\ba
\nonumber \partial_tV_i^{(1)}&+&\sup_{\pi_i}\bigg\{\sum_{j\neq i}\left(rx_j+(\mu_1-r)\pi^*_j\right)\partial_{x_j}V_i^{(1)}+\left(rx_i+(\mu_1-r)\pi_i\right)\partial_{x_i}V_i^{(1)}\\
\nonumber&+&\frac{\sigma^2}{2}\sum_{j\neq i}\sum_{k\neq i}\pi_j^*\pi^*_k\partial_{x_jx_k}V_i^{(1)}-q_1V_i^{(1)}+q_1V_i^{(2)}\\
\nonumber &+&\frac{\sigma^2}{2}\pi_i^2\partial_{x_ix_i}V_i^{(1)}+{\sigma^2}\pi_i\sum_{j\neq i}\pi_j^*\partial_{x_ix_j}V_i^{(1)}-\frac{\gamma_i\sigma^2}{2}\sum_{j\neq i}\sum_{k\neq i}{\pi_j^*\pi_k^*}\partial_{x_j}g_i^{(1)}\partial_{x_k}g_i^{(1)}\\
 &-&\frac{\gamma_i\sigma^2}{2}\pi_i^2 (\partial_{x_i}g_i^{(1)})^2-\gamma_i\sigma^2\pi_i\sum_{j\neq i}\pi_j^*\partial_{x_i}g_i^{(1)}\partial_{x_j}g_i^{(1)}\bigg\}=0 \label{V^1}
\ea
and 
\ba
\nonumber \partial_tV_i^{(2)}&+&\sup_{\pi_i}\bigg\{\sum_{j\neq i}\left(rx_j+(\mu_2-r)\pi^*_j\right)\partial_{x_j}V_i^{(2)}+\left(rx_i+(\mu_2-r)\pi_i\right)\partial_{x_i}V_i^{(2)}\\
\nonumber&+&\frac{\sigma^2}{2}\sum_{j\neq i}\sum_{k\neq i}\pi_j^*\pi^*_k\partial_{x_jx_k}V_i^{(1)}-q_2V_i^{(2)}+q_2V_i^{(1)}\\
\nonumber &+&\frac{\sigma^2}{2}\pi_i^2\partial_{x_ix_i}V_i^{(1)}+{\sigma^2}\pi_i\sum_{j\neq i}\pi_j^*\partial_{x_ix_j}V_i^{(1)}-\frac{\gamma_i\sigma^2}{2}\sum_{j\neq i}\sum_{k\neq i}{\pi_j^*\pi_k^*}\partial_{x_j}g_i^{(2)}\partial_{x_k}g_i^{(2)}\\
&-&\frac{\gamma_i\sigma^2}{2}\pi_i^2 (\partial_{x_i}g_i^{(2)})^2-\gamma_i\sigma^2\pi_i\sum_{j\neq i}\pi_j^*\partial_{x_i}g_i^{(2)}\partial_{x_j}g_i^{(2)}\bigg\}=0,  \label{V^2}
\ea
with the terminal condition $V_i^{(1)}(T,\bm x)=V_i^{(2)}(T,\bm x)=(1-\frac{\lambda^M_i}{N})x_i-\lambda^M_i\overline x_{(-i)}$, where the functions $g^{(1)}(t,\bm x)$ and $g^{(2)}(t,\bm x)$ satisfy
\begin{align}
&\partial_t g_i^{(1)}+\sum_{j=1}^N\left(rx_j+(\mu_1-r)\pi^*_j\right)\partial_{x_j}g_i^{(1)}+\frac{\sigma^2}{2}\sum_{j=1}^N\sum_{k=i}^N\pi_j^*\pi^*_k\partial_{x_jx_k}g_i^{(1)}-q_1g_i^{(1)}+q_1g_i^{(2)}=0,\label{g^1}\\
&\partial_t g_i^{(2)}+\sum_{j=1}^N\left(rx_j+(\mu_2-r)\pi^*_j\right)\partial_{x_j}g_i^{(2)}+\frac{\sigma^2}{2}\sum_{j=1}^N\sum_{k=i}^N\pi_j^*\pi^*_k\partial_{x_jx_k}g_i^{(2)}-q_2g_i^{(2)}+q_2g_i^{(1)}=0,\label{g^2}
\end{align}
with the terminal condition $g_i^{(1)}(T,\bm x)=g_i^{(2)}(T,\bm x)=(1-\frac{\lambda^V_i}{N})x_i-\lambda^V_i\overline x_{(-i)}$. By plugging the ansatz 
 \begin{equation}\label{sol. form'}
 \begin{split}
V_i^{(m)}(t,\bm x)&=A_i(t)x_i+B_i(t)\overline x_{(-i)}+C_i(t,m)\\
g_i^{(m)}(t,\bm x)&=a_i(t)x_i+b_i(t)\overline x_{(-i)}+c_i(t,m)
\end{split}
\end{equation}
into \eqref{V^1}-\eqref{g^2}, we find that the coefficients of $x_i$, the coefficients of $\overline x_{(-i)}$, and the remaining terms should all sum up to 0. By a detailed calculation, this directly yields $A_i$ and $B_i$ as in \eqref{A_i}, $a_i$ and $b_i$ as in \eqref{a_i}, $C_i(t,1)$ and $C_i(t,2)$ as in \eqref{C_1_M}-\eqref{C_2_M}, as well as
\ba
\nonumber c_i(t,1)&=&\frac{q_2\kappa_i\left(\frac{\mu_1-r}{\sigma}\right)^2+q_1\kappa_i\left(\frac{\mu_2-r}{\sigma}\right)^2}{q_1+q_2}(T-t)\\
\label{c_1_M}&&+\frac{q_1}{(q_1+q_2)^2}\bigg(\kappa_i\left(\frac{\mu_1-r}{\sigma}\right)^2-\kappa_i\left(\frac{\mu_2-r}{\sigma}\right)^2\bigg)\left(1-e^{(q_1+q_2)(T-t)}\right),\\
\nonumber c_i(t,2)&=&\frac{q_2\kappa_i\left(\frac{\mu_1-r}{\sigma}\right)^2+q_1\kappa_i\left(\frac{\mu_2-r}{\sigma}\right)^2}{q_1+q_2}(T-t)\\
\label{c_2_M}&&-\frac{q_2}{(q_1+q_2)^2}\bigg(\kappa_i\left(\frac{\mu_1-r}{\sigma}\right)^2-\kappa_i\left(\frac{\mu_2-r}{\sigma}\right)^2\bigg)\left(1-e^{(q_1+q_2)(T-t)}\right). 
\ea
That is, for all $i=1,...,N$, the extended HJB equation \eqref{V^1}-\eqref{g^2} has a solution $(V^{(m)}_i, g^{(m)}_i)$, $m=1,2$, of the form \eqref{sol. form'}, with $A_i$, $B_i$, $a_i$, $b_i$, $c_i$, and $C_i$ specified above. This in turn implies that the maximizer $\pi^*_i$ of the supremum in \eqref{V^1} (resp.\ \eqref{V^2}) is given by \eqref{NE_Markov} with $m=1$ (resp.\ $m=2$). Using the explicit forms of $V_i^{(m)}$, $g_i^{(m)}$, and $\pi^*_i$, we can follow the arguments in \cite[Theorem 5.2]{BKM2017} to show that $\pi^*_i$ satisfies \eqref{intra E''}. Finally, by \eqref{NE_Markov}, $\pi^*_i(t,m) = e^{-r(T-t)} K(m)$ for some $K(1), K(2)\in\R$. It follows that $\pi^*_i(t, M(t))$ is $\F^{\mu,S}$-progressively measurable and fulfills \eqref{integrability} (recall Remark~\ref{rem:Markov mu alternating known}), i.e., $\pi^*_i\in\mathcal A_{\mu,S}$. As this holds for all $i=1,...,N$, $\bm \pi^*=(\pi^*_1,...,\pi^*_N)\in\mathcal A^N_{\mu,S}$ is a Nash equilibrium for the $N$-player game \eqref{MV_model Z}, subject to the wealth dynamics \eqref{wealth full} with $\mu(u)$ as in \eqref{mu process}.

\section{The Probability of Warning (or Default)}\label{sec:default prob.}
For each $i=1,...,N$, let $x_i\in\R$ be the initial wealth of investor $i$ and $D_i\in\R$ with $D_i<x_i$ be the warning threshold (or the threshold of a run on the fund). Consider the warning (or default) time $\tau_i : =\inf\{t\ge 0 : X_i(t) < D_i\}$. 
For the case of a constant $\mu$ under full information (as in Section~\ref{subsec:full constant mu}), the equilibrium wealth process $X_i$ is given by \eqref{wealth full} with $\pi_i(u)= \pi^*_i(u)$ in \eqref{NE_constant}. If we further assume $r=0$ for simplicity, $\pi^*_i(u)$ simplifies to a constant $\pi^*_i = \frac{\mu}{\sigma^2}\big(\kappa_i+\frac{\lambda^V_i}{1-\overline\lambda^V}\overline  \kappa\big)$. Then, for any $T>0$, by using Girsanov's theorem as in \cite[p. 196]{Karatzas2000}, we obtain 
\[
\PP(\tau_i<T)= \int_0^T\exp\left\{\frac{\mu(D_i-x_i)}{\sigma^2 \pi_i^*}-\frac{1}{2}\left(\frac{\mu}{\sigma}\right)^2u\right\}\frac{|\frac{D_i-x_i}{\sigma \pi^*_i}|}{\sqrt{2\pi u^3}}\exp\bigg\{-\frac{1}{2u}\left(\frac{D_i-x_i}{\sigma \pi^*_i}-\frac{\mu}{\sigma}u\right)^2 \bigg\}du.
\]
Moreover, as $T\rightarrow\infty$, this implies
\begin{equation}
\nonumber\PP(\tau_i<\infty)=\exp\left\{-\frac{2\mu|D_i-x_i|}{\sigma^2\pi^*_i}\right\}=\exp\bigg\{-2 |D_i-x_i|\left(\kappa_i+\frac{\lambda^V_i}{1-\overline\lambda^V}\overline  \kappa\right)^{-1}\bigg\}. 
\end{equation}
In view of \eqref{kappa}, this indicates that $\PP(\tau_i<\infty)$ is smaller under a larger risk aversion coefficient $\gamma_i$, but larger under a larger relative performance coefficient $\lambda_i^V$. The above explicit formulas of $\PP(\tau_i<T)$ and $\PP(\tau_i<\infty)$, however, cannot be easily obtained once we consider partial information or an alternating $\mu$ (as in Sections~\ref{subsec:partial constant mu}, \ref{subsec:full alternating mu}, and \ref{subsec:partial alternating mu}). In these more general cases, the coefficients of the wealth process $X_i$ are no longer constant (even if we assume $r=0$), distinct from the case of Section~\ref{subsec:full constant mu}. By Girsanov's theorem, the original Brownian motion now becomes a Brownian motion plus a time integral of a (non-constant) stochastic process, whose distribution and first passage times do not admit explicit formulas.

\bibliographystyle{siam}
\bibliography{MV}

\end{document}